\newif\ifjpp
\jppfalse

\ifjpp
   \documentclass{jpp}
\else
   \documentclass{article}
   \usepackage[round]{natbib}
   \usepackage{geometry}
   \newcommand{\aff}[1]{\textsuperscript{#1}}
   \newcommand{\email}[1]{#1}
   \newcommand{\corresp}[1]{\thanks{Corresponding author. \email{#1}}}
   \renewcommand{\and}{\\and~}
\fi

\usepackage{graphicx}

\usepackage[utf8]{inputenc}
\usepackage[T1]{fontenc}
\usepackage{amsmath}
\usepackage{amssymb}
\usepackage{color}
\usepackage{hyperref}

\title{On the convergence of bootstrap current to the Shaing-Callen limit in stellarators}

\author{Christopher~G.~Albert\aff{1}   %
   \corresp{\email{albert@tugraz.at}}, %
   Craig~D.~Beidler\aff{2},   %
   Gernot~Kapper\aff{1},      %
   Sergei~V.~Kasilov\aff{1,3} %
\and Winfried~Kernbichler\aff{1}}

\ifjpp
   \shorttitle{On bootstrap current in stellarators}
   \shortauthor{C.~G.~Albert, C.~D.~Beidler, G.~Kapper, S.~V.~Kasilov and W.~Kernbichler}

   \affiliation{\aff{1}Fusion@ÖAW, Institute for Theoretical Physics - Computational Physics, Graz University of Technology, 8010 Graz, Austria
   \aff{2} Max-Planck-Institut für Plasmaphysik, 17489 Greifswald, Germany
   \aff{3}Institute of Plasma Physics, National Science Center ``Kharkov Institute of Physics and Technology'', 61108 Kharkov, Ukraine}
\else
   \date{
      \aff{1}Fusion@ÖAW, Institute for Theoretical Physics - Computational Physics, Graz University of Technology, 8010 Graz, Austria\\[1ex]
      \aff{2} Max-Planck-Institut für Plasmaphysik, 17489 Greifswald, Germany\\[1ex]
      \aff{3}Institute of Plasma Physics, National Science Center ``Kharkov Institute of Physics and Technology'', 61108 Kharkov, Ukraine
   }
\fi

\newcommand{\be}[1]{\begin{equation} \label{#1}}
\newcommand{\ee}{\end{equation}}
\newcommand{\bea}[1]{\begin{eqnarray} \label{#1}}
\newcommand{\eea}{\end{eqnarray}}

\newcommand{\eq}[1]{(\ref{#1})}
\newcommand{\difp}[2]{\frac{\partial #1}{\partial #2}}
\newcommand{\bE}{{\bf E}}
\newcommand{\bB}{{\bf B}}

\newcommand{\bv}{{\bf v}}

\newcommand{\rd}{{\rm d}}
\newcommand{\cV}{{\cal V}}
\newcommand{\cI}{{\cal I}}

\newcommand{\bh}{{\bf h}}
\newcommand{\nuast}{\nu_\ast}

\begin{document}

\maketitle

\begin{abstract}

Bootstrap current in stellarators can be presented as a sum of a collisionless value
given by the Shaing-Callen asymptotic formula and an off-set current, which non-trivially
depends on plasma collisionality and radial electric field.
Using NEO-2 modelling, analytical estimates and semi-analytical studies with help of a propagator method,
it is shown that the off-set current
in the $1/\nu$ regime does not converge
with decreasing collisionality $\nuast$ but rather shows oscillations over
$\log\nuast$ with an amplitude of the order of the bootstrap current in an equivalent tokamak.
The convergence to the Shaing-Callen limit appears in regimes with significant orbit precession,
in particular, due to a finite radial electric field, where the off-set current
decreases as $\nuast^{3/5}$.
The off-set current strongly increases in case of nearly aligned magnetic field maxima
on the field line where it diverges as $\nuast^{-1/2}$ in the $1/\nu$ regime and saturates
due to the precession at a level
exceeding the equivalent tokamak value by ${v_E^\ast}^{-1/2}$ where $v_E^\ast$ is the perpendicular
Mach number.  The latter off-set, however, can be minimized
by further aligning local magnetic field maxima and by fulfilling
an extra integral condition of ``equivalent ripples'' for the magnetic field.
A criterion for the accuracy of this alignment and of ripple equivalence is derived.
In addition, the possibility of the bootstrap effect at the magnetic axis caused by the
above off-set is also discussed.
%

\end{abstract}

\section{Introduction}
\label{sec:intro}

One relatively simple way to evaluate bootstrap current in stellarators is to use the long mean free path
asymptotic formula of~\cite{shaing83-3315} which contains all the information
about device geometry in a geometrical factor independent of plasma parameters. This way is especially
suited for stellarator optimization~\citep{beidler90-148,nakajima89-605}
where multiple fast estimates of bootstrap current are required.
Despite that it has been derived more than 40 years ago, the validity range of this formula still remains unclear.
Although the same result is exactly~\citep{helander11-092505} or approximately~\citep{boozer90-2408} reproduced by different
derivations, however, it is not reproduced in the $1/\nu$ regime by numerical
modelling~\citep{beidler2011-076001}. Namely, with decreasing normalized collisionality $\nuast$,
the bootstrap coefficient
does not come to a saturation at the asymptotic limit but rather shows a complicated behavior which
is quite different at magnetic surfaces with different radii~\citep{kernbichler16-104001}. In turn, in the presence
of radial electric field, numerical modelling shows a saturation to the Shaing-Callen limit with decreasing $\nuast$.
This feature of the bootstrap coefficient regularly observed in DKES~\citep{hirshman1986-2951,vanrij89-563} modelling, and
later confirmed by various codes of different types~\citep{beidler2011-076001} has been pointed out to the authors
of this paper by Henning~\cite{maasberg_private} more than 20 years ago, and, together with the problem of bootstrap
resonances~\citep{boozer90-2408}, it was the main reason to start the development of the drift kinetic equation
solver NEO-2~\citep{kernbichler2008-S1061,kasilov14-092506,martitsch16-074007,kernbichler16-104001,kapper16-112511,kapper18-122509}.
Finally, we can present here the results of this long going effort.

Some of the reasons for the anomalous behavior of the bootstrap coefficient
described above have been identified recently~\citep{beidler20} by using the
General Solution of the Ripple-Averaged Kinetic Equation, GSRAKE~\citep{beidler95-463},
to determine the Ware pinch coefficient, which is
equivalent to the bootstrap coefficient due to Onsager symmetry.  Although
instructive, these results are largely of a qualitative nature, and one of
the main purposes of this paper is to treat this problem in a more analytical
manner, with subsequent verification by numerical modelling using NEO-2 and
a simplified propagator method in order to provide simple scalings and
certain conditions useful for stellarator optimization.  The challenges
which such an endeavor must face will also be illustrated by considering
the extreme example of a so-called {\em anti-sigma} configuration --- the
antithesis of the model field considered in~\citep{mynick82-322}  ---
for which the bootstrap coefficient obviously diverges with decreasing
collisionality over the entire $\nu_*$ range of DKES computations.  A
second purpose of this paper is to outline a simple numerical approach,
utilizing the computations of the bootstrap coefficient in the $1/\nu$
regime, to also allow computations in regimes where particle precession
(in particular, due to the radial electric field) is important, thereby
providing an effective tool for the optimization problem mentioned above.

As shown below, for the ``anomalous'' behaviour of the bootstrap coefficient, 
the interaction between boundary layers separating co- and counter-passing particles from
trapped particles and the layers separating different trapped particle classes from each other
plays an important role. In collisionless asymptotic theories, these layers
are assumed infinitely thin and non-overlapping. This cannot
be fulfilled at irrational flux surfaces, where the number of trapped particle classes is
infinite, while the width of the layers is finite at any collisionality. Another reason is the
interaction of the trapped-passing boundary layer with itself, which happens in case of
anti-sigma configurations where the magnetic field maximum at a given flux surface is reached on a line
which splits the field line into non-equivalent segments exchanging transient particles via the boundary layer.
In both cases, an important prerequisite for anomaly is a finite bounce-averaged cross-surface drift of trapped particles,
which is absent in axisymmetric and nearly absent in sufficiently accurate quasi-symmetric configurations showing no anomalies of
the bootstrap current~\citep{landreman22-082501}.

We restrict our analysis here to the mono-energetic approach~\citep{beidler2011-076001} employing the Lorentz collision model,
since this approach is sufficient for the account of main effects related to the magnetic field geometry.
Effects of energy and momentum conservation pertinent to the full linearized collision model can be recovered then
with good accuracy from the mono-energetic solutions of the kinetic equation using various
momentum correction techniques~\citep{taguchi92-3638,sugama02-4637,sugama08-042502,massberg09-072504}.
This mono-energetic approach is briefly outlined in Section~\ref{ssec:mono} where also the basic notation is introduced.
In Section~\ref{ssec:colassol} we re-derive the Shaing-Callen formula~\citep{shaing83-3315}
within this approach in the $1/\nu$ transport regime
in order to obtain an explicit solution for trapped particle distribution accounting for various trapped particle classes.
The structure of this solution is quite demonstrative for the reasons behind the ``anomalous'' behavior of bootstrap
current mentioned above. In that section, we also extend the alternative derivation of \cite{helander11-092505}
within the adjoint approach to the next order in collisionality. Although the obtained correction has no effect
on the resulting Ware pinch (bootstrap) coefficient, it is useful for understanding a certain paradox contained in this
solution. Collisionless asymptotic solutions are compared to numerical solutions for finite plasma collisionality
in Section~\ref{ssec:fincolnum} where the cases with convergence of these solutions to the Shaing-Callen limit
at low collisionality are demonstrated. In Section~\ref{sec:bootware-closed} we examine the effect of collisional boundary
layers on the distribution function in the adjoint (Ware pinch) problem where they lead to the off-set of this
function from the value of collisionless asymptotic of \cite{helander11-092505} and respective off-set of the Ware-pinch
coefficient. In particular, in Section~\ref{ssec:propmet} we introduce
a simplified approach (propagator method) to describe this off-set in the leading order over collisionality with help of
a set of Wiener-Hopf type integral equations. This set is infinite in case of irrational field lines, and becomes
finite for closed field lines.
In Section~\ref{ssec:align}, with the help of this set, we derive
the conditions on the equilibrium magnetic field required to avoid the leading order off-set.
We obtain a simple expression for the distribution function off-set in case these conditions
are weakly violated in Section~\ref{ssec:imperfect} where we express these solutions via two discrete
functions tabulated using the numerical solutions of two respective infinite integral equation sets
resulting from the linearization of the original Wiener-Hopf type set.
In Section~\ref{sec:irr} we examine the off-set of the distribution function and Ware pinch coefficient
at irrational flux surfaces both, numerically, with help of NEO-2 code solutions at high order rational magnetic field
lines approximating the irrational surface (Section~\ref{ssec:lonfieldlines}), and using the analytical estimates
of the asymptotic behavior of the off-set with
decreasing plasma collisionality (Section~\ref{ssec:asymptotic}).
A related issue of bootstrap resonances is briefly discussed in Section~\ref{ssec:bootres}.
In Section~\ref{sec:precession} we study the effect of banana orbit precession (in particular, due to a finite radial
electric field) on the off-set of Ware pinch coefficient and formulate there a simple bounce averaged approach
for the account of this effect in computations of Ware pinch coefficient using the numerical solutions
for the distribution function in the $1/\nu$ regime. A qualitative discussion of the
off-set in the direct problem describing bootstrap coefficient is presented in Section~\ref{sec:boot_axis} where we also
examine the possibility of bootstrap effect at the magnetic axis. Finally, the results are summarized in
Section~\ref{sec:conclusion} where some implications for reactor optimization are also discussed.

\section{Asymptotical models and finite collisionality}
\label{sec:asfin}

In this section, we review asymptotical long mean free path models of bootstrap and Ware pinch effect
in the set-up where all explicit and implicit assumptions used in derivations of those models are fulfilled.
We use a standard method~\citep{galeev73,hinton76-239,helander02} to re-derive transport coefficients in both cases, with main focus on boundary
conditions in the presence of multiple trapped particle classes. We verify these models by numerical
computation, identify their applicability range and mechanisms responsible for discrepancies at
finite collisionality.

\subsection{Adjoint mono-energetic problems on a closed field line}
\label{ssec:mono}

For the present analysis of bootstrap current convergence with plasma collisionality, a mono-energetic approach
is sufficient where a Lorentz collision model and constant electrostatic potential within flux surfaces are assumed.
The linear deviation of the distribution function $f$ from the local Maxwellian $f_M$ is presented as a
superposition of thermodynamic forces $A_k$ as detailed in~\citep{kernbichler16-104001},
\be{superforces}
f-f_M = f_M \sum\limits_{k=1}^3 g_{(k)} A_k,
\ee
where
\bea{termforces}
A_1 = \frac{1}{n_\alpha}\difp{n_\alpha}{r}
-\frac{e_\alpha E_r}{T_\alpha}-\frac{3}{2T_\alpha}\difp{T_\alpha}{r},
\quad
A_2 = \frac{1}{T_\alpha}\difp{T_\alpha}{r},
\quad
A_3 = \frac{e_\alpha \langle E_\parallel B\rangle}{T_\alpha\langle B^2\rangle},
\eea
$e_\alpha$, $m_\alpha$, $n_\alpha$ and $T_\alpha$ are $\alpha$ species charge, mass,
density and temperature, respectively,
$E_r$, $E_\parallel$ are radial (electrostatic) and parallel (inductive) electric field, $B$ is the magnetic field
strength, and $\langle\dots\rangle$ denotes a neoclassical flux surface average. This reduces
the linearized drift kinetic equation in the $1/\nu$ regime to a set of independent equations,
\be{mono_Ak_norm}
\hat L g_{(k)} \equiv
\sigma \difp{g_{(k)}}{\varphi}
-\difp{}{\eta}\left(D_\eta\difp{g_{(k)}}{\eta}\right)=s_{(k)},
\qquad
D_\eta \equiv \frac{|\lambda|\eta}{l_c B^\varphi},
\ee
which differ only by source terms,
\bea{s_k}
s_{(1)} &=& - \frac{B v_g^r}{|v_\parallel| B^\varphi}
=
\difp{H_\varphi^\prime}{\eta},
\qquad
s_{(2)} = z s_{(1)},
\qquad
s_{(3)} = \frac{\sigma B^2}{B^\varphi},
\eea
where
\be{Hphiprime}
H_\varphi^\prime = \frac{|\lambda|}{3 B^\varphi}\left(3 +\lambda^2\right)|\nabla r|k_G \rho_L
\ee
will be integrated along the field line later to give bounce integrals $H_j$ in Eq.~\eq{bas1_defs}.
Here, we use a field aligned coordinate system $(r,\vartheta_0,\varphi)$ where
$r$ is a flux surface label (effective radius)
fixed by the condition $\langle |\nabla r| \rangle = 1$,
$\vartheta_0=\vartheta - \iota \varphi$ is a field line label, $\iota$ is the rotational transform
and $\vartheta$ and $\varphi$ are poloidal and toroidal angle of periodic
Boozer coordinates~\citep{boozer81-1999,dhaeseleer91}, respectively.
Variables in the velocity space are parallel velocity sign $\sigma=v_\parallel /|v_\parallel|$
and two invariants of motion, $z=m_\alpha v^2 /(2T_\alpha)$
and $\eta=v_\perp^2 /(v^2 B)$ respectively being the normalized kinetic energy (playing a role of parameter)
and perpendicular adiabatic invariant (magnetic moment).
The other notation is pitch parameter $\lambda=v_\parallel / v=\sigma \sqrt{1-\eta B}$,
the mean free path $l_c= v/(2 \nu_\perp)$ defined via deflection frequency $\nu_\perp$ (the same as $\nu$
in Ref.~\citep{beidler2011-076001}),
radial guiding center velocity
$v_g^r=\bv_g \cdot \nabla r$, Larmor radius $\rho_L=c m_\alpha v/(e_\alpha B)$
and the geodesic curvature given by
\be{geodcu}
|\nabla r|k_G = \left(\bh \times (\bh\cdot\nabla)\bh \right) \cdot \nabla r
= \frac{1}{B}\left(\bh\times\nabla B\right)\cdot\nabla r
=
\frac{\rd r}{\rd \psi}
\left(\frac{B_\vartheta}{B_\varphi+\iota B_\vartheta}\difp{B}{\varphi}-\difp{B}{\vartheta_0}\right)
\ee
with $\bh = \bB/B$ and $\psi$ being the toroidal flux normalized by $2\pi$,
and $B^\varphi$, $B_\varphi$ and $B_\vartheta$ being contra- and covariant components
of the magnetic field in Boozer coordinates.

Neoclassical transport coefficients $D_{jk}$ link thermodynamic forces $A_k$ by
\be{neocefs}
\cI_j = - n_\alpha \sum\limits_{k=1}^3 D_{jk}A_k
\ee
to thermodynamic fluxes $\cI_j$ defined via particle, $\Gamma_\alpha$, and energy, $Q_\alpha$, flux density and
parallel flow velocity $V_{\parallel\alpha}$ as follows,
\be{fluxes_def}
\cI_1=\Gamma_\alpha,
\qquad
\cI_2=\frac{Q_\alpha}{T_\alpha},
\qquad
\cI_3=n_\alpha\left\langle V_{\parallel\alpha} B\right\rangle.
\ee
These transport coefficients are obtained by energy convolution of mono-energetic coefficients~\citep{beidler2011-076001}
$\bar D_{jk}$ with a local Maxwellian,
\be{enconv}
D_{jk} = \frac{2}{\sqrt{\pi}}\int\limits_0^\infty\rd z \sqrt{z}{\rm e}^{-z} \bar D_{jk}.
\ee
Presenting neoclassical flux surface averages in the form of field line averages explicitly given in field
aligned variables by
\be{neo_mflav}
\langle a \rangle = \lim\limits_{\varphi_N\rightarrow \infty}
\left(\int\limits_{\varphi_0}^{\varphi_N}\frac{\rd \varphi}{B^\varphi}\right)^{-1}
\int\limits_{\varphi_0}^{\varphi_N}\frac{\rd \varphi}{B^\varphi} a,
\ee
the mono-energetic coefficients are given by
\be{Dmono}
\bar D_{jk}
=v \lim\limits_{\varphi_N\rightarrow \infty}
\left(\int\limits_{\varphi_0}^{\varphi_N}\frac{\rd \varphi}{B^\varphi}\right)^{-1}
\frac{1}{4}
\sum_{\sigma=\pm 1}\int\limits_{\varphi_0}^{\varphi_N}\rd \varphi \int\limits_0^{1/B}\rd \eta\; s_{(j)}^\dagger g_{(k)}
\equiv
v \lim\limits_{\varphi_N\rightarrow \infty}
\int\limits_{\Omega_N} \rd\Omega\; s_{(j)}^\dagger g_{(k)},
\ee
where
\be{adj}
a^\dagger(\varphi,\eta,\sigma) = a(\varphi,\eta,-\sigma).
\ee
Definitions of thermodynamic forces and fluxes and, respectively, of diffusion coefficients~\eq{enconv}
here are the same as in~\citep{kernbichler16-104001} and coincide with those of~\citep{beidler2011-076001} for the reference field $B_0=1$ except for the sign of $A_3$.
The sign convention used here results in a simple Onsager symmetry for all transport coefficients,
$D_{jk}=D_{kj}$, but negative coefficient $D_{33}$ corresponding to plasma conductivity.

For the present analysis, we solve the kinetic equation on the rational surface using a ``representative''
field line where necessary conditions valid for the irrational flux surface are fulfilled. The flux surface
average~\eq{neo_mflav} at the irrational surface corresponds then to the limit of the series of
representative field lines closed after $N$ turns at respective rational surfaces, $r=r_N=M/N$,
which converge to the irrational surface, $\lim\limits_{N \rightarrow \infty}\iota(r_N) = \iota(r)$.
The requested condition is Liouville's theorem, which states that the integral over any closed surface
of the normal component of the guiding center velocity multiplied by the phase space Jacobian is
zero for fixed total energy and perpendicular adiabatic invariant used as phase space variables.
For the magnetic surface with constant electrostatic potential, this means
$\left\langle B v_g^r / v_\parallel \right\rangle=0$, where the surface integration
is performed over regions where $v_\parallel^2 = v^2 (1-\eta B)\ge 0$ keeping invariants $z$ and $\eta$
constant. The field line average form~\eq{neo_mflav} of Liouville's theorem results in
\be{liouvpass}
\lim\limits_{\varphi_N\rightarrow \infty}
\left(\int\limits_{\varphi_0}^{\varphi_N}\frac{\rd \varphi}{B^\varphi}\right)^{-1}
\int\limits_{\varphi_0}^{\varphi_N}\frac{\rd \varphi B v_g^r}{B^\varphi v_\parallel}=0
\ee
for passing particles and in
\be{liouvtrap}
\lim\limits_{\varphi_N\rightarrow \infty}
\left(\int\limits_{\varphi_0}^{\varphi_N}\frac{\rd \varphi}{B^\varphi}\right)^{-1}
\sum\limits_{j=1}^{j_{\rm max}}\int\limits_{\varphi_j^-}^{\varphi_j^+}\frac{\rd \varphi B v_g^r}{B^\varphi v_\parallel}=0
\ee
for trapped particles where $\varphi_j^-(\eta)$ and $\varphi_j^+(\eta)$ are left and right turning points being
solutions to $\eta B(\varphi^\pm_j)=1$,
index $j$ enumerates local magnetic field maxima $B(\varphi_j)$
fulfilling $B(\varphi_j)\eta>1$ so that turning points are
contained between maximum points $\varphi_{j}$ and $\varphi_{j+1}$.
Upper summation limit $j_{\max}=j_{\max}(\eta,N)$ is the total number of such
maxima between $\varphi_0$ and $\varphi_N$.
For the ``representative'' closed field line with $\varphi_0$ at the largest maximum
we require that conditions~\eq{liouvpass}
and~\eq{liouvtrap} are fulfilled exactly for finite $\varphi_N$, i.e.
\be{liouv_reprp}
\int\limits_{\varphi_0}^{\varphi_N} \rd \varphi\; s_{(1)}=0,
\qquad
\sum\limits_{j=1}^{j_{\rm max}}\int\limits_{\varphi_j^-}^{\varphi_j^+} \rd \varphi\; s_{(1)}=0
\ee
for passing and trapped particles, respectively (see~\eq{s_k}). We will call these conditions ``quasi-Liouville's theorem''.
In the devices with stellarator symmetry
conditions~\eq{liouv_reprp} are satisfied for closed field lines passing through the magnetic field
symmetry point $\varphi_\text{sts}$ what is obvious
due to anti-symmetry of the geodesic curvature (and, respectively, of $v_g^r$ and $s_{(1)}$) with respect to
this point, $s_{(1)}(2\varphi_\text{sts}-\varphi)=-s_{(1)}(\varphi)$.
Restricting our analysis to such devices with a single global maximum per field period, the reference field
line is then the one starting from the global maximum, which must be located at one of (at least two possible)
symmetry points $\varphi_0 \in \varphi_\text{sts}$.

Representing flux surface averages~\eq{neo_mflav} and~\eq{Dmono} by the same expressions without
the limit $\varphi_N\rightarrow \infty$, one can check that so defined mono-energetic coefficients stay
Onsager-symmetric. Namely, replacing source terms in~\eq{Dmono}
via equation~\eq{mono_Ak_norm}, integrating by parts and using the periodicity of the distribution
in the passing region, $g(\varphi_0,\eta,\sigma)=g(\varphi_N,\eta,\sigma)$ and its continuity
at the turning points in the trapped region, $g(\varphi_j^\pm,\eta,1)=g(\varphi_j^\pm,\eta,-1)$,
we get
\bea{Dadj}
\bar D_{jk}
=
\int\limits_{\Omega_N} \rd \Omega\; g_{(k)} \left(\hat L g_{(j)})\right)^\dagger
=
\int\limits_{\Omega_N} \rd \Omega\; g^\dagger_{(j)} \hat L g_{(k)}
=
\int\limits_{\Omega_N} \rd \Omega\; g_{(j)}^\dagger s_{(k)}
=
\bar D_{kj}.
\eea
Thus, we can either compute the bootstrap coefficient $\bar D_{31}$ solving the direct problem driven by $s_{(1)}$ or use its
equality to the Ware pinch coefficient $\bar D_{13}$ resulting from the adjoint problem driven by $s_{(3)}$.

\subsection{Collisionless asymptotic solutions}
\label{ssec:colassol}

Omitting the drive index $(k)$,
asymptotic solutions of Eqs.~\eq{mono_Ak_norm} in the long mean free path limit $l_c \rightarrow \infty$
follow from the standard procedure where the normalized distribution function is looked for in the form
of the series expansion in $l_c^{-1}$,
\be{solform}
g(\varphi,\eta,\sigma) = g_{-1}(\eta,\sigma)+g_0(\varphi,\eta,\sigma)+g_1(\varphi,\eta,\sigma)+\dots
\ee
where the leading order term $g_{-1}$ is independent of $\varphi$, and corrections satisfy
\be{corrections_sat}
\sigma \difp{g_n}{\varphi}
=\difp{}{\eta}\left(D_\eta\difp{g_{n-1}}{\eta}\right)+\delta_{n0}s,
\qquad n \ge 0.
\ee
Eq.~\eq{corrections_sat} is integrated to
\be{gn_eq_int}
g_n(\varphi,\eta,\sigma)
=\sigma \difp{}{\eta}
\left(\int\limits_{\varphi_{\text{beg}}}^\varphi \rd \varphi^\prime D_\eta\difp{g_{n-1}}{\eta}\right)
+\sigma \delta_{n0} \int\limits_{\varphi_{\text{beg}}}^\varphi \rd \varphi^\prime s+\bar g_n(\eta,\sigma),
\ee
where $\varphi_{\text{beg}}=\varphi_j^-(\eta)$ for trapped particles and $\varphi_{\text{beg}}=\varphi_0$ for passing,
and we require that each of $g_n$ is continuous at the periodic boundary and at the turning points.
Continuity at $\varphi=\varphi_j^-$ means that $\bar g_n$ is an even function of $\sigma$ in the trapped region,
$\bar g_n=\bar g_n(\eta)$, while continuity at the periodic boundary in the passing region
and at $\varphi=\varphi_j^+$ in the trapped region results in an equation for the integration constant
$\bar g_{n-1}$ (solubility constraint for $g_n$). The leading order constraint for $g_0$ gives a bounce-averaged
equation for $g_{-1}$
\be{ba_eq}
\sum\limits_{\sigma=\pm 1}^{\text{trapped}}\left(\difp{}{\eta}
\left(
\difp{g_{-1}}{\eta}
\int\limits_{\varphi_{\text{beg}}}^{\varphi_{\text{end}}} \rd \varphi^\prime D_\eta
\right)
+\int\limits_{\varphi_{\text{beg}}}^{\varphi_{\text{end}}} \rd \varphi^\prime s\right)=0,
\ee
where $\varphi_{\text{end}}=\varphi_j^+(\eta)$ for trapped particles and $\varphi_\text{end}=\varphi_N$ for passing,
and the sum $\sum\limits_{\sigma=\pm 1}^{\text{trapped}}$ is taken for trapped particles only. Higher order
constraints result in equations for integration constants $\bar g_n$,
\be{bargn_eq}
\sum\limits_{\sigma=\pm 1}^{\text{trapped}}
\difp{}{\eta}
\left(
\difp{\bar g_n}{\eta}
\int\limits_{\varphi_{\text{beg}}}^{\varphi_{\text{end}}} \rd \varphi^\prime D_\eta
+
\int\limits_{\varphi_{\text{beg}}}^{\varphi_{\text{end}}} \rd \varphi^\prime D_\eta\difp{(g_{n}-\bar g_n)}{\eta}\right) = 0,
\qquad n \ge 0,
\ee
where $g_n-\bar g_n$ is determined by $g_{n-1}$ via~\eq{gn_eq_int}.

Since boundary conditions for the collisional flux in velocity space restrict only the whole solution~\eq{solform},
we have a freedom for setting boundary conditions for individual expansion terms. Thus, we can require that
this flux is produced by $g_{-1}$ only,
\be{moveflux}
\sum\limits_{\sigma=\pm 1}^{\text{trapped}}\int\limits_{\varphi_{\text{beg}}}^{\varphi_{\text{end}}} \rd \varphi^\prime D_\eta\difp{g}{\eta}
=
\sum\limits_{\sigma=\pm 1}^{\text{trapped}}\difp{g_{-1}}{\eta}
\int\limits_{\varphi_{\text{beg}}}^{\varphi_{\text{end}}} \rd \varphi^\prime D_\eta,
\ee
while the corrections carry no flux so that Eq.~\eq{bargn_eq} is integrated to
\be{int_bargn_eq}
\sum\limits_{\sigma=\pm 1}^{\text{trapped}}
\left(
\difp{\bar g_n}{\eta}
\int\limits_{\varphi_{\text{beg}}}^{\varphi_{\text{end}}} \rd \varphi^\prime D_\eta
+
\int\limits_{\varphi_{\text{beg}}}^{\varphi_{\text{end}}} \rd \varphi^\prime D_\eta\difp{(g_{n}-\bar g_n)}{\eta}\right) = 0,
\qquad n \ge 0.
\ee

We apply this ansatz separately to the direct problem driven by source $s_{(1)}$ and to the adjoint
problem driven by source $s_{(3)}$.

\subsubsection{Direct problem}
\label{sssec:dirprob}

\begin{figure}
\centerline{
\includegraphics[width=0.49\textwidth]{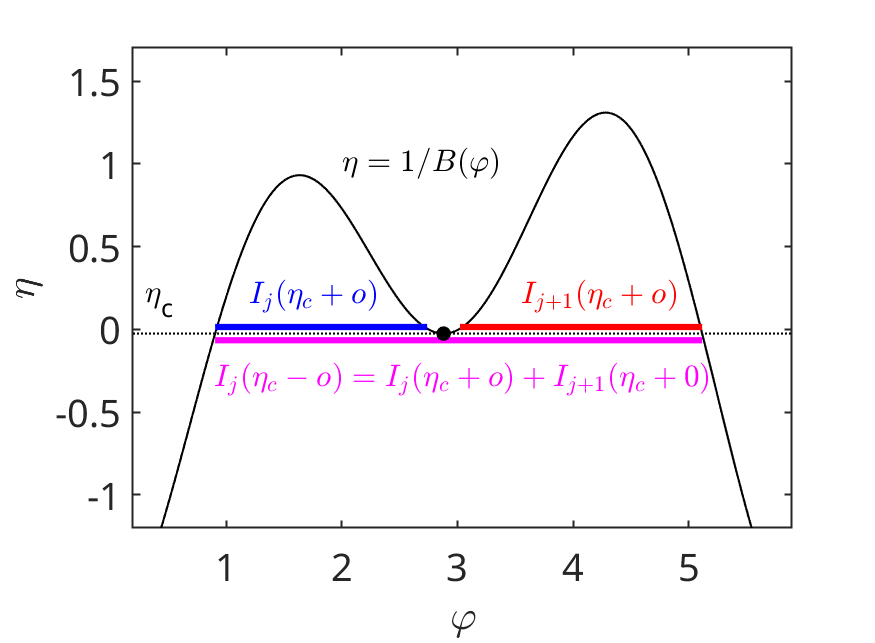}
}
\caption[]{
Example of class transition boundary introduced by local field maximum $B_{\max}^{\text{loc}}$ (black dot)
where three types of trapped particles meet (two ``single-trapped'' and one ``double trapped'').
Boundary conditions~\eq{fluxconsbou} are fulfilled by~\eq{formsol_s1} due to
$H_j(\eta_c-o)=H_j(\eta_c+o)+H_{j+1}(\eta_c+o)$.
}
\label{fig:clasbous}
\end{figure}

Due to the first condition~\eq{liouv_reprp}, the bounce averaged equation~\eq{ba_eq} for passing particles
is homogeneous,
\be{dirba_pass}
\difp{}{\eta} \left(\difp{g_{-1}}{\eta}\int\limits_{\varphi_0}^{\varphi_N} \rd \varphi^\prime D_\eta\right)=0.
\ee
Integrating it once and applying the boundary condition $\partial g_{-1} / \partial \eta = 0$ at the strongly
passing boundary $\eta=0$ we get $g_{-1}=\text{const.}$ in the passing region. Since $g_{-1}$ is continuous at the global
maximum point $\varphi=\varphi_0$ at the trapped-passing boundary $\eta=1/B_{\max}$, the function $g_{-1}$ can only be
even. Since any constant satisfies the homogeneous mono-energetic equation in the whole phase space making no
contribution to transport coefficients but only re-defining the moments of equilibrium Maxwellian, we fix
$g_{-1}=0$ in the passing region.
In the trapped region, Eq.~\eq{ba_eq} is
\be{ba_eq_s1}
\difp{}{\eta}
\left(\frac{\eta I_j}{l_c}
\difp{g_{-1}}{\eta}+H_j
\right)=0,
\ee
where we denoted
\be{bas1_defs}
I_j(\eta) = \int\limits_{\varphi_j^-(\eta)}^{\varphi_j^+(\eta)} \rd \varphi \frac{|\lambda|}{B^\varphi},
\qquad
H_j(\eta) = \int\limits_{\varphi_j^-(\eta)}^{\varphi_j^+(\eta)} \rd \varphi H^\prime_\varphi,
\ee
see definitions~\eq{mono_Ak_norm}--\eq{Hphiprime}.
Since $\partial g_{-1} /\partial \eta =0$ at the bottoms of local magnetic wells
$\eta=1/B_{\min}^{\text{loc}}$ where $H_j=0$ due to $\varphi_j^+=\varphi_j^-$ we can integrate~\eq{ba_eq_s1} to
\be{formsol_s1}
\difp{g_{-1}(\eta)}{\eta} =-\frac{l_c H_j}{\eta I_j}.
\ee
This solution automatically satisfies collisional flux conservation relations in all boundary layers
separating different trapped particle classes,
\be{fluxconsbou}
\left[I_j\difp{g^{(j)}_{-1}}{\eta}\right]_{\eta=\eta_c-o}
=
\left[I_j\difp{g^{(j)}_{-1}}{\eta}\right]_{\eta=\eta_c+o}
+
\left[I_{j+1}\difp{g^{(j+1)}_{-1}}{\eta}\right]_{\eta=\eta_c+o},
\ee
where $\eta_c=1/B_{\max}^{\text{loc}}$, see Fig.~\ref{fig:clasbous}, $o$ denotes an infinitesimal number,
and the superscript $(j)$ on $g_{-1}$
denotes particle type trapped between the reflection points $\varphi_j^\pm$.
Derivative~\eq{formsol_s1} satisfies also the flux conservation through the trapped-passing boundary $\eta=\eta_b$
turning there to zero, because only a single type of trapped particles exists near this boundary,
and $H_1(\eta_b+o)=0$ follows then from the second
condition~\eq{liouv_reprp}.

Solution~\eq{formsol_s1} is sufficient for the computation of the effective ripple~\citep{nemov99-4622}
$\varepsilon_{\text{eff}}$
which determines device geometry effect on the mono-energetic diffusion coefficient
$\bar D_{11}$ (and all other $\bar D_{jk}$ for $j,k \le 2$ trivially related to $\bar D_{11}$).
Substituting in Eq.~\eq{Dmono} $g_{(1)}=g_{-1}$ and $s_{(1)}$ via~\eq{s_k}, integrating the result
by parts over $\eta$ and using the continuity of $g_{-1}$  through all boundary layers, changing the integration
order over $\varphi$ and $\eta$ results in
\be{dmono_epseff}
\bar D_{11}
= \frac{v l_c}{2} \left(\int\limits_{\varphi_0}^{\varphi_N}\frac{\rd \varphi}{B^\varphi}\right)^{-1}
\int\limits_{\eta_b}^{\eta_m}\rd \eta\sum\limits_{j=1}^{j_{\max}(\eta,N)} \frac{H_j^2}{\eta I_j}
=\frac{4\sqrt{2}}{9\pi}\frac{v l_c \rho_L^2 B^2}{R_0^2 B_0^2}\varepsilon_{\text{eff}}^{3/2},
\ee
where $\eta_b=1/B_{\max}$ and $\eta_m=1/B_{\min}$ are defined by global field maximum and minimum, respectively.
Last equality~\eq{dmono_epseff} where $R_0$ and $B_0$ are reference values of major radius and magnetic field,
respectively,
defines $\varepsilon_{\text{eff}}^{3/2}$ in the same way as Eq.~(29) of~\citep{nemov99-4622}
where quantities $\hat I_j(b^\prime)$ and $\hat H_j(b^\prime)$ are related to the present notation
by $\hat I_j = I_j$ and $\hat H_j = 3 B_0^{3/2}\eta^{1/2}(\rho_L B)^{-1} H_j$ with $b^\prime=(B_0\eta)^{-1}$.

For the next order correction $g_0$, we notice that the first two terms in the r.h.s. of Eq.~\eq{gn_eq_int}
are odd in $\sigma$. Therefore, in the trapped particle region,
equation~\eq{int_bargn_eq} for the integration constant (which can only be even there)
is homogeneous, resulting in $\bar g_0=\text{const}$. Due to the continuity of the distribution function
across the trapped-passing boundary, the integration constant in the passing region is the same. We can again
absorb this constant into the equilibrium Maxwellian, as we have already done in the previous order
when setting $g_{-1}=0$ in the passing region.
Thus, substituting in~\eq{gn_eq_int} the derivative $\partial g_{-1}/\partial \eta$ via~\eq{formsol_s1}
and the source $s=s_{(1)}$ via~\eq{s_k} and~\eq{Hphiprime} we get explicitly
\be{g0_eq_int_s1}
g_0(\varphi,\eta,\sigma)
=\sigma \difp{}{\eta}
\int\limits_{\varphi_j^-(\eta)}^\varphi \rd \varphi^\prime
\left(H_\varphi^\prime(\varphi^\prime,\eta)
-\frac{|\lambda(\varphi^\prime,\eta)| }{B^\varphi(\varphi^\prime)}
\frac{H_j(\eta)}{I_j(\eta)}
\right),
\ee
where we have exchanged the derivative over $\eta$ with integration in the first term using
$H_\varphi^\prime(\varphi_j^-(\eta),\eta)=0$.
It should be noted now that the integral over $\varphi^\prime$ is a discontinuous function of $\eta$
at the boundaries between classes $\eta=\eta_c$ where either $\varphi_j^-(\eta)$ or
$\varphi_j^+(\eta)$ jumps.
Respectively, the function $g_0$ contains a $\delta$-function $\delta(\eta-\eta_c)$,
which is required by particle conservation in the boundary layer.
Namely, integrating Eq.~\eq{mono_Ak_norm} over $\eta$ across the boundary, we get
\be{across_bou}
\sigma \int\limits_{\eta_c-o}^{\eta_c+o}\rd \eta \difp{g}{\varphi}=
\left. D_\eta(\varphi,\eta)\difp{g(\varphi,\eta)}{\eta}\right|_{\eta=\eta_c-o}^{\eta=\eta_c+o}.
\ee
Substituting here $g=g_{-1}+g_0$ and ignoring $g_0$ in the r.h.s. where its contribution is
linear in collisionality we get
\be{across_bou_g0}
\sigma\difp{}{\varphi} \int\limits_{\eta_c-o}^{\eta_c+o}\rd \eta\; g_0(\varphi,\eta,\sigma)
=
\frac{|\lambda(\varphi,\eta_c)|}{B^\varphi(\varphi)}
\left(\frac{H_j(\eta_c-o)}{I_j(\eta_c-o)} - \frac{H_j(\eta_c+o)}{I_j(\eta_c+o)}\right)
\ee
which is an identity for $g_0$ in the form~\eq{g0_eq_int_s1}.

In the passing region where $g_{-1}=0$, Eq.~\eq{gn_eq_int} results in
\be{g0_eq_int_s1_pass}
g_0 = \sigma \difp{}{\eta}
\int\limits_{\varphi_0}^\varphi \rd \varphi^\prime
H_\varphi^\prime(\varphi^\prime,\eta)
+
\bar g_0,
\ee
where
\be{dergbar}
\difp{\bar g_0}{\eta}=
-\frac{\sigma}{\langle |\lambda| \rangle}
\left\langle
|\lambda|
\difp{^2}{\eta^2}
\int\limits_{\varphi_0}^\varphi \rd \varphi^\prime
H_\varphi^\prime(\varphi^\prime,\eta)
\right\rangle
\ee
follows from~\eq{int_bargn_eq} and $\langle\dots\rangle$ denotes a field line average
(flux surface average~\eq{neo_mflav} with $\varphi_N$ kept finite).
This derivative has an integrable singularity at the
trapped-passing boundary, $\partial \bar g_0 /\partial \eta \propto (\eta_b-\eta)^{-1/2}$
which follows from $|\nabla r| k_G \propto (\varphi -\varphi_0)$ near the global maximum $\varphi=\varphi_0$
(and $\varphi=\varphi_N$ since the innermost integral is a single valued (periodic)
function of $\varphi$ on the closed field line).
Therefore, $\bar g_0$ is continuous at the trapped-passing boundary, where it is connected
to $\bar g_0=0$ in the trapped region. Moreover, the derivative of the full solution $\partial g_0 / \partial \eta$
has no
singularity at this boundary (in contrast to $\partial \bar g_0 / \partial \eta$).
One can also check that this derivative
does not depend on the lower integration limit $\varphi_0$ in~\eq{g0_eq_int_s1_pass} and~\eq{dergbar} which,
therefore, needs not to be the global maximum point.

We can formally combine expressions~\eq{g0_eq_int_s1} and~\eq{g0_eq_int_s1_pass} into
\be{g0_eq_int_s1_comb}
g_0(\varphi,\eta,\sigma)
=\sigma \difp{}{\eta}
\int\limits_{\varphi_{\text{beg}}}^\varphi \rd \varphi^\prime
\left(H_\varphi^\prime(\varphi^\prime,\eta)
-\frac{|\lambda(\varphi^\prime,\eta)| }{B^\varphi(\varphi^\prime)}
\frac{H_j(\eta)}{I_j(\eta)}
\right)+\bar g_0(\eta,\sigma),
\ee
valid for the whole phase space with $I_j=I_0$ and $H_j=H_0=0$ in the passing region
where they are given by Eqs.~\eq{bas1_defs} with the limits $\varphi_0$ and $\varphi_N$ instead
of $\varphi_j^\pm$.
Due to the linear scaling of $g_0$ with velocity module $v$, we can evaluate energy integral
in the expression for the parallel current density of the $\alpha$ species
substituting in~\eq{superforces} $g=g_0$ which is the only term contributing in the leading order,
\be{pacd_species}
j_{\parallel\alpha}= e_\alpha \int\rd^3 v v_\parallel (f-f_M)=
C_\alpha B \sum\limits_{\sigma=\pm 1}\sigma\int\limits_0^{1/B}\rd\eta\; g_0,
\ee
where
$$
C_\alpha = \frac{3 c}{4 B\rho_L}\left(\difp{p_\alpha}{r}-e_\alpha n_\alpha E_r\right),
\qquad
p_\alpha = n_\alpha T_\alpha,
$$
and we omitted the inductive current by setting $A_3=0$. Since $g_0/\rho_L$ is independent
of particle species, total current density is independent of $E_r$ in
quasi-neutral plasmas,
\be{pacd}
j_{\parallel}= \sum_\alpha j_{\parallel\alpha}=
C_\parallel B \sum\limits_{\sigma=\pm 1}\sigma\int\limits_0^{1/B}\rd\eta\; g_0
=C_\parallel B \sum\limits_{\sigma=\pm 1}\sigma\int\limits_0^{1/B}\rd\eta\; \left(g_0 - \bar g_0
- \eta \difp{\bar g_0}{\eta}\right),
\ee
where
\be{C_parallel}
C_\parallel = C_\parallel(r) = \frac{1}{\rho_L}\sum_{\alpha^\prime} C_{\alpha^\prime} \rho_L^\prime
= \frac{3 c}{4 B \rho_L} \difp{p}{r},
\qquad
p=\sum_\alpha p_\alpha,
\ee
$\rho_L^\prime$ denotes the Larmor radius of $\alpha^\prime$ species,
and we used the condition $\bar g_0=0$ at the boundary $\eta=1/B$ when integrating by parts in the last
expression~\eq{pacd}. According to~\eq{g0_eq_int_s1_comb}, the term $g_0-\bar g_0$ is a derivative which contributes
only at the lower limit $\eta=0$ where using explicitly~\eq{Hphiprime} one gets parallel equilibrium current
density as
\be{pacd_int}
j_{\parallel}
=
- 2 c B \difp{p}{r}\int\limits_{\varphi_0}^{\varphi} \rd \varphi^\prime
\frac{|\nabla r| k_G}{B B^\varphi}
-C_\parallel B \sum\limits_{\sigma=\pm 1}\sigma\int\limits_0^{\eta_b}\rd\eta\;
\eta \difp{\bar g_0}{\eta}=j_{\text{PS}}+j_b.
\ee
Here, as well as in Eq.~\eq{dergbar}, the point $\varphi_0$
must be at a global maximum because integration over $\eta$ in the first term required the continuity
of $\varphi_{\text{beg}}$ at the trapped-passing boundary.
Eq.~\eq{pacd_int} naturally agrees with the ideal MHD radial force balance which determines the Pfirsch-Schl\"uter
current density $j_{\text{PS}}$ up to the arbitrary constant times $B$.
Due to quasi-Liouville's theorem~\eq{liouv_reprp}, the current density~\eq{pacd_int} is
periodic, $j_\parallel(\varphi_N)=j_\parallel(\varphi_0)$,
so that the last expression~\eq{geodcu} for the geodesic curvature leads to
\be{closure}
\difp{}{\vartheta_0}\int\limits_{\varphi_0}^{\varphi_N} \frac{\rd \varphi}{B^\varphi}=0,
\ee
which is a ``true'' rational surface condition~\citep{solovev70}.
Using the explicit form of $H^\prime_\varphi$ in~\eq{dergbar} and
taking the flux ``surface average''~\eq{neo_mflav} of $j_\parallel B$
thus eliminating the Pfirsch-Schl\"uter current which is driven solely by charge separation potential and,
therefore, must satisfy $\langle j_{\text{PS}} B\rangle = 0$,
we finally obtain the bootstrap current density $j_b$
which scales with $B$ as
\be{bootaver}
\langle j_\parallel B\rangle = j_b \frac{\langle B^2\rangle}{B} = - c \lambda_{bB}\difp{p}{r}.
\ee
Here,
$\lambda_{bB}$ is the dimensionless geometrical factor $\lambda_b$ given by Eq.~(9) of~\citep{nemov04-179}
in case of normalization field $B_0 = \langle B^2\rangle^{1/2}$,
\be{lambda_bB}
\lambda_{bB}
=
\left\langle
2 B^2
\int\limits_{\varphi_0}^\varphi \rd \varphi^\prime
\frac{|\nabla r| k_G}{B B^\varphi}
+
\frac{3 \langle B^2\rangle}{8}
\int\limits_0^{\eta_b} \rd\eta
\frac{\eta^2 |\lambda|}{\langle|\lambda|\rangle}
\int\limits_{\varphi_0}^\varphi\rd \varphi^\prime
\frac{|\nabla r|k_G B^2} {|\lambda|^3 B^\varphi}
\right\rangle,
\ee
with $\langle \dots \rangle$ given for a closed field line by~\eq{neo_mflav} with finite $\varphi_N$.
It is convenient to express it via the mono-energetic bootstrap coefficient~\eq{Dmono},
\be{D31_lambb}
\lambda_{bB}
=
\frac{3 \bar D_{31}}{v \rho_L B},
\ee
in order to use it also for finite collisionality.

The separate contribution of trapped particle region to $\lambda_{bB}$ is obtained by replacement of the lower integration
limit in~\eq{pacd} from 0 to $\eta_b+o$,
\be{lambda_bB_tr}
\lambda_{bB}^{tr}
=
\left\langle
B^2
\int\limits_{\varphi_0}^\varphi \rd \varphi^\prime
\frac{|\nabla r| k_G}{2 B B^\varphi} \lambda_b (3+\lambda_b^2)
\right\rangle,
\qquad
\lambda_b = \sqrt{1-\eta_b B}.
\ee
In the usual case of small magnetic field modulation amplitude $\varepsilon_M \ll 1$, this contribution is
of the order $\varepsilon_{M}^{1/2}$ as compared to the first term in~\eq{lambda_bB} and is of the order
$\varepsilon_{M}$ as compared to the second term.  Therefore, it can be ignored~\citep{boozer90-2408}
as long as an exact compensation of bootstrap current, $\lambda_{bB}=0$, is not looked for.

The factor~\eq{lambda_bB} matches the result of Shaing and Callen,
what is a natural consequence of using the same method. Namely, $\lambda_{bB} = (1-f_c) \langle \tilde G_b \rangle / S$
where $S = \rd V / \rd r$ is the flux surface area,
fraction of circulating particles $f_c$ is given by Eq.~(56) of~\cite{shaing83-3315},
and geometrical factor $\tilde G_b$ is given there by Eqs.~(75b) and~(60c). In case of negligible toroidal equilibrium current,
$B_\vartheta=0$, the result of~\cite{boozer90-2408} is also approximately recovered,
$\lambda_{bB} = \Delta_0 B_\varphi \rd r / \rd \psi + O(\varepsilon_{\text{M}})$, where
the quantity $\Delta_0$ is given by Eqs.~(51), (53) and~(54) of~\cite{boozer90-2408}
ignoring small trapped particle contribution and other corrections linear in $\varepsilon_{M}$
(in case of a circular tokamak, $\lambda_{bB} \approx A\Delta_0$ where $A$ is the aspect ratio).

Note that the trapped particle distribution function here is formally
the same as the one given in the implicit form by Eq.~(54) of~\cite{shaing83-3315}
omitting there the ripple plateau contribution absent in our case where we ignored banana precession
(cross-field drift within flux surfaces) already in the starting equation~\eq{mono_Ak_norm}.
On the other hand, we obtained also the explicit expression for its odd part~\eq{g0_eq_int_s1_comb} which is
demonstrative for the distribution of parallel equilibrium current in the velocity space.
It can be seen that a significant part of this current (essentially of the Pfirsch-Schl\"uter current)
flows in the boundary layers between trapped particle classes (but not in the main, trapped-passing boundary layer).
The origin of these localized currents is the same as the origin of Pfirsch-Schl\"uter current
in all devices, i.e. they remove charge separation introduced by the radial particle drift.
However, in contrast to tokamaks (and tokamak-like part of the current in
stellarators)
where compensation currents are produced in a long mean free path regime
within a single turn of particle bounce motion due to the finite
radial displacement during this time, localized
currents serve to compensate charge separation by finite bounce-averaged drift,
which accumulates on much longer collisional detrapping time.
Interpreting the source term $s_{(1)}$, Eq.~\eq{s_k}, in the regions where it is positive as a source of ``particles''
and where it is negative as a source of ``anti-particles'' whose total amount generated at the flux surface
is the same as the amount of ``particles'' being a consequence of quasi-Liouville's theorem~\eq{liouv_reprp},
we see that particles (or anti-particles) accumulated in the local ripple wells due to the finite bounce averaged
drift, $H_j \ne 0$, can only leave the wells due to collisional scattering flux through the class boundaries
(see Fig.~\ref{fig:clasbous}). Since local collisional flux density generally needs not to be continuous at this boundary,
$\partial g_{-1}(\eta -o) /\partial \eta \ne \partial g_{-1}(\eta +o) /\partial \eta$, a significant amount of particles
(anti-particles) is re-distributed through the boundary layers, what is manifested by the $\delta$-like behavior
of $g_0$, Eq.~\eq{g0_eq_int_s1_comb}, being, up to a factor, a parallel flow density in velocity space.

The asymptotic series expansion~\eq{solform} can be continued to the next order, leading to the correction $g_1$. This
correction, however, has a non-integrable singularity at class transition boundaries,
$g_1 \propto (\eta - \eta_c)^{-2}$ (in contrast to integrable singularity of $g_0 \propto \log|\eta - \eta_c|$),
which is a consequence of finite geodesic curvature at respective local field maxima. Therefore, contributions
of different order corrections to the collisional flux density in the matching conditions~\eq{across_bou} become
comparable
at finite collisionality for $|\eta -\eta_c| =\delta\eta \propto l_c^{-1/2} \propto \nu^{1/2}$ where
$\partial g_{0}(\eta) /\partial \eta \sim \delta\eta^{-1}$ and
$\partial g_{1}(\eta) /\partial \eta \sim l_c^{-1} \delta\eta^{-3}$. In other words, the expansion~\eq{solform}
breaks down at the edge of the boundary layer of width $\delta\eta$. Respectively, matching interval $o$
in Eq.~\eq{fluxconsbou},
which is an infinitesimal for vanishing collisionality, should satisfy $\delta \eta \ll o \ll \eta$ for finite
collisionality, which means that the error of the asymptotic solution scales with $\sqrt{\nu}$ rather than with $\nu$.
Since particle -- anti-particle re-distribution
flux manifested by $\delta$-functions in asymptotic expression~\eq{g0_eq_int_s1_comb} for $g_0$ is independent
of collisionality, we can estimate the odd part of the distribution function which carries this flux in the
class-transition boundary layers
as
$g^{\text{odd}} \propto \delta\eta^{-1} \propto l_c^{1/2} \propto \nu^{-1/2}$.

As shown below,
the presence of these strong parallel flows in class transition boundary layers is actually responsible for the ``anomalous''
behavior of
bootstrap current at finite collisionality.

\subsubsection{Adjoint problem}
\label{ssec:adjprob}

The asymptotic solution of the adjoint problem has been derived by~\cite{helander11-092505}
omitting in the correction $g_0$, which determines particle flux, the $\varphi$ independent
part $\bar g_0$ which does not contribute to this flux and, respectively, is not needed for
the Ware pinch coefficient.
Here, we extend this derivation in order to obtain the complete even part of this correction
useful for the interpretation of numerical examples in the following sections.
The leading order asymptotic solution
is odd in $\sigma$ and is given by the bounce-averaged equation~\eq{ba_eq} for the source $s_{(3)}$, see Eq.~\eq{s_k},
with the same boundary conditions
as previously
as
\be{ware_0}
g_{-1}
=
\sigma
\int\limits_{\varphi_0}^{\varphi_N}\rd \varphi\;\frac{B^2}{B^\varphi}
\int\limits^{\eta_b}_\eta \rd \eta^\prime\;\eta^\prime\;
\left(\int\limits_{\varphi_0}^{\varphi_N}\rd \varphi\;D_\eta\right)^{-1}
=\sigma l_c
\int\limits^{\eta_b}_\eta \rd \eta^\prime\;
\frac{\left\langle B^2\right\rangle}
{\left\langle |\lambda|\right\rangle}
\ee
in the passing region, and $g_{-1}=0$ in the trapped region.
The next order correction results from Eqs.~\eq{gn_eq_int} in the trapped region in
\be{g0_adj_tr}
g_0(\varphi,\eta)=\sigma \int\limits_{\varphi_j^-}^\varphi\rd \varphi^\prime s_{(3)}+\bar g_0(\eta)
=\int\limits_{\varphi_j^-}^\varphi\rd \varphi^\prime \frac{B^2}{B^\varphi}+\bar g_0(\eta)
=\int\limits_{\varphi_b}^\varphi\rd \varphi^\prime \frac{B^2}{B^\varphi}\equiv g_0^t(\varphi),
\ee
where we integrated Eq.~\eq{int_bargn_eq},
$$
\difp{\bar g_0}{\eta}=\frac{B^2(\varphi_j^-)}{B^\varphi(\varphi_j^-)}\difp{\varphi_j^-}{\eta},
$$
from the trapped-passing boundary and included an arbitrary integration constant into
the lower integration limit $\varphi_b$ being, therefore, an arbitrary constant too.
This constant is the same for all classes due to continuity of $g_0$ at class transition boundaries.

In the passing region, solutions of Eqs.~\eq{gn_eq_int} and~\eq{int_bargn_eq}
result in
\be{g0_adj_pass}
g_0(\varphi,\eta)
=
\int\limits_{\varphi_b}^\varphi\rd \varphi^\prime
\left( \frac{B^2}{B^\varphi}
-
\difp{}{\eta}
\frac{\eta \langle B^2 \rangle}{\langle|\lambda|\rangle}
\frac{|\lambda|}{B^\varphi}
\right)
+
\bar g_0(\eta),
\ee
where
\be{g0_adj_pass_ba}
\bar g_0(\eta)
=
\int\limits_0^\eta \frac{\rd \eta^\prime}{\langle|\lambda|\rangle}
\left\langle
|\lambda|
\difp{^2}{{\eta^\prime}^2}
\frac{\eta^\prime \langle B^2 \rangle}{\langle|\lambda|\rangle}
\int\limits_{\varphi_b}^\varphi\rd \varphi^\prime \frac{|\lambda|}{B^\varphi}
\right\rangle
+C_\sigma,
\ee
and where $C_\sigma$ are different integration constants of Eq.~\eq{int_bargn_eq} for $\sigma=\pm 1$.
Similar to the solution of the direct problem,
derivative $\partial g_0/\partial \eta$ of the function~\eq{g0_adj_pass} does not depend on $\varphi_b$ in the passing region.
Actually, the lower integration limit $\varphi_b$ in Eqs.~\eq{g0_adj_pass} and~\eq{g0_adj_pass_ba} enters $g_0$ via a
periodic function of $\varphi_b$ which adds up to the integration constant $C_\sigma$ (this follows from explicit
evaluation of $\partial g_0 /\partial \varphi_b =  \left(\langle B^2 \rangle - B^2(\varphi_b)\right) /B^\varphi(\varphi_b)$
whose integral over the whole range $\varphi_0<\varphi_b<\varphi_N$ is zero).
Function $g_0$
has an integrable logarithmic singularity
at the trapped passing boundary, $g_0 \propto \log(\eta_b-\eta)$, in particular, due to such singularity of
$\partial \langle|\lambda|\rangle / \partial \eta$ which, up to a constant factor, is a bounce time.
Next order correction shows again that series expansion breaks down in the boundary layer,
$|\eta_b-\eta | \le \delta\eta \propto l_c^{-1/2}$.

{
In order to fully determine $g_0$ in the whole phase space we must express three integration constants,
$\varphi_b$ and $C_\sigma$, via a single constant which is the only degree of freedom re-defining an
equilibrium Maxwellian. For that, we must match the solutions through the trapped-passing boundary where $g_0$ is
singular, but first it is convenient to split $g_0$ into even and odd parts in $\sigma$, $g_0=g_0^{\text{even}}+g_0^{\text{odd}}$,
where $g_0^{\text{even}}$ is given by Eqs.~\eq{g0_adj_tr}-\eq{g0_adj_pass_ba} with replacement of $C_\sigma$ by a constant
$C_{\text{even}}$ and $g_0^{\text{odd}}=\sigma C_{\text{odd}}\Theta(\eta_b-\eta)$
where $\Theta(x)$ is a Heaviside step function,
and $C_{\text{odd}}$ is another constant.

Function $g_0^{\text{odd}}$ is of no interest in the following
since it does not contribute to the Ware pinch but only provides a correction to the leading order solution~\eq{ware_0}.
As for $g_0^{\text{even}}$, we still need to determine it formally in the boundary layer identifying there its most singular
part. For this purpose,
we take the odd part of Eq.~\eq{mono_Ak_norm} and,
similar to~\eq{across_bou_g0}, integrate it across the boundary layer ignoring the contribution
of the source term and retaining only the leading order solution~\eq{ware_0} in the collisional flux
through the boundaries of integration domain,
\be{across_bou_g0_tp}
\difp{}{\varphi} \int\limits_{\eta_b-\Delta\eta}^{\eta_b+\Delta\eta}\rd \eta\; g_0^{\text{even}}(\varphi,\eta)
=
\left(\frac{\eta \langle B^2 \rangle}{\langle|\lambda|\rangle}
\frac{|\lambda|}{B^\varphi}\right)_{\eta=\eta_b-\Delta\eta}
\rightarrow
\left(\frac{\eta \langle B^2 \rangle}{\langle|\lambda|\rangle}
\frac{|\lambda|}{B^\varphi}\right)_{\eta=\eta_b}.
\ee
Here, $\delta\eta \ll \Delta\eta \ll \eta_b$, $\delta\eta$ is the boundary layer width and
the last expression corresponds to the collisionless limit $\Delta\eta\rightarrow 0$.
Thus $g_0^{\text{even}}$ can be formally presented near the trapped passing boundary as
\be{across_bou_g0_tp_int}
g_0^{\text{even}}=\delta(\eta_b-\eta)\left(\int\limits_{\varphi_b}^\varphi\rd \varphi^\prime
\frac{\eta \langle B^2 \rangle}{\langle|\lambda|\rangle}
\frac{|\lambda|}{B^\varphi}+C_\text{bou} \right)+o(\delta\eta^{-1}),
\ee
where $C_\text{bou}$ is yet another integration constant,
and the last term provides a vanishing contribution to the $\eta$-integral in~\eq{across_bou_g0_tp}.
Similar to the passing particle solution~\eq{g0_adj_pass} which is defined for the whole
closed field line and which is a periodic function of $\varphi$ with period
$\varphi_N-\varphi_0$, boundary layer solution~\eq{across_bou_g0_tp_int} is formally defined
for the whole field period too and must be a periodic function as well.
In order to be periodic, function~\eq{across_bou_g0_tp_int} must be discontinuous
over the variable $\varphi$, and the only point where a jump restoring the periodicity
is possible, as we see below,
is the global maximum point $\varphi=\varphi_0$ (and $\varphi=\varphi_N$).
This discontinuity requirement leads to a certain paradox because according to Eq.~\eq{mono_Ak_norm}
the distribution function is continuous with $\varphi$ everywhere in the range where $\eta B(\varphi)<1$,
which includes the whole trapped-passing boundary approached from the passing side, $\eta=\eta_b-0$,
what in turn seems to contradict the discontinuity of the function~\eq{across_bou_g0_tp_int}.
To resolve this paradox, we note that a formal representation with a $\delta$-function actually assumes
a small but finite width of the boundary layer, which is partly located in the passing particle
region where the distribution is strictly continuous with $\varphi$, and partly in the trapped particle
region where it is continuous too except a small forbidden region with $\eta B(\varphi)>1$ located near
the global maximum. Discontinuity of $g_0^\text{even}$ within this trapped part of the boundary layer
is sufficient to provide the required jumps restoring the periodicity.

Similarly to the boundary layer,
the periodicity argument means for the trapped particle distribution that its integral
form~\eq{g0_adj_tr} determined via $\varphi_b$ from the interval $\varphi_0<\varphi_b<\varphi_N$
can be used only in the same interval, $\varphi_0<\varphi<\varphi_N$. Thus, an extension of
Eqs.~\eq{g0_adj_tr} and~\eq{across_bou_g0_tp} to the infinite $\varphi$ range is obtained replacing
there the lower integration limit $\varphi_b$ with $\varphi_b + (\varphi_N-\varphi_0)
\left[(\varphi-\varphi_0)/(\varphi_N-\varphi_0)\right]$ where $[\dots]$ denotes the integer part.
As already mentioned above, the passing particle solution~\eq{g0_adj_pass} is valid for the infinite
$\varphi$ range as is. We can actually combine all these periodic solutions for $g_0^\text{even}$
in one
multiplying so extended trapped particle solution~\eq{g0_adj_tr} with $\Theta(\eta-\eta_b)$,
adding an even part of the
passing particle solution~\eq{g0_adj_pass} multiplied with $\Theta(\eta_b-\eta)$
and then adding an extended boundary layer solution~\eq{across_bou_g0_tp_int}.

In order to eliminate from such a combined solution,
all but one unknown constants out of three,
$\varphi_b$, $C_\text{even}$ and $C_\text{bou}$, we use the stellarator symmetry of our closed
field line such that $B(2\varphi_s-\varphi)=B(\varphi)$ and
$B^\varphi(2\varphi_s-\varphi)=B^\varphi(\varphi)$ where $\varphi_s$ is a symmetry point.
We use the fact that stellarator symmetric part of the generalized Spitzer function $g=g_{(3)}$ given by
$g_\text{sym}(\varphi,\eta,\sigma)\equiv \left(g(\varphi,\eta,\sigma)+g^\ast(\varphi,\eta,\sigma)\right)/2$
with $g^\ast(\varphi,\eta,\sigma) \equiv g(2\varphi_s-\varphi,\eta,-\sigma)$
can only be a constant.
Namely, it can be checked that $g^\ast$
satisfies the same Eq.~\eq{mono_Ak_norm} but with an opposite sign of the source $s_{(3)}$ and,
therefore, $g^\ast=-g + C$ where $C$ is an arbitrary constant. Thus, $g_\text{sym}=C/2$ in the whole phase space.
Introducing now a stellarator-antisymmetric part of the distribution function,
$g_\text{asym}(\varphi,\eta,\sigma)\equiv \left(g(\varphi,\eta,\sigma)-g^\ast(\varphi,\eta,\sigma)\right)/2$,
we can split our solution $g_0^\text{even}$ into stellarator-symmetric and stellarator-antisymmetric parts,
$g_0^\text{even} = g_\text{sym}^\text{even}+g_\text{asym}^\text{even}$. Thus, we obtain that stellarator-antisymmetric
part of the combined $g_0^\text{even}$ does not contain any unknown and is determined by
\be{g0_adj_comb}
g_\text{asym}^{\text{even}}(\varphi,\eta)
=
\int\limits_{\varphi_s}^\varphi\rd \varphi^\prime
\left( \frac{B^2}{B^\varphi}
-
\difp{}{\eta}
\left(
\Theta(\eta_b-\eta)
\frac{\eta \langle B^2 \rangle}{\langle|\lambda|\rangle}
\frac{|\lambda|}{B^\varphi}
\right)
\right),
\ee
where $\varphi_s$ is fixed now to a second stellarator symmetry point
$\varphi_s=(\varphi_0+\varphi_N)/2$ which necessarily exists besides the global maximum point $\varphi_0$.
Solution~\eq{g0_adj_comb} is given here only for $\varphi_0<\varphi<\varphi_N$ and can be extended
to the infinite $\varphi$ range replacing there $\varphi_s$ with
$\varphi_s + (\varphi_N-\varphi_0)
\left[(\varphi-\varphi_0)/(\varphi_N-\varphi_0)\right]$.
Such an extended expression is stellarator-antisymmetric with respect to
symmetry points of both kinds (in particular, anti-symmetry with respect to field maximum points 
$\varphi_0 + k(\varphi_N-\varphi_0)$ where $k$ are integers
is enabled by the discontinuity of the extended lower integration limit at those points)
and is independent of an arbitrary constant $\varphi_b$.

In turn, all the unknown constants enter the stellarator-symmetric part of the combined solution $g_0^\text{even}$
which is of the form
\bea{gevensym}
g_\text{sym}^{\text{even}}
&=&
\int\limits_{\varphi_b}^{\varphi_s}\rd \varphi^\prime \frac{B^2}{B^\varphi}
+
\Theta(\eta_b-\eta) \left(C_{\text{even}}
-\int\limits_{\varphi_b}^{\varphi_s}\rd \varphi^\prime \frac{\langle B^2\rangle}{B^\varphi}\right)
\nonumber \\
&+&
\delta(\eta_b-\eta)\left(\int\limits_{\varphi_b}^{\varphi_s}\rd \varphi^\prime
\frac{\eta \langle B^2 \rangle}{\langle|\lambda|\rangle}
\frac{|\lambda|}{B^\varphi}+C_\text{bou} \right).
\eea
Since this function can only be a constant, we must set $C_\text{even}$ and $C_\text{bou}$ so
that they annihilate respective brackets multiplying Heaviside function and $\delta$-function
which permits setting $\eta=\eta_b$ in the sub-integrand in the latter bracket. Finally, the remaining first term
can be, as usual, absorbed into the equilibrium Maxwellian so that the whole final solution
$g_0^\text{even}=g_\text{asym}^{\text{even}}$ is given by Eq.~\eq{g0_adj_comb}.
The only difference of this solution from Eq.~(12) of~\cite{helander11-092505}
is the lower integration limit $\varphi_s$ which is the second
symmetry point (usually global minimum)
instead of a global maximum point $\varphi_0$.

Substituting~\eq{ware_0} and~\eq{g0_adj_comb} for $g_{(3)}=g_{-1}+g_0$ in Eq.~\eq{Dmono}
we obtain the mono-energetic Ware-pinch coefficient as
\be{ware-mono}
\bar D_{13} = \frac{1}{3} v \rho_L B \lambda_{bB}^\dagger,
\ee
with the geometrical factor given by
\be{lambb-ware}
\lambda_{bB}^\dagger
=\frac{1}{2}
\left\langle
\int\limits_0^{1/B} \rd\eta
\;
g_0^{\text{even}}
\frac{|\nabla r|k_G} {B}
\difp{}{\eta}\left(3|\lambda|+|\lambda|^3\right)
\right\rangle.
\ee
Exchanging in this expression the integration order over $\eta$ and $\varphi$ and integrating
by parts over $\varphi$ within the motion domains $\varphi_{\text{beg}}(\eta) < \varphi < \varphi_{\text{end}}(\eta)$
we use conditions~\eq{liouv_reprp} to eliminate contribution from the integration limits. Thus,
only the derivative $\partial g_0 / \partial \varphi$ can contribute.
Integrating the result by parts over $\eta$ in the term with $\Theta(\eta_b-\eta)$, one finally obtains
$\lambda_{bB}^\dagger=\lambda_{bB}$ given by Eq.~\eq{lambda_bB}.

It should be noted that in contrast to the solution of direct problem,
solution of the adjoint problem~\eq{g0_adj_comb}
is regular at all class transition boundaries and has a  $\delta$-like behavior
only in the boundary layer at the trapped-passing boundary
where finite collisionality scaling of $g^{\text{even}}$ is
$g^{\text{even}}\sim l_c^{1/2} \sim \nu^{-1/2}$.
As shown below, interaction of this boundary layer with locally trapped particle domains
is responsible for the ``anomalous'' behavior of Ware pinch coefficient at finite collisionalities.

\subsection{Finite collisionality, numerical examples}
\label{ssec:fincolnum}

For the numerical tests, we use the drift kinetic equation solver
NEO-2~\citep{kernbichler2008-S1061,kasilov14-092506,martitsch16-074007,kernbichler16-104001,kapper16-112511,kapper18-122509}
which, generally, employs a full linearized collision operator including the relativistic
effects~\citep{kapper18-122509}. Here, we restrict its collision operator to the Lorentz model only.
The magnetic field model corresponds to a circular tokamak with concentric flux surfaces ($\beta=0$) and with a
toroidal ripple-like perturbation, given in Boozer coordinates by
\be{magfield_mod}
B(r,\vartheta,\varphi) = B_0(r,\vartheta)
\left(1+\varepsilon_M \frac{B_0^2(r,\vartheta)}{B_{00}^2(r)}\cos(n\varphi)\right)^{-1/2}.
\ee
Here, $B_0(r,\vartheta)$ is the unperturbed field, $B_{00}(r)$ is the $(m,n)=(0,0)$ harmonic of this field and
$\varepsilon_M$ is the modulation amplitude. It can be checked that this field fulfills ``true surface''
condition~\eq{closure} at all rational flux surfaces, which is a property useful for the studies of bootstrap resonances
briefly discussed in Section~\ref{ssec:lonfieldlines}.
We fix $n=3$ and
$\varepsilon_M=0.1$
and consider two cases
of the rotational transform $\iota=1/q$
with $\iota=1/4$ and $\iota=2/5$ where the field line is closed after one and two poloidal turns, respectively.

\begin{figure}
\centerline{
\includegraphics[width=0.49\textwidth]{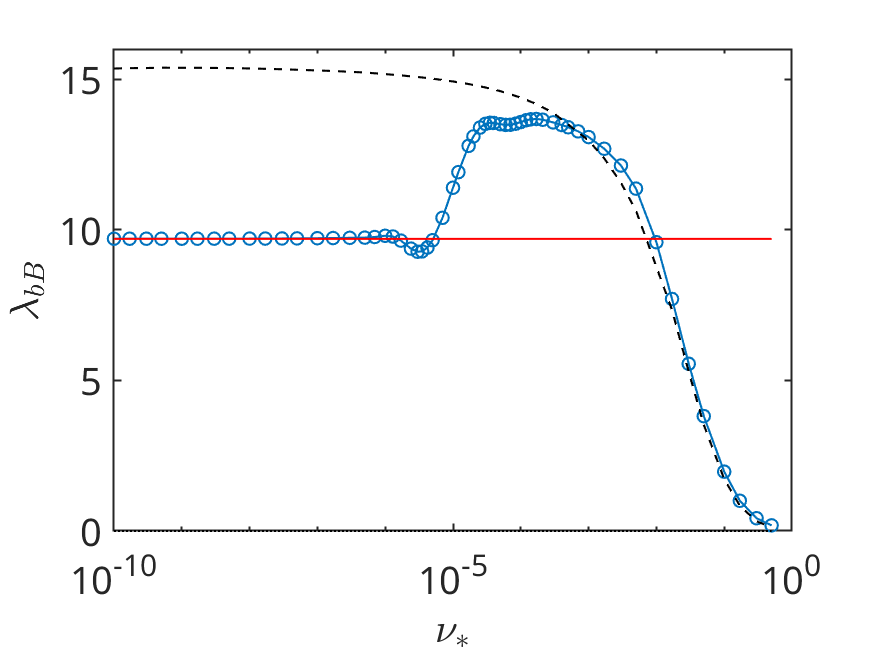}
\includegraphics[width=0.49\textwidth]{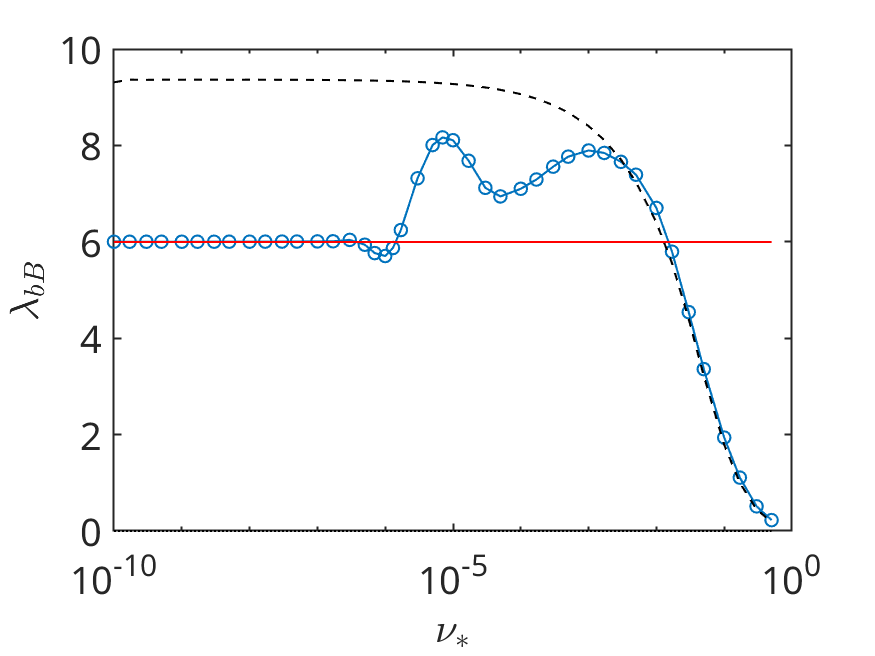}
}
\caption[]{Geometrical factor $\lambda_{bB}$ computed by NEO-2 (blue) for different normalized collisionalities $\nuast$
and computed from the Shaing-Callen limit~\eq{lambda_bB} evaluated by NEO (red) for $\iota=1/4$ (left) and $\iota=2/5$ (right).
Black dashed lines show the result of NEO-2 for axisymmetric fields  ($\varepsilon_{M}=0$).
}
\label{fig:lambda_bB_test_newtest}
\end{figure}
The
normalized bootstrap coefficient~\eq{D31_lambb} computed by NEO-2 for the above two cases as function of the normalized
collisionality $\nuast = \pi R \nu_\perp / v = \pi R /(2 l_c) = \pi \iota \nu^\ast$
where $R$ is the major radius of the magnetic axis and $\nu^\ast$ is the definition
of~\cite{beidler2011-076001}
is compared to Shaing-Callen limit~\eq{lambda_bB} in Fig.~\ref{fig:lambda_bB_test_newtest}.
(The latter asymptotical limit is computed here using the code NEO~\citep{nemov99-4622,nemov04-179}
which should not be confused with the drift kinetic code of Belli and Candy~\citep{belli15-054012} having the same name.)
It can be seen that the asymptotic value is abruptly reached in both cases at rather low collisionalities
$\nuast \sim 10^{-6}$ with the collisional $\lambda_{bB}$ dropping briefly below the Shaing-Callen limit prior
to convergence.
Naturally, the normalized Ware pinch coefficient $\lambda_{bB}^\dagger$, Eq.~\eq{ware-mono}, differs
from bootstrap coefficient only by small numerical errors and cannot be distinguished in this plot.
For the studies of the off-set of collisional bootstrap (Ware-pinch) coefficient from the
asymptotic value seen at low but finite collisionality we examine the distribution function
computed by NEO-2 in direct and adjoint problem separately.

\subsubsection{Direct problem}
\label{sssec:dirprob_num}

An odd part of the distribution function perturbation
$g^{\text{odd}}$
responsible for bootstrap current in the direct problem is shown in Fig.~\ref{fig:g_odd_direct}.
\begin{figure}
\centerline{
\includegraphics[width=0.49\textwidth]{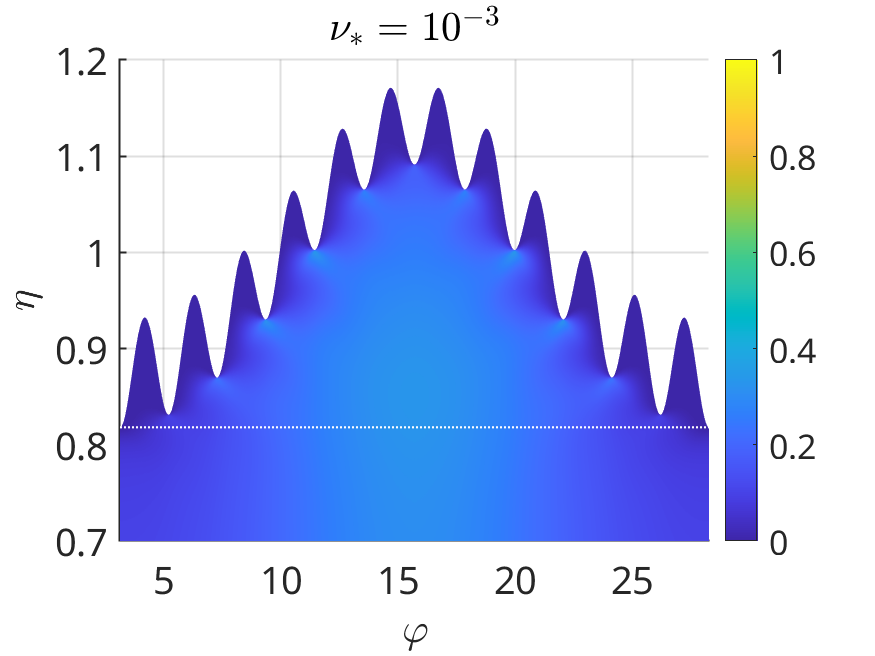}
\includegraphics[width=0.49\textwidth]{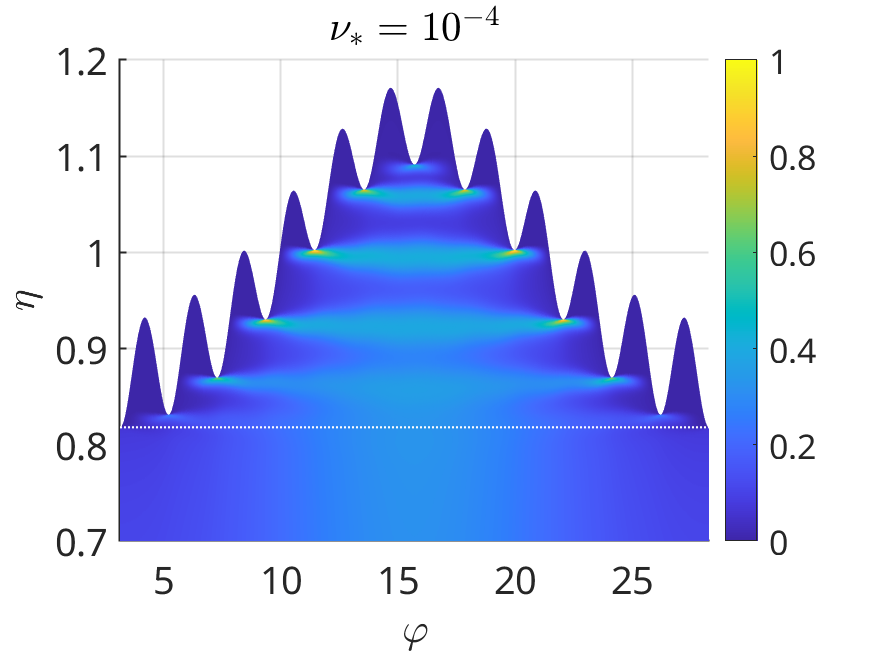}
}
\centerline{
\includegraphics[width=0.49\textwidth]{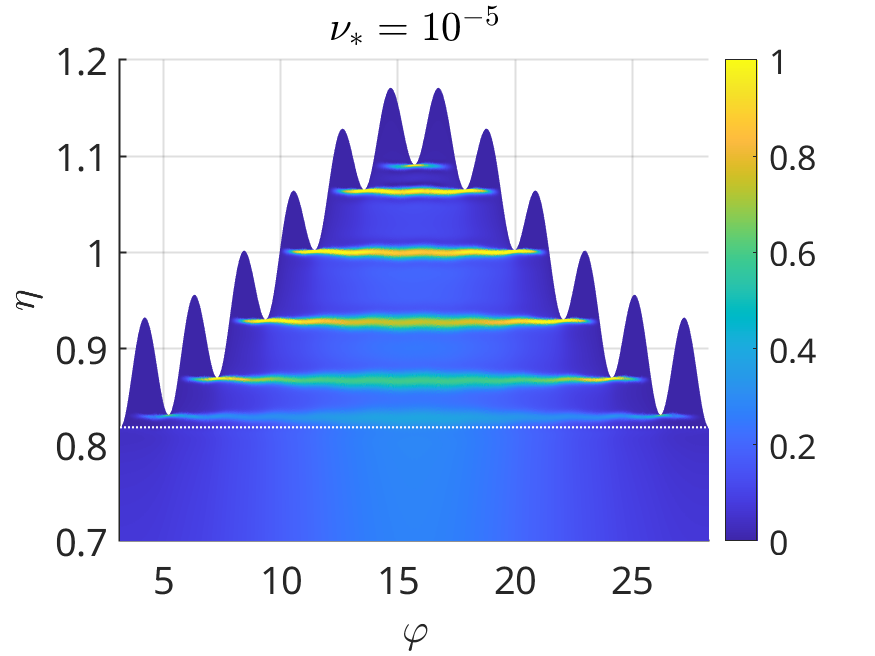}
\includegraphics[width=0.49\textwidth]{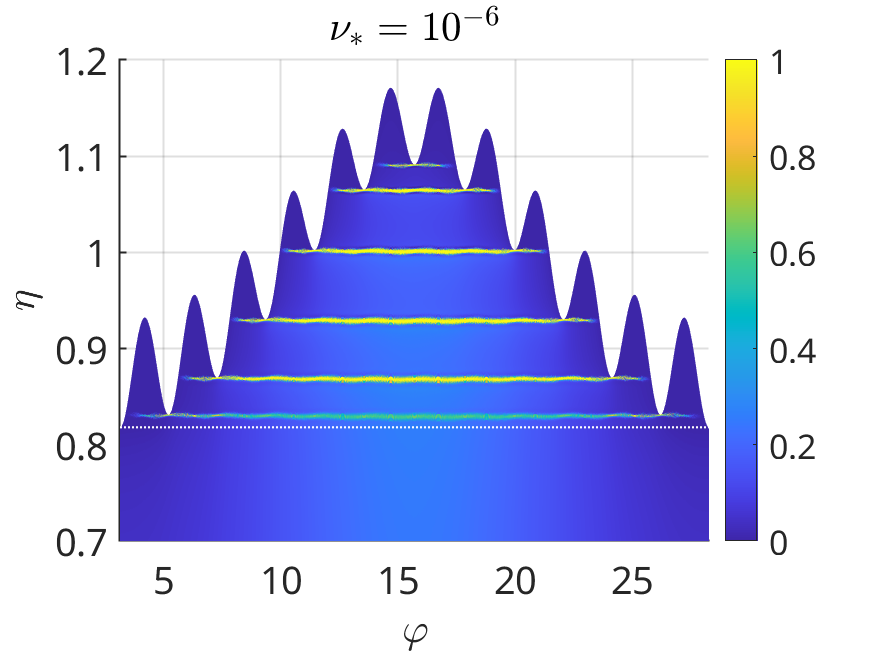}
}
\caption[]{Odd part of the distribution function,
$g^{\text{odd}}(\varphi,\eta)=(g(\varphi,\eta,1)-g(\varphi,\eta,-1))/2$,
driven by $s_{(1)}$ for different normalized collisionalities $\nuast$ (see the legend)
in case $\iota=1/4$.
The trapped-passing boundary is shown by a white dotted line.
}
\label{fig:g_odd_direct}
\end{figure}
One can observe the appearance of boundary layers around class transition boundaries
(but not at the trapped-passing boundary) with decreasing collisionality.
It can be seen that at lowest collisionality, $\nuast=10^{-6}$, where the asymptotic limit is reached
in Fig~\ref{fig:lambda_bB_test_newtest}, the highest class transition boundary layer
corresponding to two highest local maxima becomes clearly separated from the trapped-passing boundary.

The distribution of parallel equilibrium current in velocity space is better seen from Fig.~\ref{fig:intga}
where an integral
$\int_\eta^{1/B} \rd \eta^\prime\; g^{\text{odd}}$
is shown at the middle of the field line
($\varphi=\varphi_b=5\pi$). According to Eq.~\eq{pacd} value of this integral at $\eta=0$ up to a factor
equals parallel equilibrium current density. It can be seen that the contribution of the trapped
particle region to the parallel current is independent of collisionality despite the varying width
of class transition boundary layers which carry a significant amount of this current.
\begin{figure}
\centerline{
\includegraphics[width=0.49\textwidth]{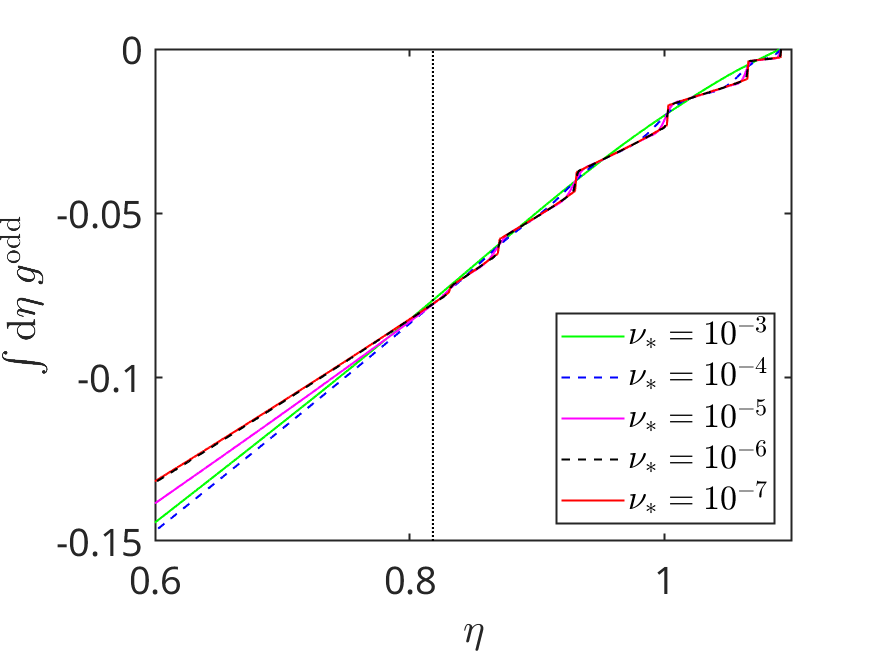}
\includegraphics[width=0.49\textwidth]{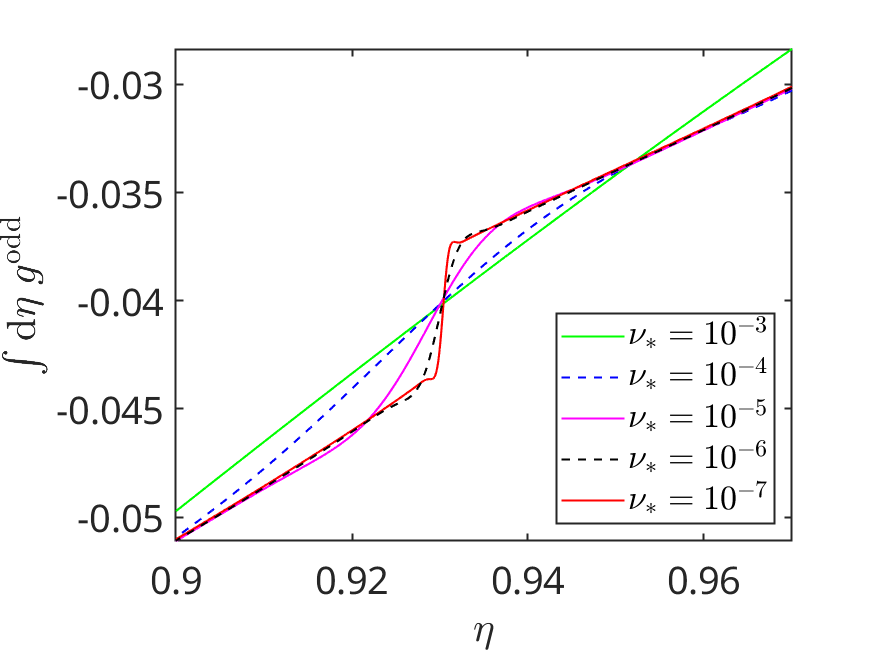}
}
\caption[]{
Integral
$\int_\eta^{1/B} \rd \eta^\prime\; g^{\text{odd}}$
as function of the lower limit (left),
and its zoom near one of class transition boundaries (right)
for various plasma collisionalities (see the legend) at $\varphi=5\pi$.
The trapped-passing boundary is shown by a vertical dotted line.
}
\label{fig:intga}
\end{figure}
Off-set of the current which depends on collisionality is produced in the passing particle region.
Solution
$g^{\text{odd}}$
in this region is determined up to arbitrary constant, which should match then
the solution in the trapped-passing boundary layer.
\begin{figure}
\centerline{
\includegraphics[width=0.49\textwidth]{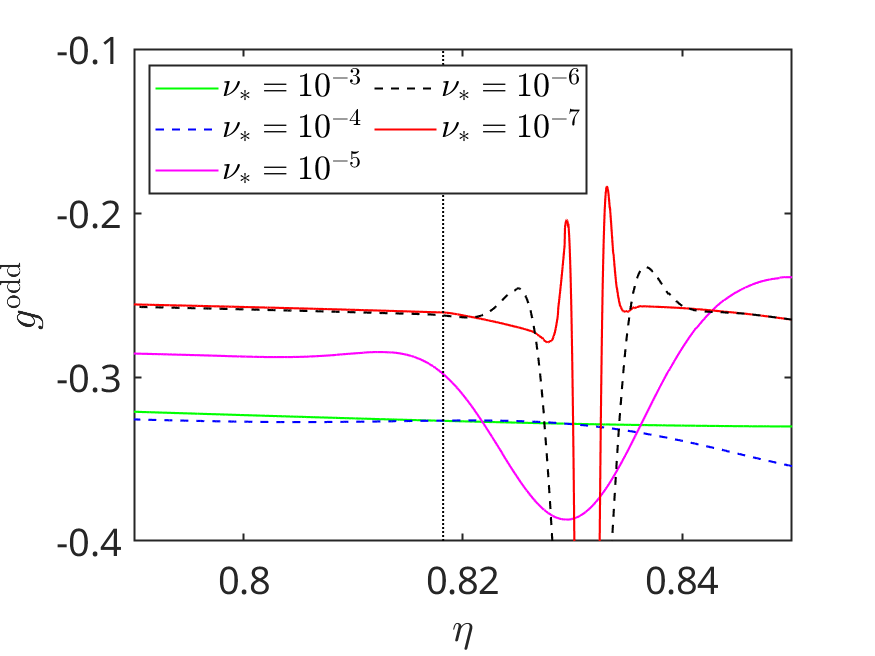}
}
\caption[]{
Odd part of the distribution function
$g^{\text{odd}}$
near the trapped-passing boundary (vertical dotted line)
at $\varphi=5\pi$ for various collisionalities (see the legend).
}
\label{fig:ga_nearbou}
\end{figure}
At sufficiently low collisionalities ($\nuast < 10^{-5}$), where bootstrap current is independent of collisionality
the nearest class transition boundary layer
is well separated from the trapped-passing boundary (see Fig.~\ref{fig:ga_nearbou}), and the boundary condition
for passing particles is fully determined by the last trapped orbit nearest to the boundary between
trapped and passing particle domains~\citep{boozer90-2408}. The rest of trapped particle domain has no effect on passing
distribution, because ``particles'' generated by bounce averaged drift in the first local ripple fully annihilate
in class transition boundary layer with ``anti-particles'' generated in the last local ripple where the sign
of geodesic curvature is opposite to the sign in the first ripple.
At higher collisionalities ($\nuast \ge 10^{-5}$), where the trapped-passing boundary layer crosses the class transition boundary,
some part of ``particles'' generated in the first ripple (where source $s_{(1)}>0$) and leaving this ripple with
positive velocity (in the direction of a local maximum) when changing due to collisions the trapping class,
can later cross the trapped-passing boundary without mirroring and enter a co-passing particle domain.
Similarly, ``anti-particles'' from the last ripple enter the counter-passing domain.
Both create a current off-set of the same sign. Naturally, they cannot annihilate because they enter different
domains.

Note that at some intermediate collisionality off-set changes sign (see Fig.~\ref{fig:lambda_bB_test_newtest}).
In this case, the class transition layer is almost isolated from the trapped-passing boundary, so that the probability
to enter the passing domain after a single pass along the boundary layer is lower than the probability to do that
after the first reflection near the global maximum. In such case, most ``particles'' enter counter-passing domain
(``anti-particles'' - co-passing) what results in negative current off-set.
\begin{figure}
\centerline{
\includegraphics[width=0.49\textwidth]{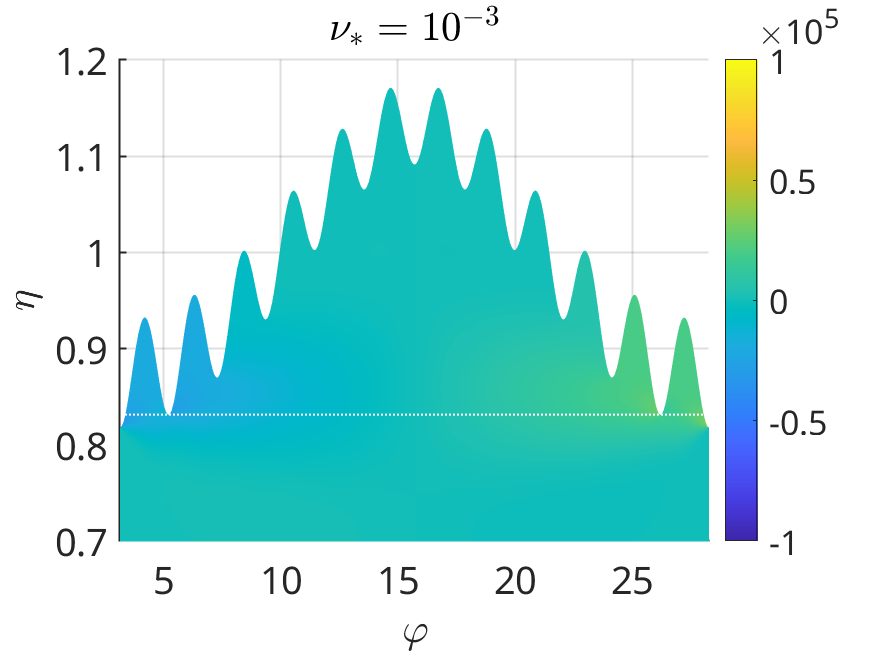}
\includegraphics[width=0.49\textwidth]{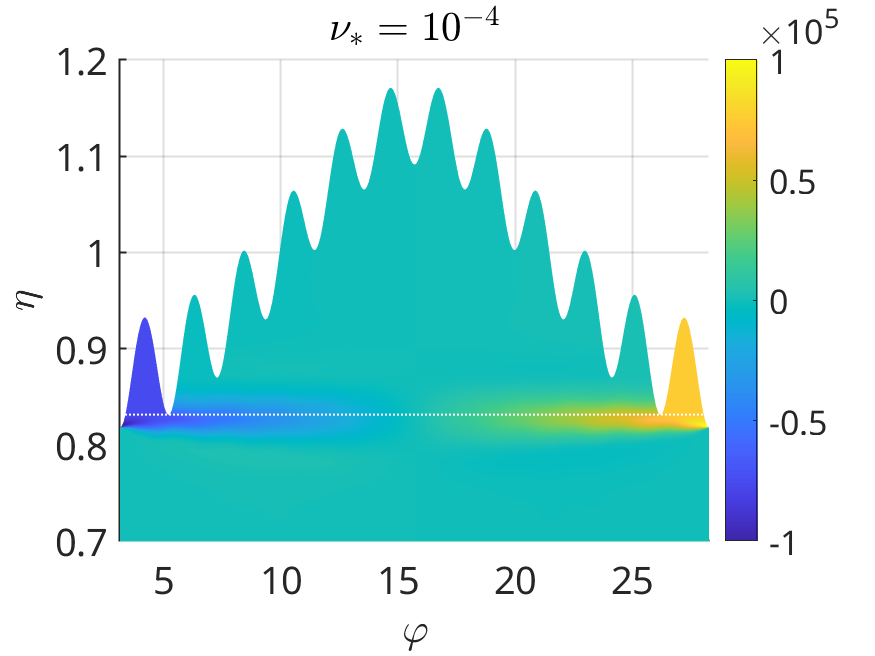}
}
\centerline{
\includegraphics[width=0.49\textwidth]{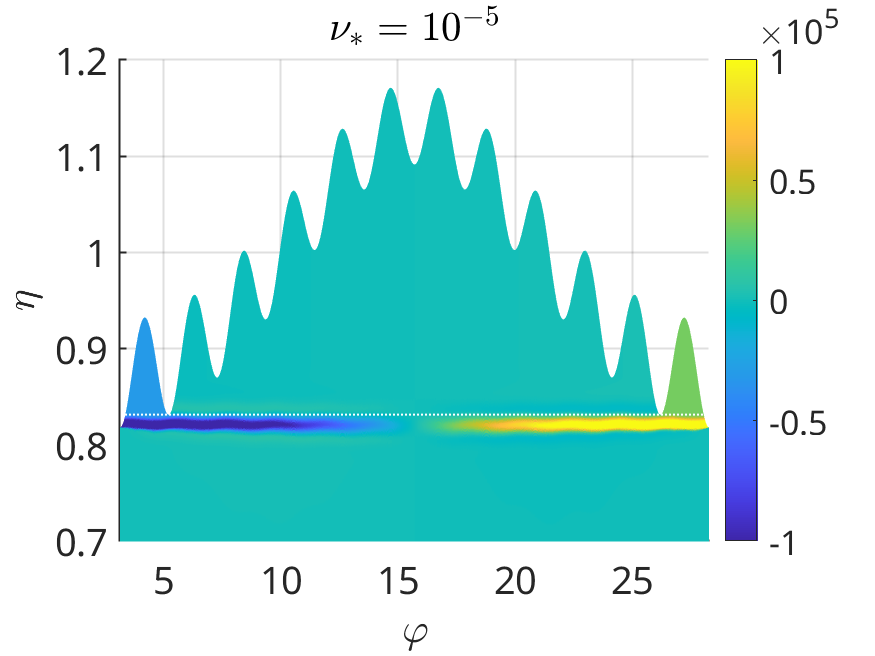}
\includegraphics[width=0.49\textwidth]{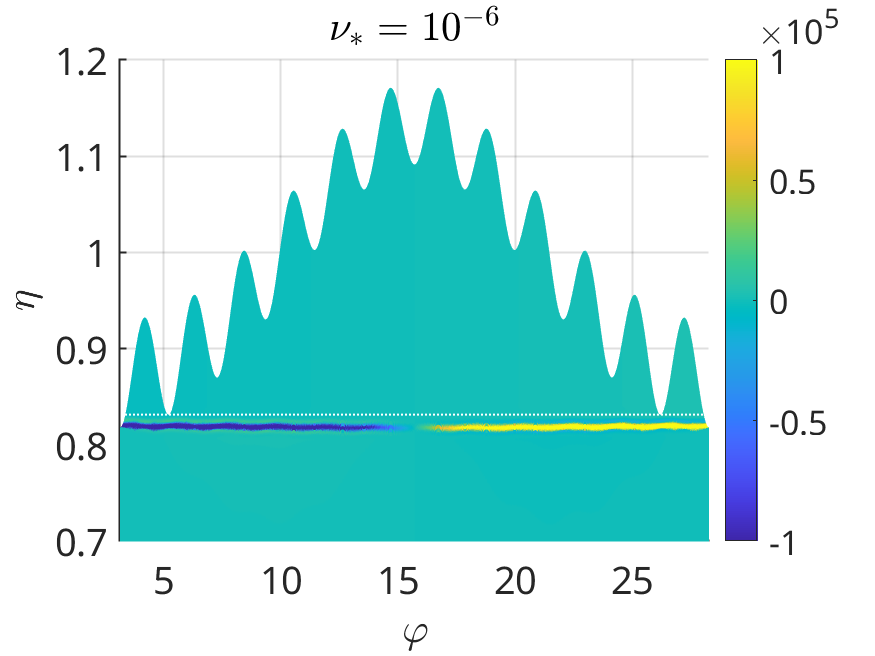}
}
\caption[]{
Even part of the distribution function,
$g^{\text{even}}(\varphi,\eta)=(g(\varphi,\eta,1)+g(\varphi,\eta,-1))/2$,
driven by $s_{(3)}$ for the same cases as in Fig.~\ref{fig:g_odd_direct}.
The transition boundary between the two highest trapping classes is shown by a white dotted line.
}
\label{fig:spitf_surf}
\end{figure}
\subsubsection{Adjoint problem}
\label{sssec:adjoint_num}
An even part of the distribution function perturbation
$g^{\text{even}}$
responsible for particle flux (Ware pinch)
in the adjoint problem is shown in Fig.~\ref{fig:spitf_surf} for the same cases as in Fig.~\ref{fig:g_odd_direct}.
As expected, boundary layer appears at low collisionality only at the trapped-passing boundary. The structure
of this layer well agrees with the complete analytical solution~\eq{g0_adj_comb}, i.e.
$g^{\text{even}}$
is an anti-symmetric
function of $\varphi$ with respect to the middle stellarator symmetry point $\varphi_b$.

\begin{figure}
\centerline{
\includegraphics[width=0.50\textwidth]{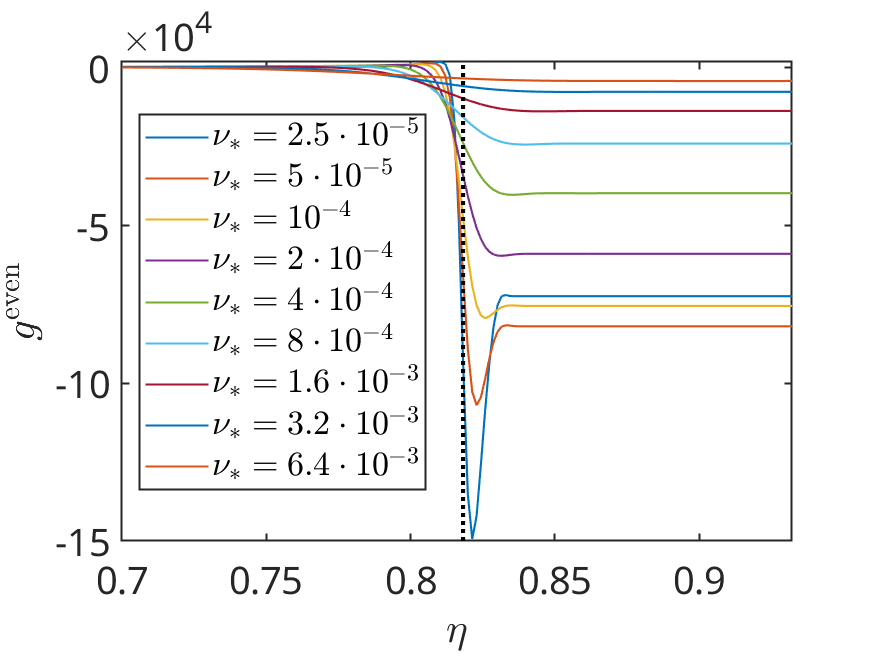}
\includegraphics[width=0.49\textwidth]{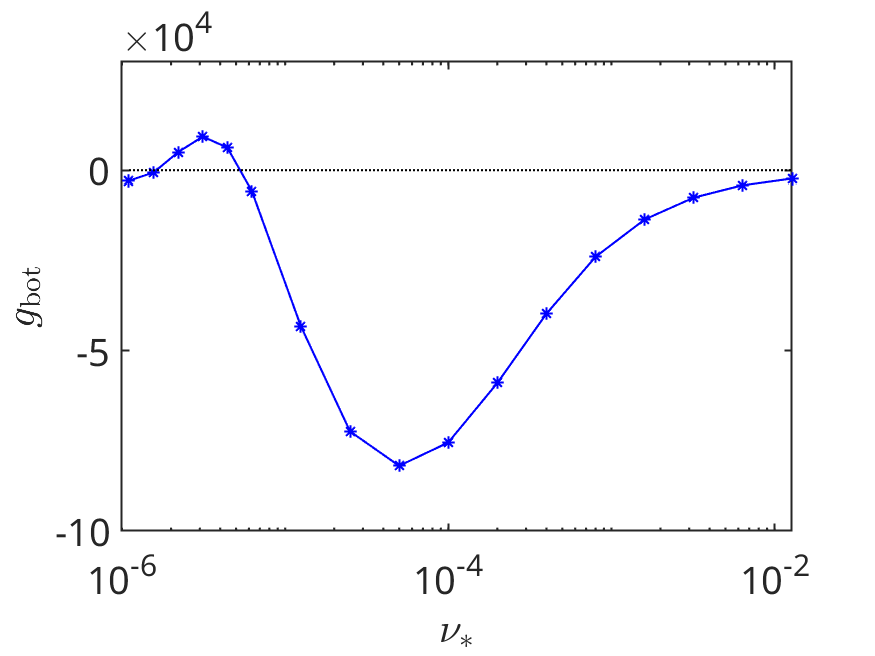}
}
\caption[]{Even part of the distribution function
$g^{\text{even}}$
as function of $\eta$ in the middle of the left off-set well,
$\varphi=4 \pi /3$ for various collisionalities (left) and $\iota=1/4$.
The trapped-passing boundary is shown by a vertical dotted line.
The value of
$g^{\text{even}}$
at the off-set well bottom, $g_\text{bot}$, is shown as a function of $\nuast$ in the right plot.
}
\label{fig:offset_neo2_pure}
\end{figure}
Similar to the direct problem, if the trapped-passing boundary layer is well separated from the nearest class
transition boundary (lowest collisionality case, $\nuast = 10^{-6}$)
the Ware pinch coefficient agrees with the asymptotic limit.
At higher collisionalities where this boundary layer enters a class-transition boundary, one can observe
increased
$g^{\text{even}}$
in the local ripple wells whose lowest maximum determines this transition boundary.
This accumulation of the particles in local ripples drives the off-set of Ware pinch since
$g^{\text{even}}$
there
is anti-symmetric
with respect to mid-point $\varphi_b=5\pi$,
what is true also for geodesic curvature and, respectively, for bounce-averaged radial
drift velocity in these local ripples. Therefore, particles in both ripples drive the flux of the same sign.
In the following, we will call such ripples ``off-set wells''.
As one can see from Fig.~\ref{fig:offset_neo2_pure},
$g^{\text{even}}$
is constant in most of the off-set well, and changes only in the vicinity of class transition boundary.
Therefore, we can characterize
$g^{\text{even}}$
in the whole off-set well by its value
at the off-set well bottom, which will be denoted as $g_{\text{off}}$. It can be seen that $g_{\text{off}}$ has similar feature
to $\lambda_{bB}$, i.e. it changes sign with reducing collisionality before going to zero.

\begin{figure}
\centerline{
\includegraphics[width=0.49\textwidth]{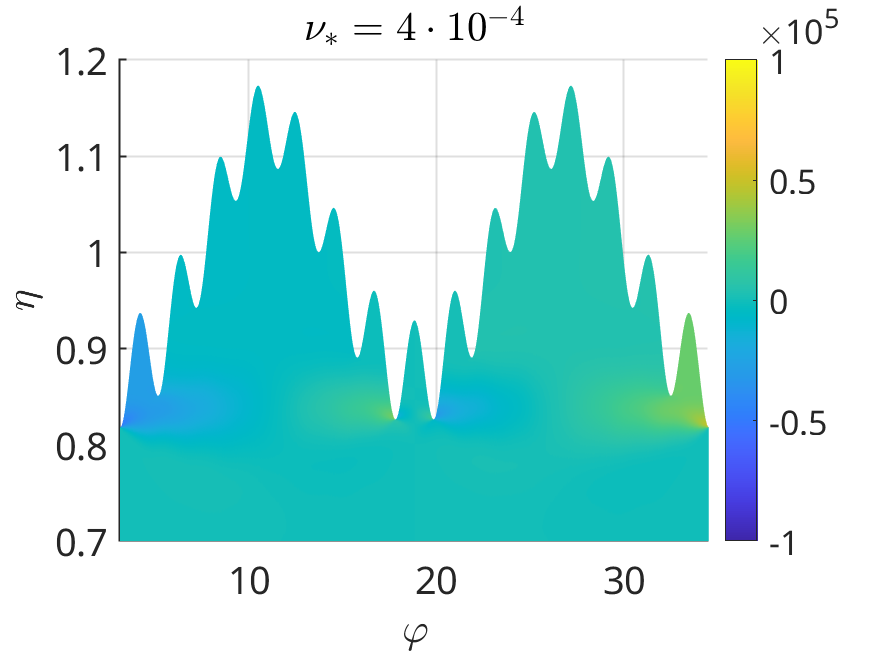}
\includegraphics[width=0.49\textwidth]{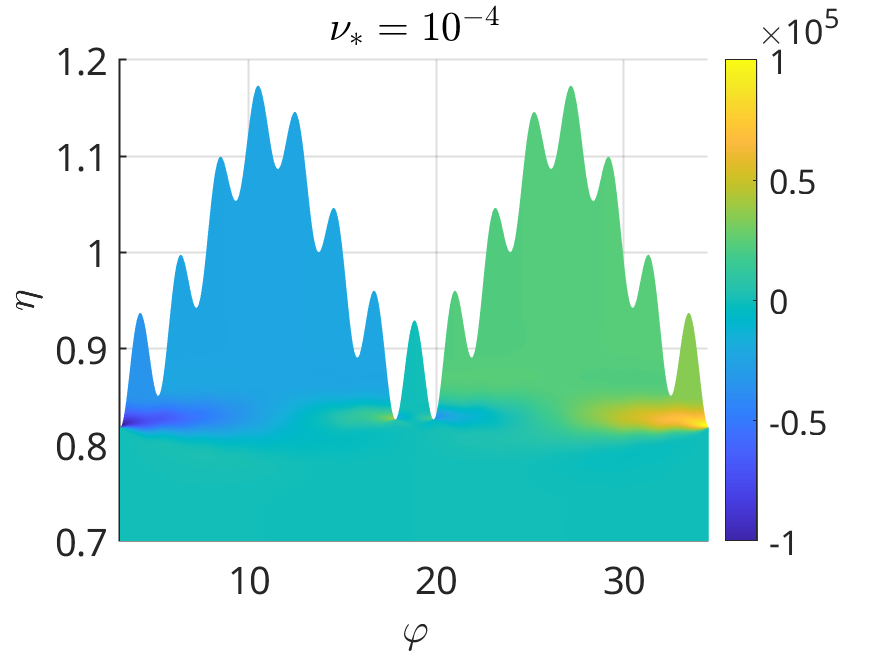}
}
\centerline{
\includegraphics[width=0.49\textwidth]{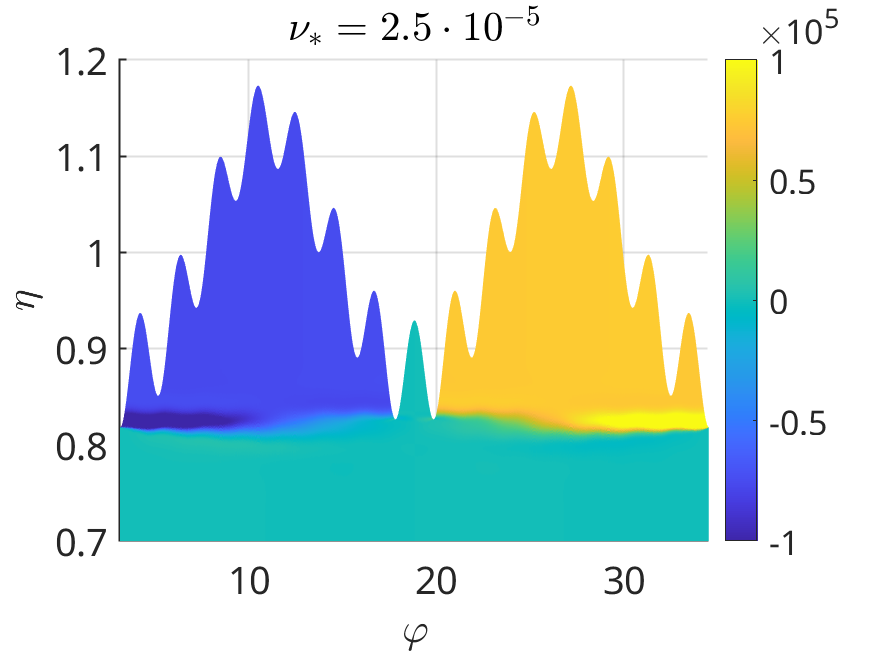}
\includegraphics[width=0.49\textwidth]{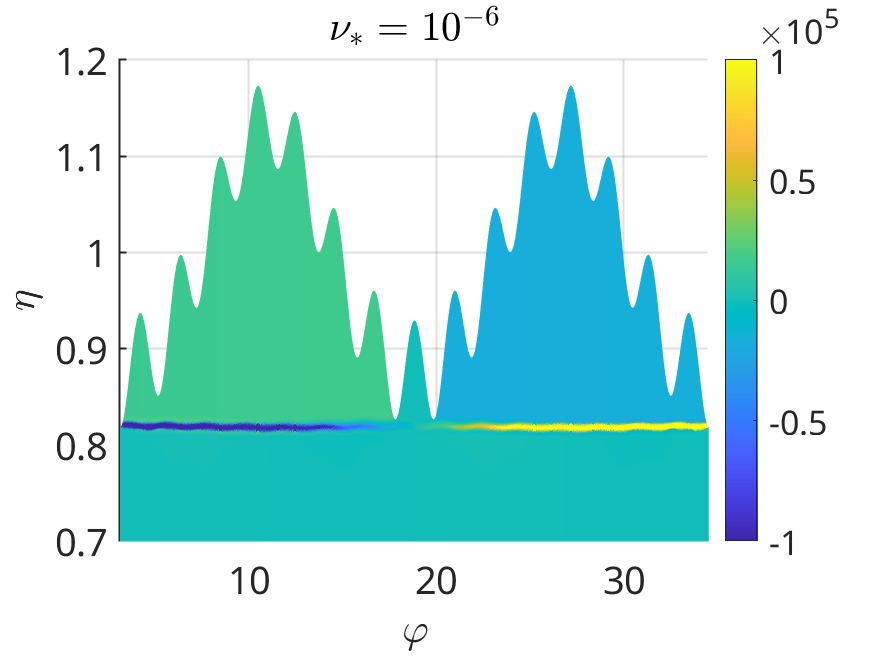}
}
\caption[]{The same as in Fig.~\ref{fig:spitf_surf} for $\iota=2/5$ and different collisionalities (see the legend).
}
\label{fig:spitf_surf_mod}
\end{figure}
Properties of this distribution function off-set, which
was also observed in GSRAKE modelling~\citep{beidler20}, will be studied in
more detail in the next section.
Here, we only show an extra example in Fig.~\ref{fig:spitf_surf_mod}
where more than one type of off-set wells is present simultaneously with different types of off-sets
dominating at different collisionalities. In this example, both, off-set wells containing only the lowest
class of trapped particles who traverse a single minimum during their bounce period (as in Fig.~\ref{fig:spitf_surf})
and an off-set wells containing many classes are shown. Note that trapped-passing boundary layer which
has a standard (collisionless) form
at lowest collisionality $\nuast=10^{-6}$ appears to be split in two independent layers at the intermediate
collisionalities $\nuast \le 10^{-4}$ where mismatch of two local maxima located near the middle of field
line with the global maximum becomes small compared to the typical boundary layer width (see the
discussion of few global maxima case after Eq.~\eq{g0_adj_comb}).
Thus, off-set phenomenon has rather high complexity even in the simple
case of a closed field line which is, nevertheless, illustrative for the realistic configurations.
In particular, switching off-sets can be observed in Fig.~3 of \citep{kernbichler16-104001}
where bootstrap coefficient has been computed for W-7X down to collisionalities $\nuast=10^{-9}$.

\section{Bootstrap / Ware-pinch off-set for a  closed field line}
\label{sec:bootware-closed}

Let us study the bootstrap / Ware-pinch off-set phenomenon in more details. Since both effects are quantitatively
the same due to the Onsager symmetry of transport coefficients, we focus now on the Ware pinch effect only.

\subsection{Boundary layer e-folding}
\label{ssec:efolding}

As we have seen from asymptotic solutions and numerical examples in both, direct and adjoint problem,
distribution function $g_0$ which is independent of collisionality in most phase space includes huge
contributions localized in boundary layers where they scale as $l_c^{1/2} \sim \nu^{-1/2}$ and formally
become infinite in the collisionless limit manifested by $\delta$-functions in the asymptotic
solutions.
To estimate the decay of these localized contributions outside the boundary layer, we consider the
homogeneous kinetic equation~\eq{mono_Ak_norm} and use the ansatz 
which is only slightly different from the ansatz of \cite{helander11-092505} and leads to the same approximation at the end.
Namely, we replace in the trapped particle domain the independent variable  $\varphi$ with
\be{canpol}
\theta_H = \theta_H(\varphi,\eta,\sigma)
= \frac{\pi\sigma}{I_j}
\int\limits_{\varphi^-_j}^\varphi \frac{\rd \varphi^\prime |\lambda|}{ B^\varphi},
\ee
where the integral $I_j=I_j(\eta)$ is defined in Eq.~\eq{bas1_defs}.
Thus, particles with $\lambda > 0 $ are described by $0 < \theta_H < \pi$, particles
with $\lambda < 0 $ are described by $-\pi < \theta_H < 0$, and continuity of the distribution function
at both turning points is enabled by periodicity of $g$ with $\theta_H$.
Formally presenting the homogeneous equation~\eq{mono_Ak_norm} in tensor form
$$
\difp{}{z^i} {\cal J}\left({\cal V}^i g- {\cal D}^{ij}\difp{g}{z^j}\right)=0,
$$
where $z^i=(\varphi,\eta)$, and transforming the contra-variant components of the phase space velocity ${\cal V}^i$ and 
diffusion tensor ${\cal D}^{ij}$
and the Jacobian $\cal J$ to the new phase space coordinates $(\theta_H,\eta)$ using tensor algebra rules,
this equation takes the form
\be{four_inv_cons}
\difp{g}{\theta_H}
=
\left(\difp{}{\eta} +\difp{}{\theta_H} \difp{\theta_H}{\eta}\right)
D_H
\left(\difp{g}{\eta} + \difp{\theta_H}{\eta}\difp{g}{\theta_H}\right),
\ee
where differential operators $\partial/\partial \eta$ and $\partial/\partial \theta_H$ act on everything to the right, and
\be{Dcan}
D_H = D_H(\eta)
= \frac{\eta I_j}{\pi l_c}.
\ee
Expanding the distribution function in Fourier series over $\theta_H$,
\be{canseries}
g(\theta_H,\eta)=\sum\limits_{m=-\infty}^\infty g_{[m]}{\rm e}^{i m \theta_H},
\ee
Eq.~\eq{four_inv_cons} turns into a coupled ODE set for Fourier amplitudes $g_{[m]}(\eta)$.
It should be noted that terms with $\partial \theta_H / \partial \eta$ which lead to the coupling
of Fourier modes are negligible small in most of phase space except for the close vicinity
of the boundary (sub-layer~\citep{helander11-092505}) where due to the scaling
$\partial^2 \theta_H / \partial \eta^2 \sim \eta^{-1} (\eta-\eta_b)^{-1}$ their contribution is
comparable
with the left-hand side which means the sub-layer width $\eta-\eta_b \sim \nuast \ll \sqrt{\nuast}$.
\begin{figure}
\centerline{
\includegraphics[width=0.49\textwidth]{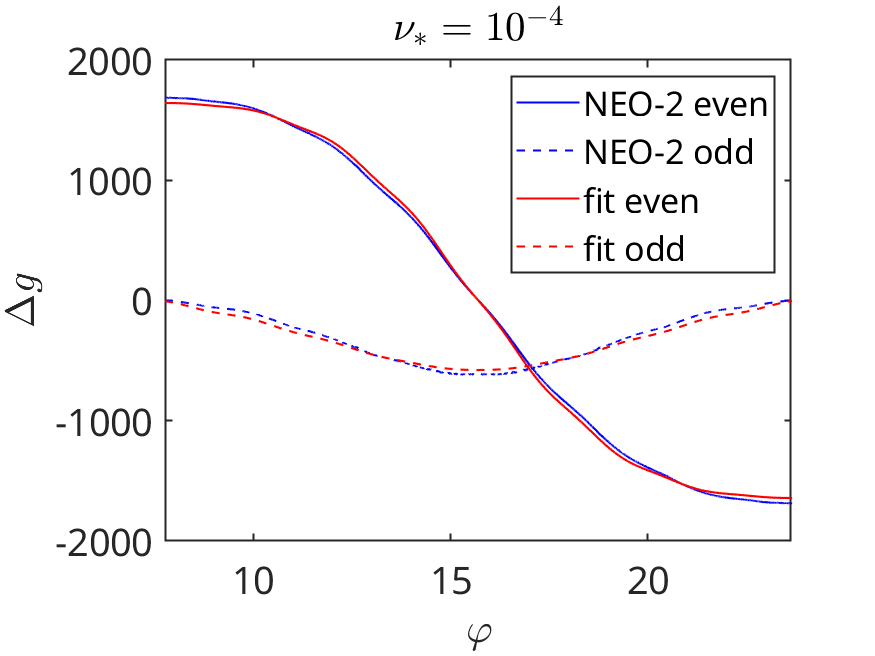}
\includegraphics[width=0.49\textwidth]{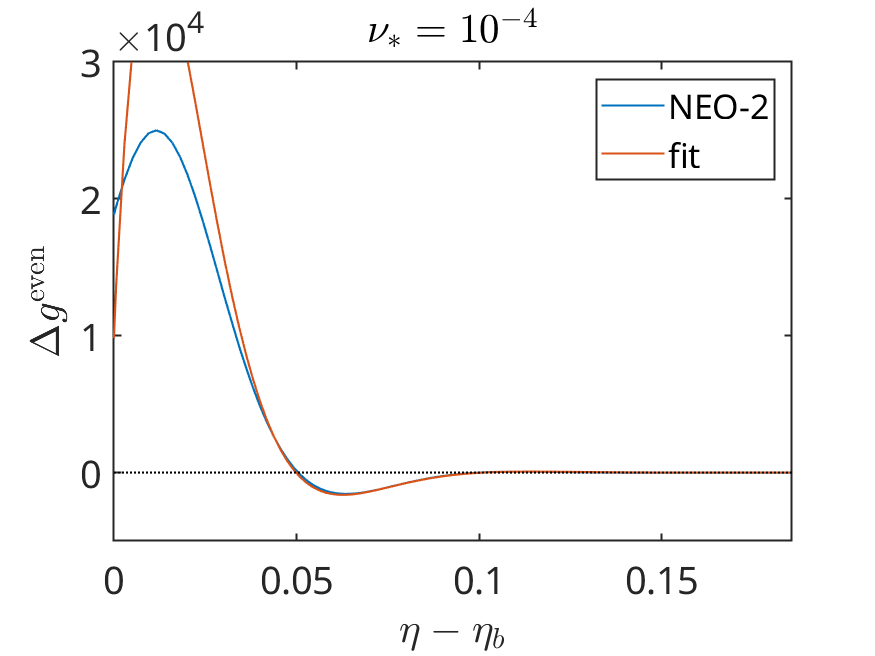}
}
\centerline{
\includegraphics[width=0.49\textwidth]{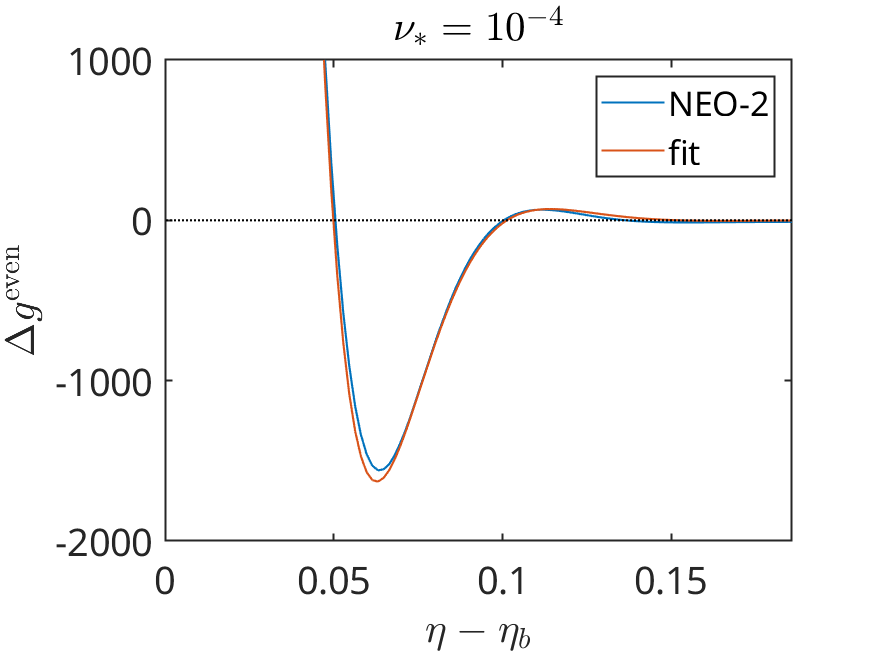}
\includegraphics[width=0.49\textwidth]{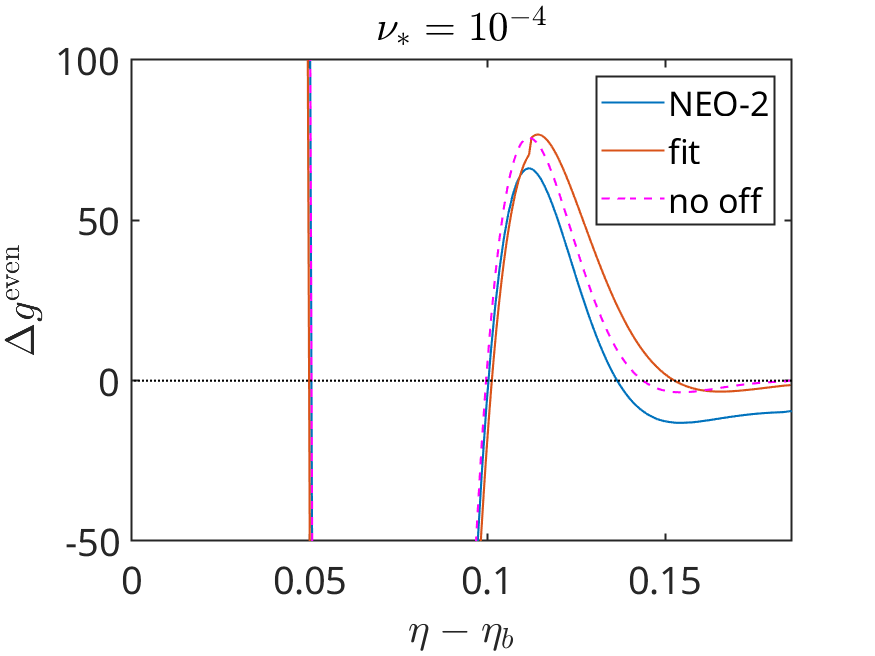}
}
\caption[]{
Difference $\Delta g$ and the fit $\Delta g_\text{fit}$, Eq.~\eq{gmdecay}
for the case $\nuast=10^{-4}$ in Fig.~\ref{fig:spitf_surf}.
Upper left - even and odd parts as functions of $\varphi$ for $\eta-\eta_b=0.07$.
Upper right - even parts as functions of $\eta$ for $\varphi=19 \pi /3$ (7-th local maximum).
Lower panel - zooms of the upper right plot over Y-axis.
Dashed line in the last zoom corresponds to NEO-2 result for
$g^\text{even}-g_{\lambda=0}$ where $g_{\lambda=0}$ is $g^\text{even}$ for standing
particles with $\eta=1/B$. Function $g_{\lambda=0}$ differs here from $g_0=g_0^t$ by an exponentially small off-set.
}
\label{fig:efolding}
\end{figure}
Thus, one can ignore the right-hand side terms with derivatives over $\theta_H$ outside the sub-layer,
and the set decouples into
\be{fourcons}
im g_{[m]}
=
\difp{}{\eta}D_H\difp{g_{[m]}}{\eta}.
\ee
Harmonic
\be{g0harm}
g_{[0]} = \frac{1}{2 I_j}\sum_{\sigma = \pm 1}
\int\limits_{\varphi^-_j}^{\varphi^+_j}\rd \varphi\frac{|\lambda|g}{B^\varphi}
\ee
is actually zero in our case because
$\delta$-like terms in the direct
problem are odd, and the solution of the adjoint problem is stellarator-antisymmetric.
Ignoring in~\eq{fourcons} variation of $D_H$ in the main boundary layer,
$D_H(\eta) \approx D_H(\eta_b)$,
which brings this equation in agreement with the result of the ``rectangular well'' ansatz of \cite{helander11-092505},
we look for solutions in the form $g_{[m]} \propto \exp(\kappa_m (\eta-\eta_b))$ with $m=1$
harmonic being dominant with increasing $\eta-\eta_b$. Thus, boundary layer contribution decays outside
the layer as
\be{gmdecay}
\Delta g_\text{fit} \approx a \exp\left(\frac{\eta_b-\eta}{\delta \eta_e}\right)
\left(
\cos\left(\frac{\eta_b-\eta}{\delta \eta_e}+\phi\right)
\cos\left(|\theta_H|\right)
-
\sigma \sin\left(\frac{\eta_b-\eta}{\delta \eta_e}+\phi\right)
\sin\left(|\theta_H|\right)
\right),
\ee
where $|\theta_H|=|\theta_H(\varphi,\eta)|$,
and $\delta \eta_e$ is boundary layer e-folding length,
\be{deltaeta}
\delta \eta_e = \sqrt{2 D_H(\eta_b)} = \sqrt{\frac{2 \eta_b I_j(\eta_b)}{\pi l_c}}
\sim \eta_b \left(\frac{\Delta B}{B}\right)^{1/4}
\left(\frac{\nuast\Delta \varphi}{2\pi}\right)^{1/2},
\ee
estimated in the last expression
via typical variation of $B$ in the magnetic well $\Delta B$ and
toroidal extent of the boundary layer $\Delta \varphi$.

\begin{figure}
\centerline{
\includegraphics[width=0.49\textwidth]{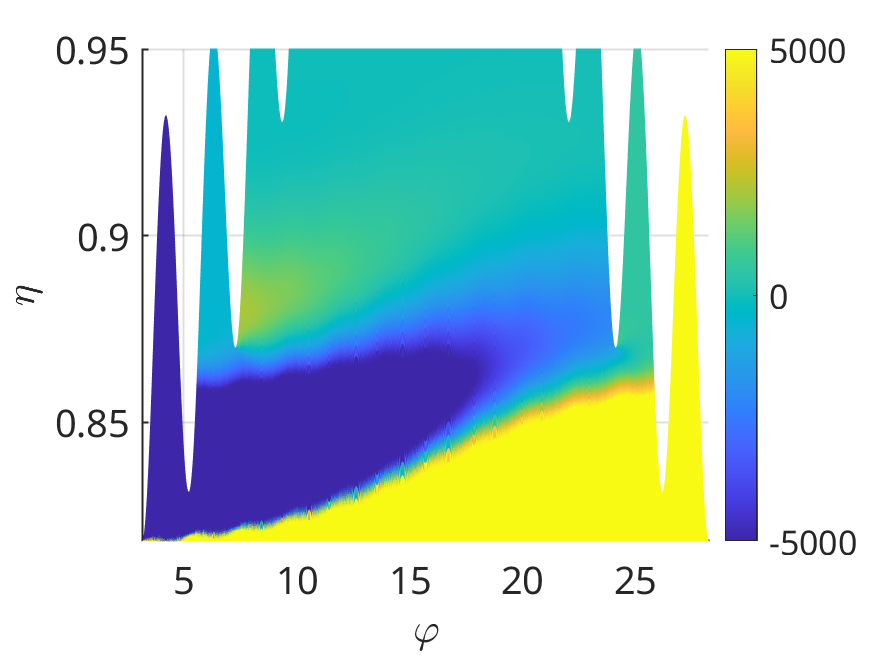}
\includegraphics[width=0.49\textwidth]{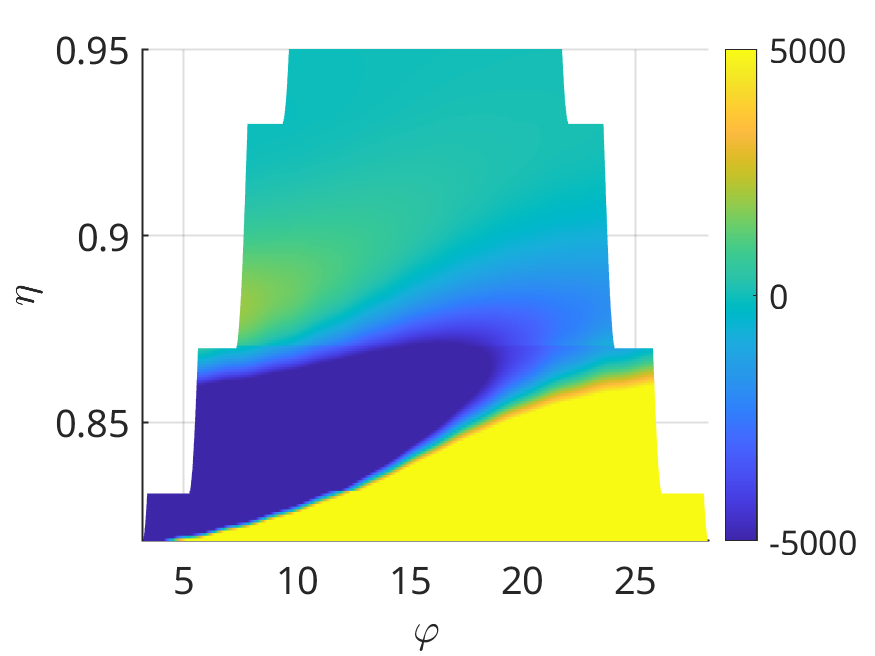}
}
\centerline{
\includegraphics[width=0.49\textwidth]{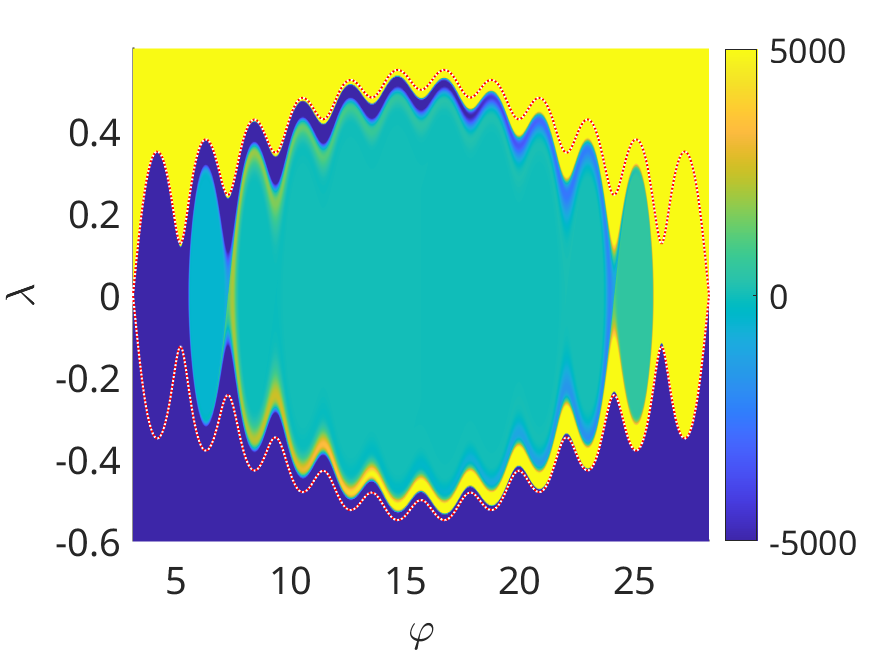}
\includegraphics[width=0.49\textwidth]{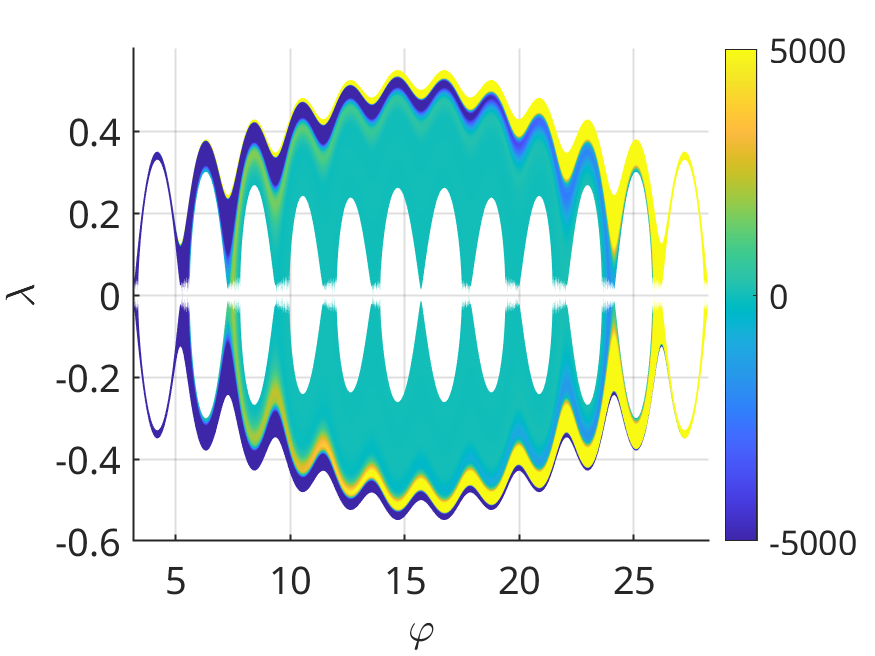}
}
\caption[]{
Difference $\Delta g$ (left) and its fit $\Delta g_\text{fit}$ (right)
as functions of $(\varphi,\eta)$ and $\sigma=1$ (upper panel) and $(\varphi,\lambda)$ (lower
panel) for the same case as in Fig.~\ref{fig:efolding}.
Saturation of color scale at $|\Delta g| \ge 5000$ occurs in the whole
passing particle region. Fit $\Delta g_\text{fit}$ is plotted excluding
passing particle domains and local trapping domains where $\Delta g_\text{fit}$ is not defined.
Red dotted line in $\Delta g(\varphi,\lambda)$ plot shows the trapped-passing boundary
where transient particles move clockwise.
}
\label{fig:efolding_2d}
\end{figure}
In Figs.~\ref{fig:efolding} and~\ref{fig:efolding_2d},
boundary layer contribution represented by the difference $\Delta g = g-g_0$
between the numerical (NEO-2) and analytical (Eq.~\eq{g0_adj_tr}) solution of the adjoint problem
is compared to the asymptotic boundary layer solution $\Delta g_\text{fit}$ given by Eq.~\eq{gmdecay} with
amplitude $a$ and phase $\phi$ fitted to match $\Delta g$
in the upper left plot of Fig.~\ref{fig:efolding}.
Oscillations of $\Delta g$ with $\eta$ in Fig.~\ref{fig:efolding}
result from interchanging dominance of ``particles' and ``anti-particles''
which enter the trapping domain and become mirrored there once (twice, etc.). Since the observation point
in the last three plots is in the region $\varphi_b<\varphi<\varphi_N$, ``particles'' dominate
there over ``anti-particles'' in the first peak (see also Fig.~\ref{fig:spitf_surf})
which they produce before the first reflection. The above mechanism results in helical structures
in $g(\varphi,\lambda)$ distribution formed in the trapped particle domain (see Fig.~\ref{fig:efolding_2d}).
They are produced by the combination of the collisionless phase space flow (which is clockwise
in this domain) and the flow driven by the collisional diffusion (directed over $\lambda$ to the trapped particle domain
from the co- and counter-passing particle domains, where sources of ``particles'' and ``anti-particles'' are located, respectively).

Note that oscillatory behavior similar to that in Fig.~\ref{fig:efolding} is seen
also in the class-transition boundary layers
in Fig.~\ref{fig:ga_nearbou}
and is reflected then in the small
bootstrap off-set
oscillation in Fig.~\ref{fig:lambda_bB_test_newtest} at low collisionality where it decreases exponentially.

\subsection{Propagator method, leading order solution}
\label{ssec:propmet}

For more detailed analysis of boundary layer effects on bootstrap / Ware pinch off-set, we use propagator method.
We look for the solution of the adjoint problem in the whole phase space in the form
\be{ansatz_adj}
g_{(3)}(\varphi,\eta,\sigma) = \Theta(\eta_b-\eta)g_{-1}(\eta)+g_0^{t}(\varphi)+g_\text{bou}(\varphi,\eta,\sigma),
\ee
where $g_{-1}$ and $g_0^{t}$ are given by Eqs.~\eq{ware_0} and~\eq{g0_adj_tr}, respectively.
The kinetic equation~\eq{mono_Ak_norm} corresponding to the source $s_{(3)}$ transforms to
\be{eq_g_bou}
\hat L g_\text{bou} = s_\text{bou} + \Delta s,
\ee
where the source terms are
\bea{sources_bou}
s_\text{bou} &=& \sigma l_c \delta(\eta_b-\eta) \frac{D_\eta\langle B^2 \rangle}{\langle |\lambda| \rangle},
\\
\Delta s
&=& -\sigma l_c  \Theta(\eta_b-\eta) \difp{}{\eta} \frac{D_\eta\langle B^2 \rangle}{\langle |\lambda| \rangle}
- \sigma l_c \difp{}{\eta}\left(
\delta(\eta_b-\eta) D_\eta\int\limits_\eta^{\eta_b} \rd \eta^\prime
\frac{\langle B^2 \rangle}{\langle |\lambda| \rangle}
\right).
\label{Delta_s_bou}
\eea
In the following, we use a formal solution of Eq.~\eq{eq_g_bou} in terms of its Green's function, which makes it evident
that the last term in Eq.~\eq{Delta_s_bou} does not contribute to the solution since $\eta$-integral of this term
weighted with any function with finite derivative over $\eta$ is zero. The remaining term, as we check later, provides
a correction of the order $\delta \eta_e$ to the solution driven by $s_\text{bou}$, Eq.~\eq{sources_bou}. Thus, we ignore
$\Delta s$ fully in the leading order solution.
Note that $g_\text{bou}$ is an aperiodic function in the passing region
due to aperiodicity of $g_o^t$ in the definition~\eq{ansatz_adj}. The leading order solution is treated nevertheless
as periodic, with the aperiodic term included in the ignored correction driven by $\Delta s$.
It should be noted that correction terms ignored here may become important in a special case where the leading order solution
tends to vanish. These cases are discussed below in Sections~\ref{ssec:align} and~\ref{ssec:offset_imperfect}.

We make the following simplifications. As follows from the analysis in Section~\ref{ssec:efolding}, solutions driven
by sources located in the passing region or at the trapped-passing boundary tend exponentially to constants of
$\eta$ at some sufficiently large distance to the trapped-passing boundary, $\eta > \eta_m$, where matching boundary
$\eta_m$ satisfies $\delta\eta_e \ll \eta_m -\eta_b \ll \eta_b$. Generally, these constants can be different
in different ripple wells in case they are bounded on both sides by the maxima fulfilling $\eta_m B(\varphi_j)> 1$,
see Figs.~\ref{fig:spitf_surf} and~\ref{fig:offset_neo2_pure}. Thus, we solve Eq.~\eq{eq_g_bou} only in the domain
$\eta < \eta_m$ imposing at the matching boundary zero flux condition,
$\partial g_\text{bou}(\varphi,\eta_m,\sigma) / \partial \eta_m=0$.
The boundary $\eta=\eta_m$ serves then as an upper boundary in most trapped particle domain except the interface regions
near global maximum and local maxima fulfilling $\eta_m B(\varphi_j) > 1$ where
$\eta_m B(\varphi) > 1$,
see
Fig.~\ref{fig:rect_domains}. Since widths of these regions $\delta\varphi$ are small at low collisionalities,
\mbox{$\delta\varphi \sim B\delta\eta_e^{1/2}\left(\partial^2 B / \partial \varphi^2\right)^{-1/2} \propto l_c^{-1/4}$},
we ignore the collision and source terms there, which leads in such a leading order solution to an error of the order
$|\lambda|\delta\varphi \sim (\delta\eta_e B)^{1/2} \delta\varphi \sim \delta\eta_e B^{3/2} \left(\partial^2 B / \partial \varphi^2\right)^{-1/2} \propto l_c^{-1/2}$,
being of the same order with an error introduced by ignoring the sub-layer in Section~\ref{ssec:efolding}.
(For this error estimate in the main boundary layer, $|\eta-\eta_b| \sim \delta\eta_e$,
we compared the variance $2\int\rd\varphi D_\eta$ in particle $\eta$ introduced by the collisions
when traversing the interface region with such variance for the whole trapping region
assuming
in the latter
for simplicity
that both, $\lambda$ and region extent in $\varphi$ are of the order one.
Error of the above model is of the order one within the sub-layers where
effect of collisions on $\lambda$ becomes comparable with the effect of mirroring force,
but the overall contribution of these narrow sub-layers is similarly small.)
\begin{figure}
\centerline{
\includegraphics[width=0.49\textwidth]{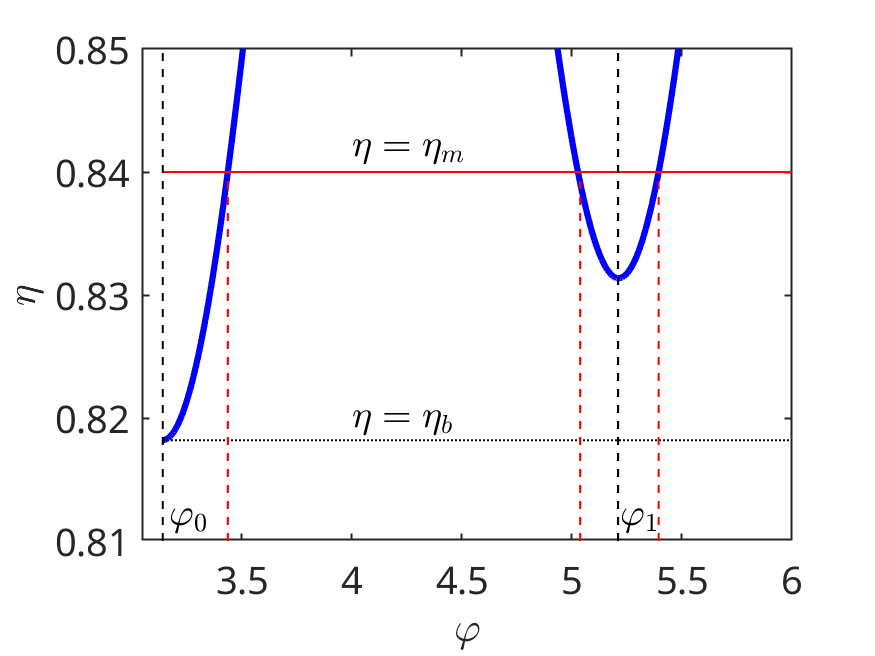}
\includegraphics[width=0.49\textwidth]{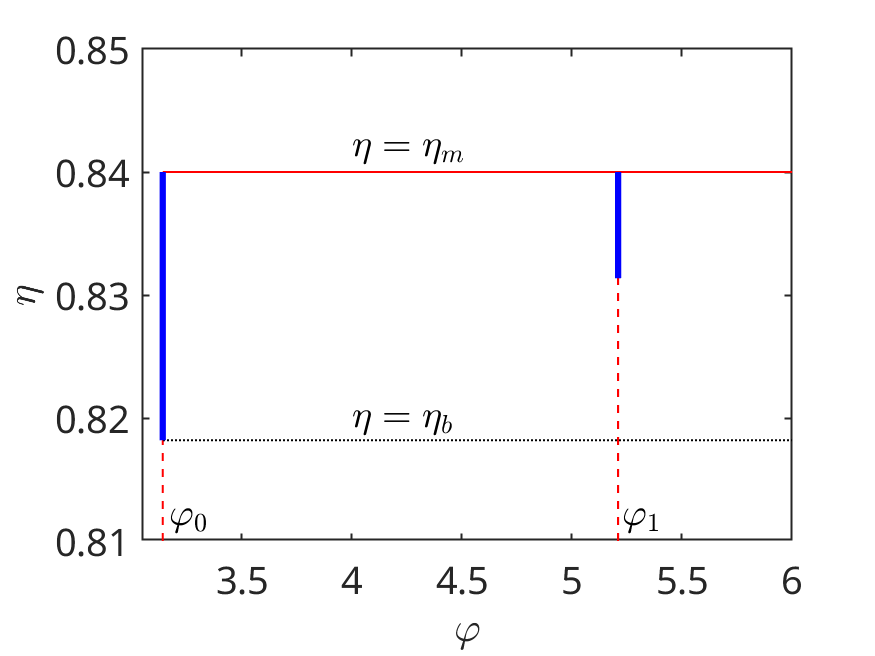}
}
\caption[]{Computation domain before (left) and after (right) simplifying transformation.
Trapped-passing boundary $\eta=\eta_b$ and
matching boundary $\eta=\eta_m$ are shown by black dotted and red solid line, respectively. Original domain boundary
$\eta=1/B(\varphi)$ (solid blue), field maxima $\varphi_j$ (dashed black) and boundaries of interface regions
$\varphi_j \pm \delta \varphi$ (dashed red) are shown in the left plot. Reflecting (solid) and transparent (dashed)
boundaries of the transformed domain are shown in the right plot.
}
\label{fig:rect_domains}
\end{figure}

The solution is trivial in the interface regions, and we can express it via mapping between the boundaries
$\varphi_j \pm \delta \varphi$ where $\varphi_j$ is maximum point,
\bea{inter_map}
g_\text{bou}(\varphi_j \pm \delta \varphi,\eta,\pm 1) &=& g_\text{bou}(\varphi_j \mp \delta \varphi,\eta,\pm 1),
\qquad \eta B(\varphi_j) < 1,
\nonumber \\
g_\text{bou}(\varphi_j \pm \delta \varphi,\eta,\pm 1) &=& g_\text{bou}(\varphi_j \pm \delta \varphi,\eta,\mp 1),
\qquad \eta B(\varphi_j) > 1.
\eea
Similar to Section~\ref{ssec:efolding}, we ignore the dependence of diffusion coefficient on $\eta$
setting $D_\eta(\varphi,\eta) \rightarrow D_\eta(\varphi,\eta_b)$ and, finally, ignore the extent of interface regions,
$\delta \varphi \rightarrow 0$, with each of these steps introducing an error of the order
$B\delta \eta_c \sim l_c^{-1/2}$, as before. Thus, we have transformed our computation domain to a set
of rectangular domains $0<\eta < \eta_m$, $\varphi_j < \varphi < \varphi_{j+1}$, limited by relevant field
maxima, $\eta_m B(\varphi_j) > 1$ (see Fig.~\ref{fig:rect_domains}).
This transformation leads to similar simplifications of the kinetic equation as the transformation
used in Section~\ref{ssec:efolding} for the main trapping region
(or for the whole field line in case of vanishing boundary layer width)
and is the same as the transformation of \cite{helander11-092505}
extended here to multiple local trapping regions.

Introducing Green's function $G_\sigma(\varphi,\eta,\varphi^\prime,\eta^\prime)$ which satisfies
with respect to $(\varphi,\eta)$ variables the homogeneous equation~\eq{eq_g_bou} with
$D_\eta=D_\eta(\varphi,\eta_b)$, boundary conditions at the strongly passing and the matching boundary,
\be{bougreen}
\left(\difp{}{\eta} G_\sigma(\varphi,\eta,\varphi^\prime,\eta^\prime)\right)_{\eta=0}
=
\left(\difp{}{\eta} G_\sigma(\varphi,\eta,\varphi^\prime,\eta^\prime)\right)_{\eta=\eta_m}
=0,
\ee
and initial condition $G_\sigma(\varphi^\prime,\eta,\varphi^\prime,\eta^\prime)=\delta(\eta-\eta^\prime)$
at the starting point $\varphi^\prime$,
we can formally express the leading order solution to Eq.~\eq{eq_g_bou} within a single trapping domain,
$\varphi_j<\varphi,\varphi^\prime < \varphi_{j+1}$, as
\be{formexprsols}
g_\text{bou}(\varphi,\eta,\sigma) = \int\limits_0^{\eta_m}\rd\eta^\prime
G_\sigma(\varphi,\eta,\varphi^\prime,\eta^\prime) g_\text{bou}(\varphi^\prime,\eta^\prime,\sigma)
+
Q_\sigma(\varphi,\varphi^\prime,\eta),
\ee
where $\varphi > \varphi^\prime$ for $\sigma=1$ and $\varphi < \varphi^\prime$ for $\sigma=-1$,
and where
\bea{Qsigma}
Q_\sigma(\varphi,\varphi^\prime,\eta)
&=&
\sigma\int\limits_{\varphi^\prime}^\varphi\rd\varphi^{\prime\prime}
\int\limits_0^{\eta_m}\rd\eta^\prime
G_\sigma(\varphi,\eta,\varphi^{\prime\prime},\eta^\prime) s_\text{bou}(\varphi^{\prime\prime},\eta^\prime,\sigma)
\nonumber \\
&=&
C_0 \int\limits_{\varphi^\prime}^\varphi\rd\varphi^{\prime\prime} D_\eta(\varphi^{\prime\prime},\eta_b)
G_\sigma(\varphi,\eta,\varphi^{\prime\prime},\eta_b)
\eea
with a constant
\be{C0}
C_0 = l_c\left(\frac{\left\langle B^2\right\rangle}
{\langle |\lambda|\rangle}\right)_{\eta=\eta_b}.
\ee
Introducing the fundamental solution in the unbounded $\eta$-region,
\be{green}
G_\infty(\delta\eta,\eta-\eta^\prime)=\frac{1}{\sqrt{2\pi} \delta \eta}
\exp\left(-\frac{(\eta-\eta^\prime)^2}{2\delta\eta^2}\right),
\ee
where
\be{delta_eta_fund}
\delta\eta = \delta \eta(\varphi,\varphi^\prime)
= \left|2\int\limits_{\varphi^\prime}^{\varphi}\rd \varphi^{\prime\prime}\;
D_\eta(\varphi^{\prime\prime},\eta_b)\right|^{1/2},
\ee
we can construct Green's function satisfying boundary conditions~\eq{bougreen} as
\bea{mirroring}
G_\sigma(\varphi,\eta,\varphi^\prime,\eta^\prime)
&=&
G_\infty\left(\delta \eta,\eta-\eta^\prime\right)
+
G_\infty\left(\delta \eta,\eta+\eta^\prime\right)
\nonumber \\
&+&
G_\infty\left(\delta\eta,\eta+\eta^\prime-2\eta_m\right)
+{\cal O}\left(\exp\left(-\frac{\eta_m^2}{2 \delta\eta^2}\right)\right),
\eea
with the last term being negligible small in the long mean free path regime where $\delta\eta \ll \eta_m$.
By our assumption, $\delta \eta < \delta\eta_e \ll (\eta_m-\eta_b)$, and, of course, $\delta\eta \ll \eta_b$.
Therefore, second and third terms in Eq.~\eq{mirroring} provide similarly exponentially small contributions
to the source term~\eq{Qsigma} which can be expressed then as
\be{QviaPhi}
Q_\sigma(\varphi,\varphi^\prime,\eta)=\frac{\sigma C_0\delta\eta(\varphi,\varphi^\prime)}{\sqrt{8}}
\Phi\left(\frac{\eta-\eta_b}{\sqrt{2}\delta\eta(\varphi,\varphi^\prime)}\right),
\ee
where
\be{Phi}
\Phi(x)=\frac{2}{\sqrt{\pi}}\int\limits_0^1 \rd t \exp\left(-\frac{x^2}{t^2}\right).
\ee
We can estimate now the next order correction to the source term $\Delta Q$
driven by $\Delta s$, Eq.~\eq{Delta_s_bou}, in the trapped particle domain
as $\Delta Q / Q_\sigma \sim B\delta \eta \log(B\delta\eta) \ll 1$ where logarithm is due to the scaling of bounce time
with the distance to the trapped-passing boundary.

Since non-trivial behavior of the solution driven by $s_\text{bou}$ is localized in the region
$|\eta-\eta_b| \le \delta \eta_e$ while $g_\text{bou}$ tends to constant elsewhere, we can formally extend the domain
$0<\eta<\eta_m$ to an infinite domain $-\infty<\eta<\infty$ with boundary conditions
$\lim\limits_{\eta\rightarrow \pm \infty} \partial g_\text{bou}/\partial \eta = 0$. Respectively, one should
retain only the first term in Green's function~\eq{mirroring} and use infinite limits of $\eta^\prime$ integration
in Eq.~\eq{formexprsols}. Thus, mapping between distribution function $g^\pm_{j(in)}$ of particles which enter
trapping domain $\varphi_j < \varphi < \varphi_{j+1}$ and which leave this domain, $g^\pm_{j(out)}$,
\bea{notation_inout}
g^+_{j(in)}(\eta) &=& g_\text{bou}(\varphi_j,\eta,1),
\qquad
g^+_{j(out)}(\eta)=g_\text{bou}(\varphi_{j+1},\eta,1),
\nonumber \\
g^-_{j(in)}(\eta) &=& g_\text{bou}(\varphi_{j+1},\eta,-1),
\qquad
g^-_{j(out)}(\eta)=g_\text{bou}(\varphi_j,\eta,-1),
\eea
follows from~\eq{formexprsols} as
\be{in_out_map}
g^\pm_{j (out)}(\eta) = \int\limits_{-\infty}^{\infty}\rd\eta^\prime
G_\infty(\delta\eta_j,\eta-\eta^\prime) g^\pm_{j (in)}(\eta^\prime)
+
Q_{\text{off}}(\delta\eta_j,\eta),
\ee
where
\be{detaj_Qoff}
\delta\eta_j=\delta\eta(\varphi_j,\varphi_{j+1})=\left(2\eta_b I_j(\eta_b)\right)^{1/2}l_c^{-1/2},
\qquad
Q_{\text{off}}(\delta\eta_j,\eta) = Q_\sigma(\varphi_j,\varphi_{j+1},\eta),
\ee
and $I_j(\eta)$ is given by~\eq{bas1_defs} so that $\delta\eta_j = \pi^{1/2}\delta\eta_e$, see~\eq{deltaeta}.
Conditions~\eq{inter_map} with $\delta\varphi \rightarrow 0$ link the incoming distributions $g^\pm_{j (in)}$
to the outgoing $g^\pm_{j (out)}$ as follows
\bea{in_to_out}
g^+_{j (in)}(\eta<\eta_j) &=& g^+_{j-1 (out)}(\eta<\eta_j),
\qquad
g^+_{j (in)}(\eta>\eta_j)=g^-_{j (out)}(\eta>\eta_j),
\\
g^-_{j (in)}(\eta<\eta_{j+1}) &=& g^+_{j+1 (out)}(\eta<\eta_{j+1}),
\qquad
g^-_{j (in)}(\eta>\eta_{j+1})=g^+_{j (out)}(\eta>\eta_{j+1}),
\nonumber
\eea
where $\eta_j = 1/B(\varphi_j)$. Substitution of $g_{j(in)}^\pm$ into the mapping~\eq{in_out_map} via~\eq{in_to_out}
leads to a coupled set of Wiener-Hopf type integral equations for $g_{j(out)}^\pm$.
Introducing a dimensionless
variable $x$ and dimensionless distribution function $\alpha_j^\pm$ as follows
\be{define_x}
x=\frac{\eta_b-\eta}{\sqrt{2}\delta\eta_{\rm ref}},
\qquad
g_{j(out)}^\pm(\eta)=\frac{C_0 \delta\eta_{\rm ref}}{\sqrt{8}}\alpha_j^\pm(x),
\ee
where $\delta\eta_{\rm ref}$ is some reference boundary layer width which is set in case of a closed field line
to the largest $\delta\eta_j$, this set is of the form
\bea{set_for_alphas}
\alpha_j^\pm(x) &=& \frac{A_j}{\sqrt{\pi}}\int\limits_{-\infty}^\infty \rd x^\prime {\rm e}^{-A_j^2(x-x^\prime)^2}
\left(
\Theta\left(x^\mp_j-x^\prime\right)\alpha_j^\mp\left(x^\prime\right)
+
\Theta\left(x^\prime-x^\mp_j\right)\alpha_{j\mp 1}^\pm\left(x^\prime\right)
\right)
\nonumber \\
&\pm& \frac{1}{A_j}\Phi(A_j x),
\eea
where mis-alignments of relevant maxima $x_j^\pm < 0$ and aspect ratios $A_j$ are
\be{xjpm}
x_j^+=\frac{\eta_b -\eta_{j+1}}{\sqrt{2}\delta\eta_{\rm ref}},
\qquad
x_j^-=\frac{\eta_b -\eta_j}{\sqrt{2}\delta\eta_{\rm ref}},
\qquad
A_j=\frac{\delta \eta_{\rm ref}}{\delta \eta_j}.
\ee
It should be noted that solution of set~\eq{set_for_alphas} is determined up to the null-space of the original
operator $\hat L$, Eq.~\eq{mono_Ak_norm}, which is an arbitrary constant,
i.e. $\alpha^\pm_j=\text{const}$,
and which produces no particle flux or parallel current but only re-defines the equilibrium Maxwellian.

\subsection{Alignment of maxima, equivalent ripples}
\label{ssec:align}

An obvious consequence of Eq.~\eq{set_for_alphas} is the configuration where the leading order off-set
is absent. Namely, coupled set~\eq{set_for_alphas} reduces to two coupled equations
in case all trapping domain are equivalent which means that
all relevant maxima are aligned with a global maximum, $x_j^\pm=0$, and all aspect ratios
are the same (equal to one), $A_j=1$. The first condition means that global maximum is reached on a line (lines) rather than
at a point (points). The second condition means that all $\delta\eta_j$ are the same, i.e. the integral~\eq{delta_eta_fund}
between maximum points is the same for all trapping domains. In other words, this means that the value of the first
integral~\eq{bas1_defs} at the trapped-passing boundary $\eta=\eta_b$ does not depend on field line parameter $\vartheta_0$,
\be{omnigen}
\difp{I_j(\vartheta_0,\eta_b)}{\vartheta_0} = 0,
\qquad
I_j(\vartheta_0,\eta)=
\frac{1}{2}\oint \frac{\rd l}{B} \lambda =\frac{1}{2v}\oint \frac{\rd l}{B} v_\parallel.
\ee
Thus, $\alpha^\pm_j(x)=\alpha^\pm$, and the
set~\eq{set_for_alphas}, respectively, closes to the form
\be{itneq_both}
\alpha^\pm(x) = \frac{1}{\sqrt{\pi}}\int\limits_{-\infty}^\infty \rd x^\prime {\rm e}^{-(x-x^\prime)^2}
\left(
\Theta\left(x^\prime\right)\alpha^\pm\left(x^\prime\right)
+
\Theta\left(-x^\prime\right)\alpha^\mp\left(x^\prime\right)
\right)
\pm\Phi(x),
\ee
corresponding to quasi-symmetry~\citep{nuhrenberg88-113} (omnigeneity~\citep{cary97-3323,helander09-055004}).
Obviously, the even part of the solution
$\alpha^++\alpha^-$ satisfies a homogeneous equation and thus is constant, which
we set to zero. Expressing $\alpha^-=-\alpha^+$,
set~\eq{itneq_both} is reduced to a single Wiener-Hopf type integral equation
\be{WH_tok}
\alpha^+(x) = \frac{1}{\sqrt{\pi}}\int\limits_{-\infty}^\infty \rd x^\prime {\rm e}^{-(x-x^\prime)^2}
{\rm sign}\left(x^\prime\right)\alpha^+\left(x^\prime\right)
+\Phi(x).
\ee
Obviously, deep in the trapped region $x\rightarrow -\infty$ where $\Phi(x)\rightarrow 0$ and
$\alpha^+ \rightarrow const$, this
equation reduces to $\alpha^+=-\alpha^+$, i.e. it results in zero off-set $\alpha^\pm=0$.
Actually, as shown in~\citep{helander11-092505}, the ripple equivalence condition~\eq{omnigen}
which is fulfilled for a perfectly quasi-isodynamic stellarator where the magnetic field is
omnigeneous, allows reducing the boundary layer problem
to a tokamak like problem up to the same degree of accuracy as the analysis here
(essentially, we use the integral representation of the same problem as in~\citep{helander11-092505}).

\begin{figure}
\centerline{
\includegraphics[width=0.49\textwidth]{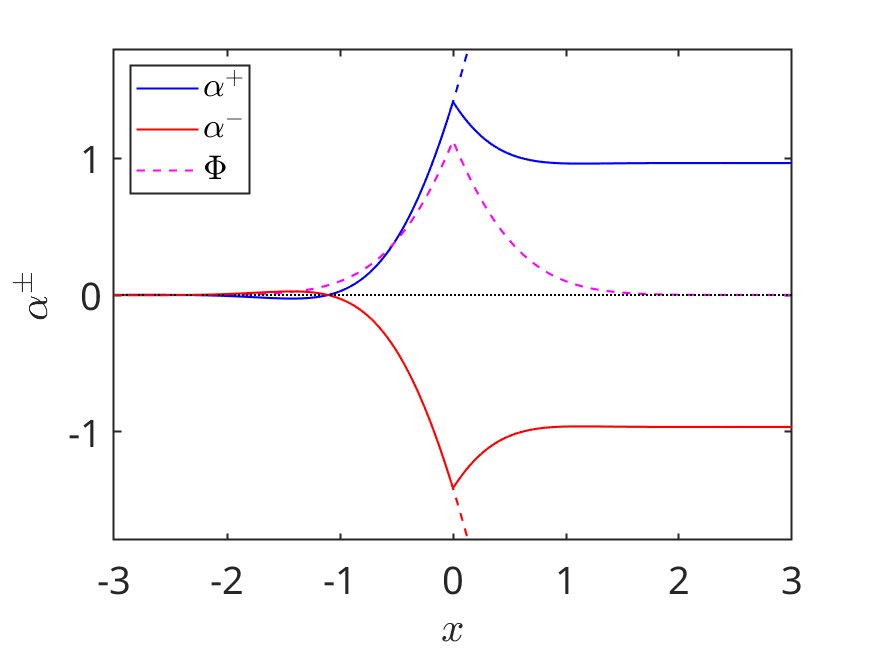}
\includegraphics[width=0.49\textwidth]{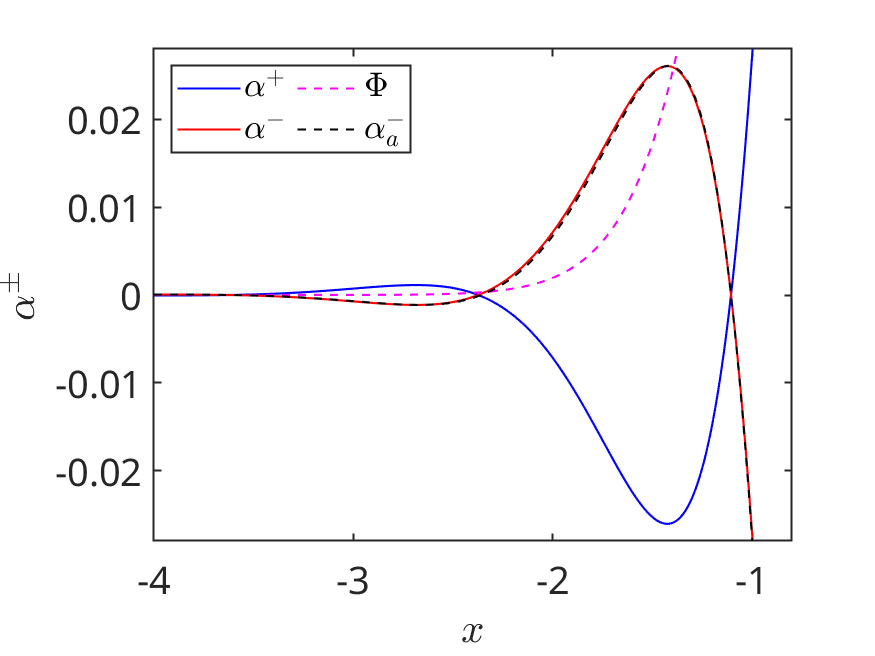}
}
\caption[]{Normalized distribution functions $\alpha^+$ (blue), $\alpha^-$ (red) and source function $\Phi$ (magenta).
Full normalized distribution functions, $\alpha^{\pm}+\alpha_{1/\nu}^\pm$, are shown with dashed lines.
The right picture is a zoom to the trapped region, where also the asymptotic solution~\eq{alpha_ass}
is shown (black dashed).
}
\label{fig:alpha_pm}
\end{figure}
Solutions to the equation set~\eq{itneq_both} are shown in Fig.~\ref{fig:alpha_pm}
together with asymptotic solution~\eq{gmdecay} expressed in terms of normalized variables
using $\delta\eta_j=\delta\eta_{\rm ref}=\sqrt{\pi}\delta\eta_e$ as
\be{alpha_ass}
\alpha^\pm_{a}(x)=\pm\alpha_0 \exp\left(\sqrt{2\pi}\; x \right)\cos\left(\sqrt{2\pi}\; x +\phi\right),
\ee
where $\alpha_0=\text{const}$. They agree well because the
latter solution must actually be the same as $\alpha^\pm$ in the
region where the source term is negligible. We can recover
the full solution if we add to $\alpha^\pm(x)$ the rescaled leading order term~\eq{ware_0} expanded to the
linear order in $\eta-\eta_b$,
\be{alpha_m1}
\alpha^\pm_{1/\nu}(x)=\frac{\sqrt{8}}{C_0 \delta\eta_{\rm ref}}\Theta(\eta_b-\eta)g_{-1}(\eta)\approx \pm 4 x \Theta(x).
\ee
Naturally, full solution $\alpha^\pm(x)+\alpha^\pm_{1/\nu}(x)$ has a continuous derivative at the trapped-passing
boundary.

Note that alignment of maxima, $x_j^\pm=0$, is not sufficient on its own, without conditions of equal aspect ratios,
$A_j=1$, for avoiding the leading order off-set. This can be seen from Fig.~\ref{fig:alpha_3ripples} where
a solution to Eqs.~\eq{set_for_alphas} is shown for the case of three relevant trapping domains
and aligned maxima but different aspect ratios, $A_0=A_2=2$ and $A_1=1$
(i.e. $2\delta\eta_0=2\delta\eta_2=\delta\eta_1 \equiv \delta\eta_{\rm ref}$).
\begin{figure}
\centerline{
\includegraphics[width=0.49\textwidth]{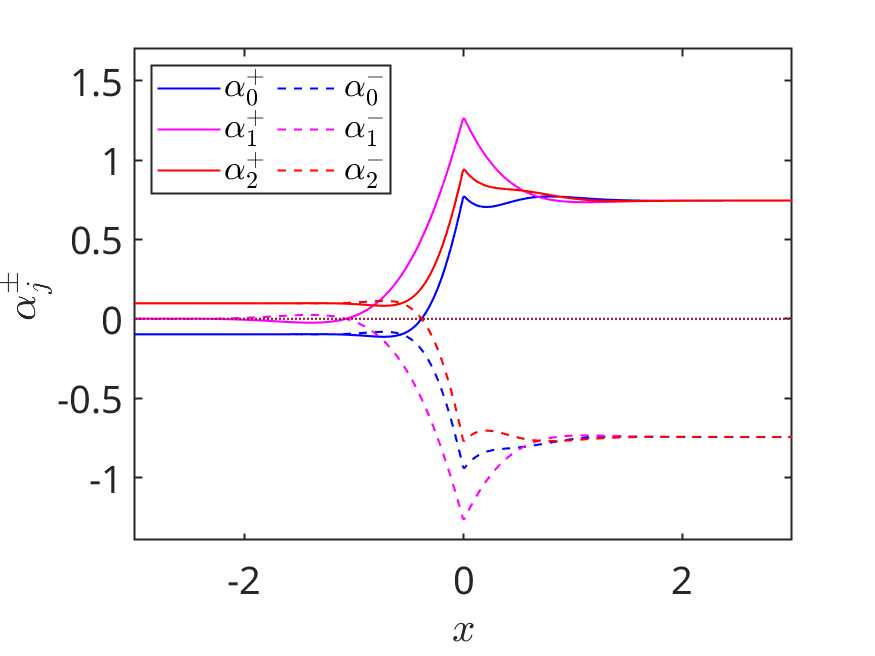}
\includegraphics[width=0.49\textwidth]{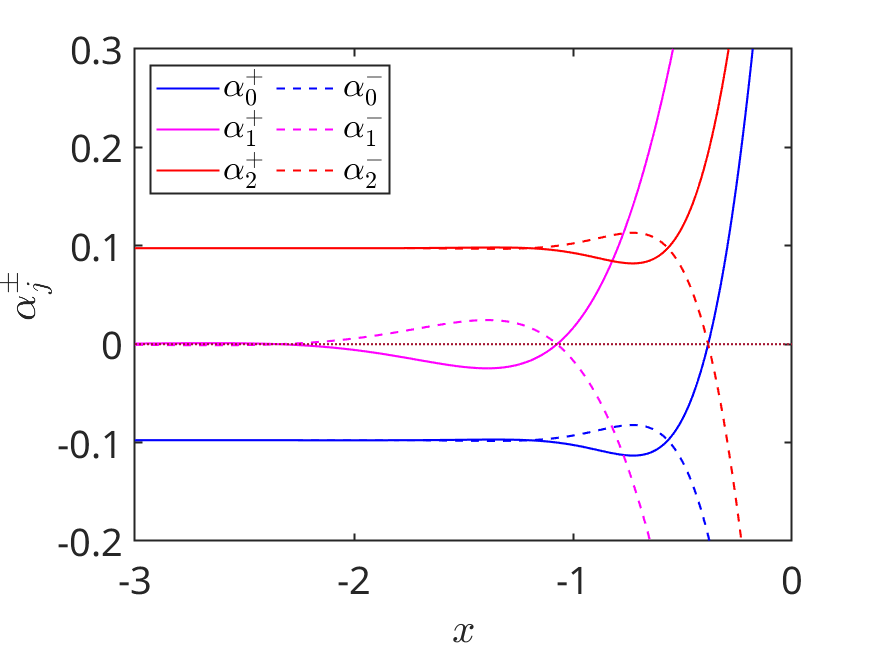}
}
\caption[]{Normalized distribution functions $\alpha_j^+$ (solid) and $\alpha_j^-$ (dashed)
in case of three relevant trapping domains per closed field line.
The right picture is a zoom to the trapped particle domain.
}
\label{fig:alpha_3ripples}
\end{figure}
A finite off-set in the first and last domains, $\alpha_2^\pm(-\infty)= - \alpha_0^\pm(-\infty)$ can be clearly seen.
In such a case, the function $g_\text{bou}$ in the trapped particle domain and, respectively, bootstrap coefficient
diverge at low collisionality as $\lambda_{bB} \sim \nu^{-1/2}$ (see the normalization~\eq{C0} and~\eq{define_x}).

A particular example of configurations where bootstrap coefficient diverges in the $1/\nu$ regime in this way
is a family of ``anti-sigma optimized'' configurations, i.e. configurations where
sigma optimization~\citep{mynick82-322,shaing83-2136} is applied to align maxima instead of minima.
\begin{figure}
\centerline{
\includegraphics[width=0.49\textwidth]{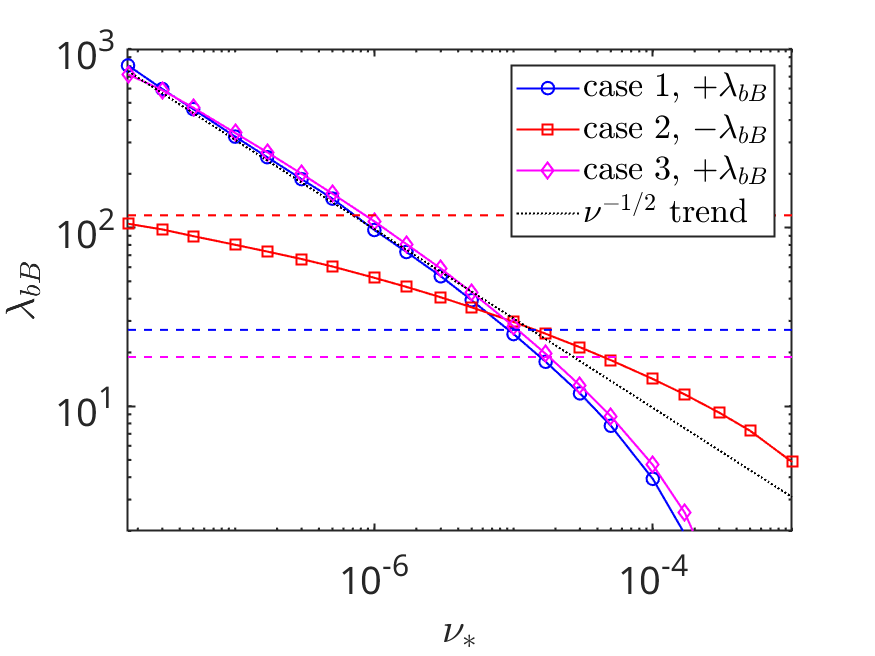}
\includegraphics[width=0.49\textwidth]{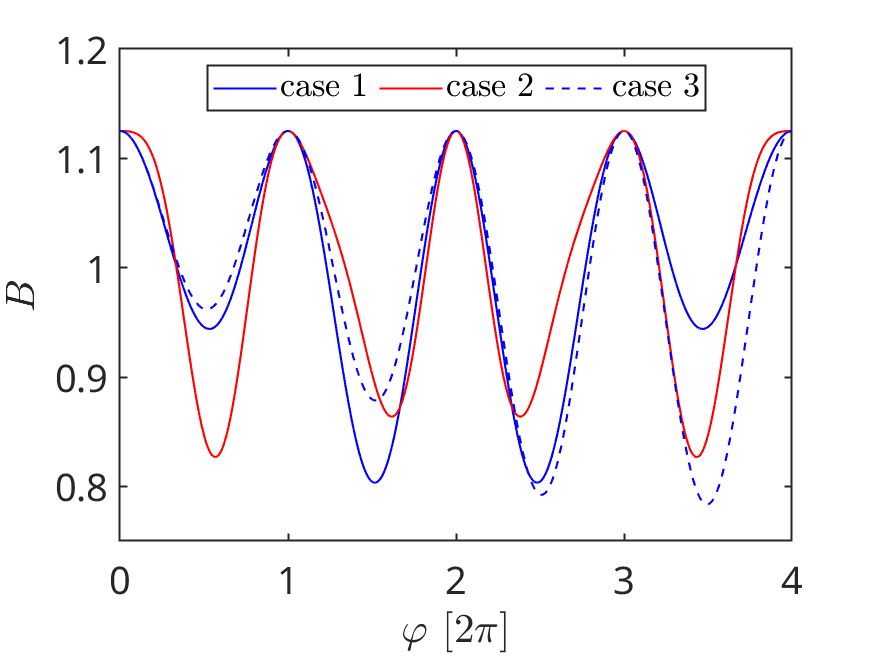}
}
\caption[]{
Left: Normalized bootstrap coefficient $\lambda_{bB}$, Eq.~\eq{D31_lambb},
computed by NEO-2 for ``anti-sigma optimized'' configurations
in cases 1, 2 and 3 (see the legend)
as function of normalized collisionality $\nuast$. 
For each case, the Shaing-Callen limit~\eq{lambda_bB} is shown with a dashed line of the respective color.
Sign of $\lambda_{bB}$ (but not of the respective asymptotic) is reversed in case 2.
Right: Magnetic field strength $B(\varphi)$, Eq.~\eq{QI-antisigma} for these cases
plotted along the field line within the first four toroidal periods.
}
\label{fig:lambb_antisigma}
\end{figure}
A magnetic field strength for a particular family of such configurations with dominant toroidal ripple is
of the form
\be{QI-antisigma}
B(\vartheta,\varphi)=B_0\left(1+\varepsilon_0 \cos\left(N_{\rm tor}\varphi\right)
+ f(\vartheta)\left(1-\cos\left(N_{\rm tor}\varphi\right)\right)\right),
\ee
where $B_0$ and $\varepsilon_0$ are constant parameters, $N_{\rm tor}$ is the number of field periods,
and $f(\vartheta)<\varepsilon_0$.
For the computations with NEO-2, we use the following simple form of this function,
\be{f_antisig}
f(\vartheta)=\varepsilon_1 \cos\vartheta + \varepsilon_3\cos(3\vartheta),
\ee
with $\varepsilon_0=0.125$, $\varepsilon_1=0.05$ and two values of $\varepsilon_3$. We set
$N_{\rm tor}=1$ and $\iota=1/4$
to get 4 relevant domains (ripples) on a closed field line passing through the point $(\vartheta,\varphi)=(0,0)$.
In case $\varepsilon_3=0$ (case 1), condition $A_j=1$ does not hold, and a strong off-set of bootstrap coefficient
$\lambda_{bB} \propto \nu^{-1/2}$ is seen in Fig.~\ref{fig:lambb_antisigma}.
In case $\varepsilon_3=0.06994425$ (case 2) all four ripples are equivalent, $A_j=1$, and the leading order
off-set is absent. However, there still remains a residual off-set driven by source $\Delta s$, Eq.~\eq{Delta_s_bou},
which includes a constant and logarithmic parts, $\delta\lambda_{bB} \rightarrow C_1+C_2 \log(\nuast)$,
see the previous section. 
It should be noted that $\nu^{-1/2}$ divergence of $\lambda_{bB}$ in case of
aligned maxima but non-equivalent ripples is not a property of a closed field line. A similar behavior
is seen also for the case $\varepsilon_3=0$ at the irrational flux surface with
\mbox{$\iota=(1+\sqrt{5})/20 \approx 0.1618$} labelled there as case 3.

\subsection{Non-aligned maxima, large aspect ratio}
\label{ssec:nonalign}

Let us consider now a more typical situation where local field maxima are not aligned.
For simplicity, we consider a configuration with short ripples located near global maximum
adjacent to a long ripple, as in the case of the first rippled tokamak configuration~\eq{magfield_mod}
with $\iota=1/4$ illustrated in Figs.~\ref{fig:g_odd_direct} and~\ref{fig:spitf_surf}.
Besides the global maximum, there are only two relevant maxima at low collisionalities. They
separate two short and shallow trapping domains called below ``off-set domains'' from
a wide domain in between called below ``main region''. Setting $\delta\eta_{\rm ref}=\delta\eta_1$ so that $A_1=1$,
aspect ratios for off-set domains are the same, $A_0=A_2\equiv A_o$.
Respective value $A_o \approx 23.7 \gg 1$ will be called below ``off-set aspect ratio'' or simply ``aspect ratio''.
To keep an example more general than configuration with $\iota=1/4$, we do not assume that local maxima
separating the main region from two off-set domains are the same (thus allowing broken stellarator symmetry).
For the present example, it is convenient to change general notation in Eqs.~\eq{set_for_alphas} as follows,
$\alpha_0^\pm \equiv \alpha_{-o}^\pm$, $\alpha_1^\pm \equiv \alpha^\pm_{mr}$ and
$\alpha_2^\pm \equiv \alpha_{+o}^\pm$ so that subscripts ``$+o$'' and ``$-o$'' denote the left and right off-set
domain, respectively, and subscript ``$mr$'' stands for the main region.
Respective normalized mis-alignments of local maxima we re-notate as follows,
$x_1^+=x_2^- \equiv x_{-o}$ and $x_3^-=x_2^+ \equiv x_{+o}$, see schematic Fig.~\ref{fig:scheme_alphas}.
Thus, we cast the closed set of six equations~\eq{set_for_alphas} to
\bea{set_for_alphas_case1}
\alpha_{mr}^\pm(x) &=& \frac{1}{\sqrt{\pi}}\int\limits_{-\infty}^\infty \rd x^\prime {\rm e}^{-(x-x^\prime)^2}
\left(
\Theta\left(x_{\mp o}-x^\prime\right)\alpha_{mr}^\mp\left(x^\prime\right)
+
\Theta\left(x^\prime-x_{\mp o}\right)\alpha_{\mp o}^\pm\left(x^\prime\right)
\right)
\pm \Phi(x),
\nonumber \\
\alpha_{\mp o}^\pm(x) &=& \frac{A_o}{\sqrt{\pi}}\int\limits_{-\infty}^\infty \rd x^\prime
{\rm e}^{-A_o^2(x-x^\prime)^2}
\left(
\Theta\left(-x^\prime\right)\alpha_{\mp o}^\mp\left(x^\prime\right)
+
\Theta\left(x^\prime\right)\alpha_{\pm o}^\pm\left(x^\prime\right)
\right)
\pm \frac{\Phi(A_o x)}{A_o},
\\
\alpha_{\pm o}^\pm(x) &=& \frac{A_o}{\sqrt{\pi}}\int\limits_{-\infty}^\infty \rd x^\prime
{\rm e}^{-A_o^2(x-x^\prime)^2}
\left(
\Theta\left(x_{\pm o}-x^\prime\right)\alpha_{\pm o}^\mp\left(x^\prime\right)
+
\Theta\left(x^\prime-x_{\pm o}\right)\alpha_{mr}^\pm\left(x^\prime\right)
\right)
\pm \frac{\Phi(A_o x)}{A_o},
\nonumber
\eea
where notation $\alpha_{\mp o}^\pm$ means that either $\alpha_{-o}^+$ or $\alpha_{+o}^-$ should be taken,
and $\alpha_{\pm o}^\pm$ means that either $\alpha_{+o}^+$ or $\alpha_{-o}^-$ should be taken.
\begin{figure}
\centerline{
\includegraphics[width=0.49\textwidth]{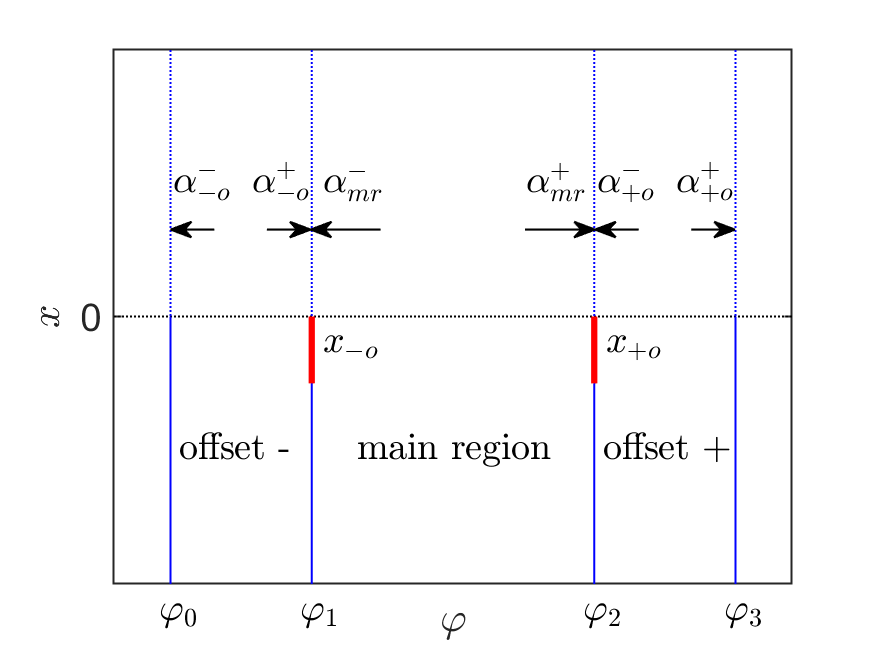}
}
\caption[]{
Locations of normalized outgoing particle distributions $\alpha_{\pm o}^\pm$ and $\alpha_{mr}^\pm$
in the respective off-set domains and the main region. Positions of relevant maxima $\varphi_j$ are shown with blue lines.
Normalized misalignments $x_{\pm o}$ are shown with red lines.
}
\label{fig:scheme_alphas}
\end{figure}

Due to $A_o \gg 1$, source terms $\Phi(A_o x)/A_o$ can be ignored in the last two pairs of equations
since $\alpha_{\mp o}^\pm \sim \alpha_{\pm o}^\pm \sim \alpha_{mr}^\pm \sim \Phi \gg \Phi/A_o$.
Since $\alpha^\pm_{mr}(x)$ which varies on the scale $x\sim 1$
changes little compared to the exponent in the sub-integrand of the last pair of equations,
we can set there $\alpha^\pm_{mr}\left(x^\prime\right) \approx \alpha^\pm_{mr}\left(x\right)$. Thus,
the last pair of Eqs.~\eq{set_for_alphas_case1} takes the approximate form
\be{last_pair}
\alpha_{\pm o}^\pm(x) \approx \frac{A_o}{\sqrt{\pi}}\int\limits_{-\infty}^{x_{\pm o}} \rd x^\prime
{\rm e}^{-A_o^2(x-x^\prime)^2}
\alpha_{\pm o}^\mp\left(x^\prime\right)
+
\tilde \Theta\left(A_o(x-x_{\pm o})\right)
\alpha_{mr}^\pm\left(x\right),
\ee
where we introduced a ``smooth step'' function as
\be{Theta_tilde}
\tilde \Theta(x)\equiv \frac{1}{2} \left(1+{\rm erf}(x) \right),
\qquad
\lim\limits_{A_o\rightarrow\infty} \tilde \Theta(A_o x)=\Theta(x).
\ee

Further analysis is strongly simplified if we restrict ourselves to the most probable case of strong
mis-alignment of local maxima where boundary layers in both off-set domains associated with these maxima
are well separated from the trapped-passing boundary,
$\delta\eta_0 \ll |\eta_b-\eta_1|$ and $\delta\eta_2 \ll |\eta_b-\eta_2|$.
In normalized variables, these conditions mean $A_o |x_{\pm o}| \gg 1$.
Since the first term in~\eq{last_pair} is exponentially small in the region $A_o (x-x_{\pm o}) \gg 1$
where also $\Theta\left(A_o(x-x_{\pm o})\right) \approx 1$, Eq.~\eq{last_pair} results there in
$\alpha_{\pm o}^\pm(x) \approx \alpha_{mr}^\pm\left(x\right)$. Thus, the second pair of
Eqs.~\eq{set_for_alphas_case1} with the source term ignored approximately is
$$
\alpha_{\mp o}^\pm(x) \approx \frac{A_o}{\sqrt{\pi}}\int\limits_{-\infty}^\infty \rd x^\prime
{\rm e}^{-A_o^2(x-x^\prime)^2}
\left(
\Theta\left(-x^\prime\right)\alpha_{mr}^\mp\left(x^\prime\right)
+
\Theta\left(x^\prime\right)\alpha_{mr}^\pm\left(x^\prime\right)
\right),
\qquad
A_o (x-x_{\mp o}) \gg 1.
$$
Moreover, due to slow variation of $\alpha_{mr}^\pm(x^\prime)$ compared to the exponent one can
further simplify this expression to
\be{furthersimp}
\alpha_{\mp o}^\pm(x) \approx \Theta\left(-x\right)\alpha_{mr}^\mp\left(x\right)
+
\Theta\left(x\right)\alpha_{mr}^\pm\left(x\right),
\ee
which is valid in the region $x>x_{\mp o}$ except for small vicinities of class boundary,
$|x-x_{\mp o}| \le 1/A_o$, and trapped-passing boundary $|x| \le 1/A_o$ which contain
local boundary layers of the off-set domains (layers of the width $\delta\eta_0 \sim \delta\eta_2 \ll \delta\eta_1$).
Contribution of these vicinities to the integral in the first pair of Eqs.~\eq{set_for_alphas_case1}
is of the order of $1/A_o$, and, therefore, this pair can be closed as follows,
\be{eqs_main}
\alpha_{mr}^\pm(x) = \frac{1}{\sqrt{\pi}}\int\limits_{-\infty}^\infty \rd x^\prime {\rm e}^{-(x-x^\prime)^2}
\left(
\Theta(x^\prime) \alpha_{mr}^\pm\left(x^\prime\right)
+
\Theta\left(-x^\prime\right)\alpha_{mr}^\mp\left(x^\prime\right)
\right)
\pm \Phi(x) + O(A_o^{-1}).
\ee
Obviously, in the leading order over $A_o^{-1}$, these equations are the same
as Eqs.~\eq{itneq_both} for equivalent ripples.

For the solution in local trapping regions of the off-set domains, $x \le x_{\pm o}$, Heaviside functions
in the sub-integrands of the second pair of Eqs.~\eq{set_for_alphas_case1} can be replaced by their values
at $x^\prime < 0$ which turns this pair (with the source term ignored) into
$$
\alpha_{\mp o}^\pm(x) = \frac{A_o}{\sqrt{\pi}}\int\limits_{-\infty}^\infty \rd x^\prime
{\rm e}^{-A_o^2(x-x^\prime)^2}
\alpha_{\mp o}^\mp\left(x^\prime\right).
$$
Substituting this result in Eq.~\eq{last_pair} we note also that
variation of the last term in Eq.~\eq{last_pair} is determined in the region of interest by $\tilde\Theta$ while
the factor $\alpha_{mr}^\pm$ varies slowly and can be replaced by its value at $x = x_{\pm o}$.
Introducing the new independent variable $y=A_o(x-x_{\pm o})$, equation set for $\alpha^\pm_{\pm o}$
reduces to two independent equations
\be{last_pair_ofsreg}
\alpha_{\pm o}^\pm\left(x(y)\right) \approx \frac{1}{\pi}\int\limits_{-\infty}^{0} \rd y^\prime
\int\limits_{-\infty}^\infty \rd y^{\prime \prime}
{\rm e}^{-(y-y^\prime)^2-(y^\prime-y^{\prime\prime})^2}
\alpha_{\pm o}^\pm\left(x(y^{\prime\prime})\right)
+
\tilde \Theta\left(y\right)
\alpha_{mr}^\pm\left(x_{\pm o}\right),
\ee
which have constant solutions,
\be{last_pair_ofsreg_sols}
\alpha_{\pm o}^\pm\left(x(y)\right) = \alpha_{mr}^\pm\left(x_{\pm o}\right),
\qquad
\alpha^\mp_{\pm o}=\alpha^\pm_{\pm o}.
\ee
Thus, we obtained an intuitively clear result that
\begin{itemize}
\item
The distribution function in the main region (long ripple) is weakly affected by off-set domains (short ripples)
which provide only a correction of the order of $A_o^{-1}$;
\item
The distribution function in the off-set well
takes the value of the distribution function in the main region at the local minimum point $(\varphi_j,\eta_j)$
separating the main region and the off-set domain (see also Fig.~\ref{fig:spitf_surf}).
\end{itemize}
The last conclusion can be easily generalized to the case where a short off-set domain is separated from
two long main regions by local maxima on both sides.
Assuming that solutions in both main regions are known,
the off-set is determined by the solution in the main region separated from the off-set domain
by the lower local maximum (this local maximum determines the class transition boundary $\eta_c$
limiting the region $\eta > \eta_c$ with particles trapped in the off-set domain).

The large aspect ratio approximation, $A_o=23.7$, is checked in Fig.~\ref{fig:offset_prop_realasp} where
the left picture corresponds to the stellarator symmetric case of Fig.~\ref{fig:spitf_surf} and
in the right picture symmetry of local maxima is destroyed by setting $x_{-o}=2 x_{+o}$.
\begin{figure}
\centerline{
 \includegraphics[width=0.49\textwidth]{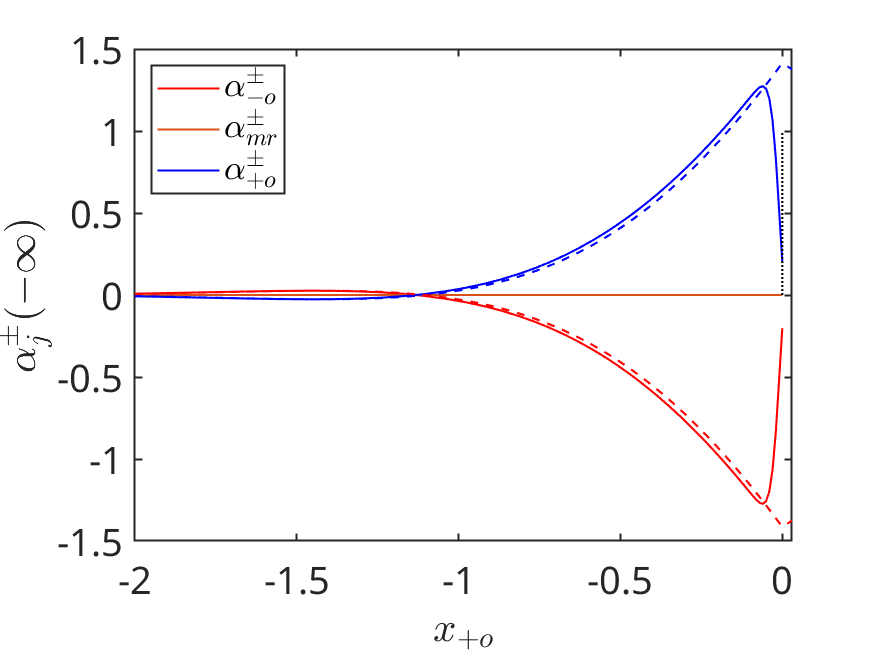}
 \includegraphics[width=0.49\textwidth]{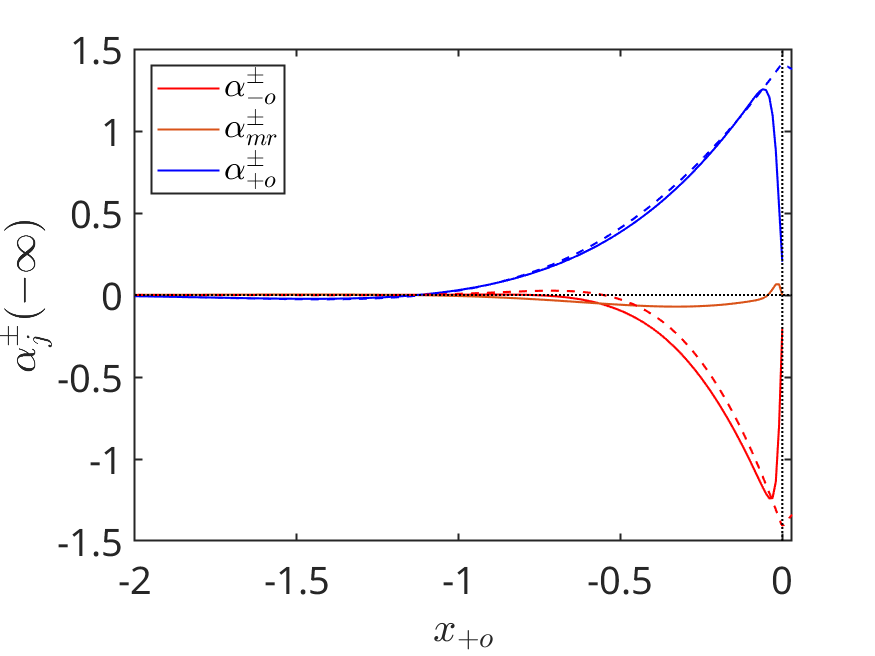}
}
\caption[]{Off-sets in three local wells as functions of the normalized mis-match parameter $x_{+o}$
for the symmetric case, $x_{-o}=x_{+o}$, (left) and for the destroyed stellarator symmetry, $x_{-o}=2 x_{+o}$, (right).
Solid - results of direct solution of Eqs.~\eq{set_for_alphas}, dashed - approximation
Eq.~\eq{last_pair_ofsreg_sols}.
}
\label{fig:offset_prop_realasp}
\end{figure}
It can be seen that approximation~\eq{last_pair_ofsreg_sols} is good in its validity domain and is
violated for small mis-match parameter values $x_{\pm o} \le A_o^{-1}$. The finite off-set value
for aligned maxima, $x_{\pm o}=0$, is the result of $A_o \ne 1$.
Note that left Fig.~\ref{fig:offset_prop_realasp} corresponds to collisionality dependence of the off-sets
in case of fixed magnetic field geometry. This dependence for the first local ripple is shown for the dimensional
distribution function~\eq{define_x} in Fig.~\ref{fig:offset_fits_realdim} where it is compared to the NEO-2 result.
\begin{figure}
\centerline{
 \includegraphics[width=0.49\textwidth]{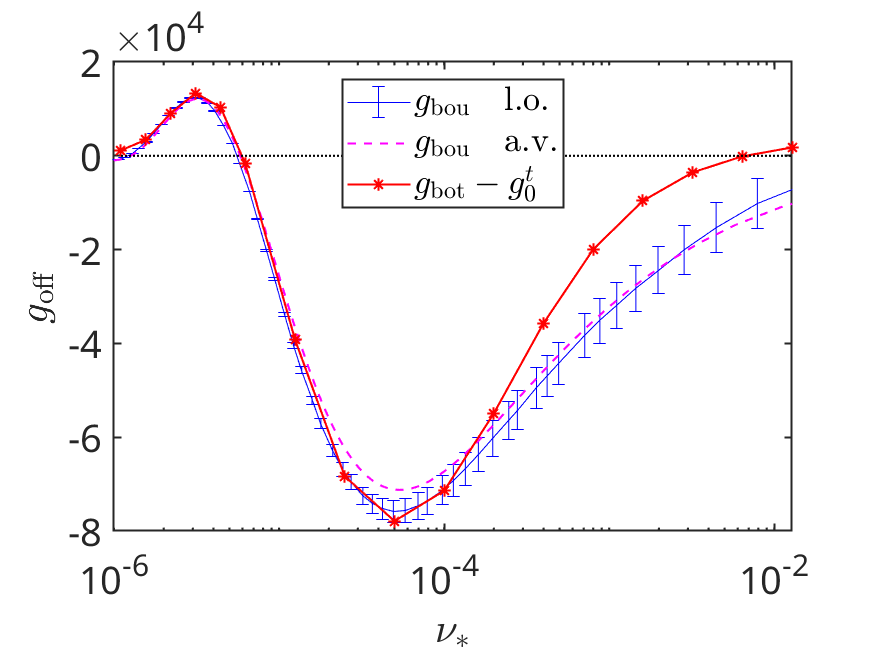}
}
\caption[]{Distribution function off-set $g_{\text{off}} = g_{(3)} - g_0^t$ at the bottom of the first local well, $j=0$,
as function of collisionality parameter $\nuast$ for configuration
in Fig.~\ref{fig:offset_neo2_pure}.
Blue - leading order solution for $g_\text{bou}$ via Eqs.~\eq{set_for_alphas}, magenta - its large aspect ratio
approximation Eq.~\eq{last_pair_ofsreg_sols}, red - off-set $g_{\text{off}}=g_\text{bot}-g_0^t$
via the result of NEO-2 shown in Fig.~\ref{fig:offset_neo2_pure}.
}
\label{fig:offset_fits_realdim}
\end{figure}
Since $g_\text{bou}$ is given there by propagator method in the leading order only, we added error bars
to the result of Eq.~\eq{set_for_alphas} showing the order of magnitude estimate of the omitted correction
term $\Delta g_\text{bou}$ driven by the source $\Delta s$ in Eq.~\eq{eq_g_bou}. 
According to Eq.~\eq{sources_bou}, this correction term is driven by the sources in the passing particle region
and is of the order of $g_0^t$ at the trapped passing boundary (and in the whole boundary layer) because 
$\Delta g_\text{bou}$ must restore periodicity in the passing region. Since there are no sources in the trapped particle
region, $\Delta g_\text{bou}$ decays exponentially outside the boundary layer, as follows from Eq.~\eq{gmdecay}
and from the solution $\alpha^+(x)$ for equivalent ripples, Eq.~\eq{WH_tok}, where $x$ is defined in Eq.~\eq{define_x}. 
Therefore, we extend the estimate of the correction term to the whole trapped particle region by using 
$\Delta g_\text{bou} \sim g_0^t \alpha^+ (x)$.  This simple estimation gives an error bar $g_0^t \alpha^+ (x_{\pm o})$
that decreases when $x_{\pm o} \propto \nu_\ast^{-1/2}$ increases.

\subsection{Weakly non-equivalent ripples}
\label{ssec:imperfect}

An important case is that of weakly non-equivalent ripples where the conditions of local maxima alignment, $\eta_j=\eta_b$,
and of equal aspect ratios, $A_j=1$, are only slightly violated.
Assuming infinitesimal
mis-alignments $\eta_j-\eta_b$ and aspect ratio perturbations $\Delta A_j = A_j-1$
and looking for the solution of Eqs.~\eq{set_for_alphas} in the form
$\alpha^\pm_j = \alpha^\pm +\Delta \alpha^\pm_j$ where $\alpha^+=-\alpha^-$ are the solution to (quasi-symmetric) Eq.~\eq{itneq_both},
equation for the linear order correction $\Delta \alpha^\pm_j$ is
\bea{pert_set_for_alphas}
\Delta \alpha_j^\pm(x) &=& \frac{1}{\sqrt{\pi}} \int\limits_{-\infty}^\infty \rd x^\prime {\rm e}^{-(x-x^\prime)^2}
\left(
\Theta\left(-x^\prime\right)\Delta \alpha_j^\mp\left(x^\prime\right)
+
\Theta\left(x^\prime\right)\Delta \alpha_{j\mp 1}^\pm\left(x^\prime\right)
\right)
\nonumber \\
&\mp&
x^\mp_j\; \Phi_B(x) \pm \Delta A_j\; \Phi_A(x).
\eea
Here, the first source term is due to small mis-alignment of maxima $|x^\mp_j| \ll 1$ with
\bea{Xmisalign}
\Phi_B(x) = \mp\lim_{|x^\mp_j| \rightarrow 0}\frac{2}{\sqrt{\pi}x^\mp_j}
\int\limits_{x^\mp_j}^0 \rd x^\prime {\rm e}^{-(x-x^\prime)^2}
\alpha^\pm\left(x^\prime\right)
=
\frac{2 \alpha^+(0)}{\sqrt{\pi}} {\rm e}^{-x^2},
\eea
and the second source term is due to small differences in boundary layer widths $|\Delta A_j| \ll 1$ with
\be{justtosee}
\Phi_A(x) = \frac{1}{\sqrt{\pi}}\int\limits_{-\infty}^\infty \rd x^\prime
{\rm e}^{-(x-x^\prime)^2}
\left(1-2 (x-x^\prime)^2\right)
{\rm sign}\left(x^\prime\right)
\alpha^+\left(x^\prime\right)
-\frac{2}{\sqrt{\pi}}{\rm e}^{-x^2}.
\ee

Let us assume now an infinite number of ripples, $-\infty < j < \infty$, with only the first maximum mis-aligned,
$\eta_1-\eta_b=\Delta\eta \ne 0$, while the rest are aligned perfectly. Then the non-zero normalized mis-alignments
are $x_0^+=x_1^-=-\Delta x \equiv - \Delta \eta /(\sqrt{2}\delta\eta_{\rm ref})$ and the rest $x_j^\pm=0$.
In the absence of $\Delta A_j=0$, solution to Eqs.~\eq{pert_set_for_alphas} is anti-symmetric with respect
to the mis-aligned maximum, $\Delta \alpha_0^\pm = - \Delta \alpha_1^\mp$, $\Delta \alpha_{-1}^\pm = - \Delta \alpha_2^\mp$, etc.
Thus, ripples $j \le 0$ are eliminated from the set~\eq{pert_set_for_alphas}, where the only inhomogeneous equation is
\be{onlyinhomeq_B}
\Delta \alpha_1^+(x) = \frac{1}{\sqrt{\pi}} \int\limits_{-\infty}^\infty \rd x^\prime {\rm e}^{-(x-x^\prime)^2}
\left(
\Theta\left(-x^\prime\right)\Delta \alpha_1^-\left(x^\prime\right)
-
\Theta\left(x^\prime\right)\Delta \alpha_{1}^+\left(x^\prime\right)
\right)+\Delta x\; \Phi_B(x),
\ee
while equations for $\Delta\alpha^-_1$ and $\Delta\alpha_j^\pm$ with $j>1$ are given by the homogeneous set~\eq{pert_set_for_alphas}
with $x_j^\pm=\Delta A_j=0$.

In the other case of perfectly aligned maxima, $x_j^\pm=0$, and only one perturbed aspect ratio, $\Delta A_0 \ne 0$ and
$\Delta A_j=0$ for $j \ne 0$, solutions are anti-symmetric with respect to ripple $j=0$, i.e.
$\Delta\alpha_{-j}^\pm = - \Delta \alpha_j^\mp$. Thus, solutions with $j<0$ are again eliminated from the set where
the only inhomogeneous equation is
\be{onlyinhomeq_A}
\Delta \alpha_0^+(x) = -\frac{1}{\sqrt{\pi}} \int\limits_{-\infty}^\infty \rd x^\prime {\rm e}^{-(x-x^\prime)^2}
\left(
\Theta\left(-x^\prime\right)\Delta \alpha_0^+\left(x^\prime\right)
+
\Theta\left(x^\prime\right)\Delta \alpha_{1}^-\left(x^\prime\right)
\right)+\Delta A_0\; \Phi_A(x),
\ee
while the rest $\Delta\alpha_j^\pm$ are given by the homogeneous set~\eq{pert_set_for_alphas}
for $j>0$. 

Off-set factors $\Delta_j^{A,B}=\Delta\alpha_j^+(-\infty)=\Delta\alpha_j^-(-\infty)$
as functions of ripple index $j$ are shown in Fig.~\ref{fig:linoff_B}.
Factors $\Delta_j^B$ are due to mis-alignment of a single local maximum between ripples 0 and 1
($\Delta x =1$ and $\Delta A_0=0$) and $\Delta_j^A$ are due to violation of aspect ratio in ripple 0
($\Delta x =0$ and $\Delta A_0=1$).
\begin{figure}
\centerline{
 \includegraphics[width=0.49\textwidth]{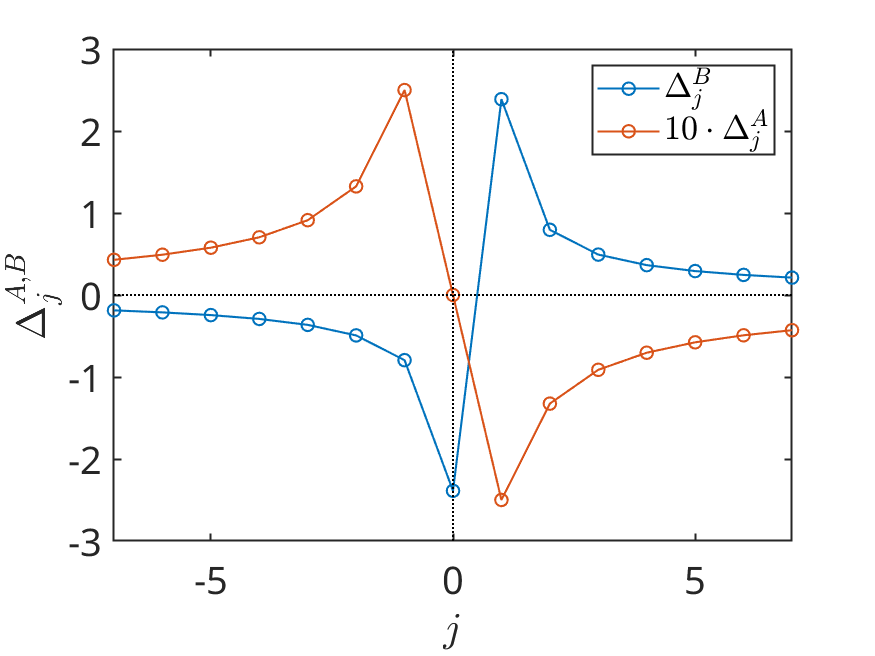}
}
\caption[]{Factors $\Delta^B_j$ (blue) and $\Delta^A_j$ (red) as functions of ripple index $j$.
}
\label{fig:linoff_B}
\end{figure}
Off-set $\Delta^B_j$ is mainly produced in ripples 0 and 1 adjacent to the mis-aligned maximum, but it also propagates
via passing particle region to the distant maxima, decaying there as $\propto j^{-1}$. The effect on adjacent ripples can be understood
in terms of ``particles'' (produced by inductive field in the passing region with positive velocities and having positive weight)
and ``anti-particles'' (negative velocities and weight). Part of ``particles'' entering the trapped region in ripple 0 from the
passing region penetrate through the gap $\Delta\eta$ to the trapped region in ripple 1 thus producing a positive off-set there.
Similarly, part of ``anti-particles'' entering ripple 1 produce a negative off-set in ripple 0.
The effect on distant ripples is because part of particles which penetrated the trapped region in ripple 1 return to the passing region
and thus create the asymmetry of the boundary layer in those distant ripples.
For the off-set $\Delta^A_{j}$ only the second mechanism is possible: If boundary layer width in ripple 0 is larger than in other
ripples (longer or deeper ripple), ``particles'' which enter this ripple from the passing region have more chance to return
thus dominating the ``anti-particles'' in the boundary layers of ripples with $j>0$. Respectively, ``anti-particles'' dominate
in ripples with $j<0$. Due to the symmetry, none are prevailing in ripple 0 where off-set is absent in this case.
Note that $\Delta A_0 > 0$ means more narrow boundary layer, which is the reason for the negative $\Delta_j^A$ at $j>0$ in
Fig.~\ref{fig:linoff_B}.

The corresponding dimensional off-set for the ripple $j$ related in the
general case to the dimensionless off-set via~\eq{define_x} as follows,
\be{dimoffset_def}
g_\text{off}^{(j)} \equiv g_{j(out)}^+(\infty) = g_{j(out)}^-(\infty)
= \frac{C_0\delta\eta_\text{ref}}{\sqrt{8}}\alpha_j^\pm(-\infty),
\ee
is expressed in the present case via the off-set factors as
\be{dimoff_B}
g_{\text{off}}^{(j)}=\frac{l_c}{4}\left(\frac{\langle B^2\rangle}{\langle|\lambda|\rangle}\right)_{\eta=\eta_b}
\left(\Delta^B_j\Delta\eta+\sqrt{2}\Delta^A_j \Delta A_o\delta\eta_{\rm ref}\right).
\ee
One can see that off-set due to maxima scales as $1/\nu$ and off-set due to aspect ratios scales as $1/\sqrt{\nu}$,
i.e., with reducing collisionality role of the first one increases. Note that linear approximation employed
in this section requires $\Delta\eta \ll \delta\eta_{\rm ref}$. If the opposite limit is reached, ripples
0 and 1 combine into a ripple with $\Delta A_{0+1} \sim 1$ which strongly violates the alignment of aspect ratios,
leading to much stronger off-set typical for anti-sigma configurations discussed in Section~\ref{ssec:align}.

\subsection{Off-set of Ware pinch coefficient}
\label{ssec:wareoff}
We define the off-set in the normalized Ware pinch (bootstrap) coefficient as follows,
\mbox{$\lambda_{\text{off}}\equiv \lambda_{bB}-\lambda_{bB}^\infty$}, where $\lambda_{bB}=\lambda_{bB}^\dagger$
is defined by Eq.~\eq{ware-mono} via mono-energetic $\bar D_{13}$ for finite collisionality and
$\lambda_{bB}^\infty$ denotes its asymptotic ($l_c \rightarrow \infty $) value~\eq{lambda_bB}.
Thus, $\lambda_{\text{off}}$ is given by Eq.~\eq{ware-mono} with the replacement
$g_{(3)} \rightarrow g_{\text{off}} = g_{(3)}-g_{(3)}^\infty$ in the definition~\eq{Dmono} of $\bar D_{13}$,
where $g_{(3)}^\infty$ is the asymptotic solution $g_0$ derived in Section~\ref{ssec:adjprob}.
Splitting in~\eq{Dmono} the integration interval $[\varphi_0,\varphi_N]$ into sub-intervals
$[\varphi_j,\varphi_{j+1}]$ separated by relevant maxima $\varphi_j$ (points of highest local maxima
$\eta_j B(\varphi_j)=1$
contained within the matching boundary $\eta_m$, $\eta_j < \eta_m$, see Section~\ref{ssec:propmet}),
these intervals correspond to off-set domains or main regions where the even part of the distribution
function $g_{\text{off}}$ is essentially different from zero in the trapped particle domain and is close to
a constant $g_{\text{off}}^{(j)}$ approximated using propagator techniques by Eq.~\eq{dimoffset_def}
in each interval $\varphi_j<\varphi<\varphi_{j+1}$ where we set
$g_{\text{off}}(\varphi,\eta) \approx \Theta(\eta-\eta_\text{loc}^{(j)}) 
\Theta(\eta B(\varphi) - 1)\; g_{\text{off}}^{(j)}$
with $\eta_{\text{loc}}^{(j)}=\max(\eta_j,\eta_{j+1})$, we obtain the off-set coefficient
as
\be{lambda_off}
\lambda_{\text{off}} \approx \sum_j g_{\text{off}}^{(j)} w_{\text{off}}^{(j)}.
\ee
Here, weighting factors are
\be{w_off}
w_{\text{off}}^{(j)} =
\frac{3}{2\rho_L B}
\left(\int\limits_{\varphi_0}^{\varphi_N}\frac{\rd \varphi}{B^\varphi}\right)^{-1}
\int\limits_{\varphi_{(j)}^-}^{\varphi_{(j)}^+}\rd \varphi \int\limits_{\eta_{\text{loc}}^{(j)}}^{1/B}\rd \eta\;
s_{(1)}^\dagger
= - \frac{3}{2\rho_L B}
\left(\int\limits_{\varphi_0}^{\varphi_N}\frac{\rd \varphi}{B^\varphi}\right)^{-1}H_j\left(\eta_\text{loc}^{(j)}\right),
\ee
$H_j(\eta)$ is given by~\eq{bas1_defs}, 
$\varphi_{(j)}^\pm = \varphi_j^\pm(\eta_{\text{loc}}^{(j)})$ are coordinates of banana tips 
(turning points, see the definition of $\varphi_j^\pm(\eta)$ below Eq.~\eq{liouvtrap}) of the transient orbit
separating locally trapped orbits in the offset well from the rest phase space (one of tips $\varphi_{(j)}^\pm$ is located
at the point of lowest local maximum, $\varphi_j$ or $\varphi_{j+1}$),
and we assume here the closed field line (i.e. finite $\varphi_N$).

Using for $s_{(1)}$ the first definition~\eq{mono_Ak_norm} and exchanging the integration order
we can express weighting factors in terms of bounce averaged radial drift velocity
and bounce time, 
\be{bavr}
\langle v_g^r \rangle_{bk} = \frac{2}{\tau_{bk}}
\int\limits_{\varphi_k^-(\eta)}^{\varphi_k^+(\eta)} \frac{\rd \varphi B v_g^r}{|v_\parallel| B^\varphi},
\qquad
\tau_{bk} = 2\int\limits_{\varphi_k^-(\eta)}^{\varphi_k^+(\eta)}\frac{\rd \varphi B}{|v_\parallel| B^\varphi},
\ee
where $\varphi^\pm_k$ are the turning points defined below Eq.~\eq{liouvtrap},
\be{w_off_viabavr}
w_{\text{off}}^{(j)} =
-\frac{3}{2\rho_L B}
\left(\int\limits_{\varphi_0}^{\varphi_N}\frac{\rd \varphi}{B^\varphi}\right)^{-1}
\int\limits_{\eta_\text{loc}^{(j)}}^{1/B_{\min}^{(j)}}\rd \eta
\sum_{k \in {\rm off}(j)} \tau_{bk} \langle v_g^r \rangle_{bk}.
\ee
Here, $B_{\min}^{(j)}$ is the minimum value of $B$ in the segment $[\varphi_j,\varphi_{j+1}]$, and the notation
$k \in {\rm off}(j)$ means all segments satisfying
$[\varphi_k^-(\eta),\varphi_k^+(\eta)] \in [\varphi_j,\varphi_{j+1}]$. As follows from Eq.~\eq{w_off_viabavr},
large off-set in the distribution function does not necessarily mean large off-set in bootstrap/Ware pinch coefficient.
In the devices with minimized bounce averaged radial drift such as, e.g., devices optimized for quasi-symmetry,
the off-set of bootstrap current is also minimized.

Less demonstrative but more practical expressions are obtained by evaluating the integral over
$\eta$ in~\eq{w_off} with help of the second definition~\eq{mono_Ak_norm} of $s_{(1)}$ and Eq.~\eq{Hphiprime}. Thus,
weighting factors are expressed directly via the geodesic curvature,
\be{w_off_eval}
w_{\text{off}}^{(j)} =
-\frac{1}{2}
\left(\int\limits_{\varphi_0}^{\varphi_N}\frac{\rd \varphi}{B^\varphi}\right)^{-1}
\int\limits_{\varphi_{(j)}^-}^{\varphi_{(j)}^+}\rd \varphi \frac{|\nabla r|k_G}{BB^\varphi}
\left(4-\eta_\text{loc}^{(j)} B\right) (1-\eta_\text{loc}^{(j)} B)^{1/2}.
\ee
In field aligned Boozer coordinates,
we replace $B^\varphi=B^2 (\iota B_\vartheta+B_\varphi)^{-1}$ with covariant field components
being the flux functions, and use the last expression~\eq{geodcu} for $|\nabla r|k_G$ to get
\be{w_off_booz}
w_{\text{off}}^{(j)}
=
\frac{\rd r}{\rd \psi}
\left(\int\limits_{\varphi_0}^{\varphi_N}\frac{\rd \varphi}{B^2}\right)^{-1}
I_{\cV}^{(j)},
\qquad
I_{\cV}^{(j)} =
\int\limits_{\varphi_{(j)}^-}^{\varphi_{(j)}^+}
\rd \varphi \cV,
\ee
where
\be{cVdef}
\cV =
\frac{1}{2 B^3} \difp{B}{\vartheta_0}
\left(4-\eta_\text{loc}^{(j)} B\right) (1-\eta_\text{loc}^{(j)} B)^{1/2}
=-\left[\difp{}{\vartheta_0} \frac{(1-\eta B)^{3/2}}{B^2}\right]_{\eta=\eta_\text{loc}^{(j)}}.
\ee

It should be noted that the off-set~\eq{lambda_off} ignores the contribution of the trapped-passing boundary layer,
where the distribution function is the largest and strongly depends on $\varphi$. This kind of off-set,
which scales with the width of this layer, $\propto \nu^{1/2}$, is present also in axisymmetric devices
and has been studied by \cite{helander11-092505}.

The result of Eqs.~\eq{lambda_off} and~\eq{w_off_booz} with $g^{(j)}_{\text{off}}$ obtained by propagator
method (Fig.~\ref{fig:offset_fits_realdim}) is compared to the NEO-2 result in Fig.~\ref{fig:lambda_off}.
\begin{figure}
\centerline{
 \includegraphics[width=0.49\textwidth]{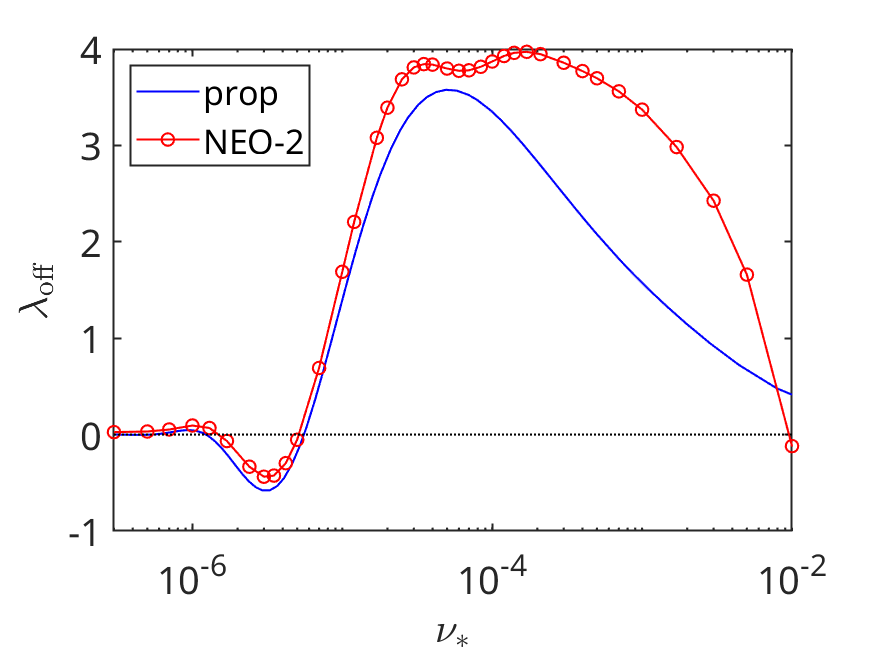}
 \includegraphics[width=0.49\textwidth]{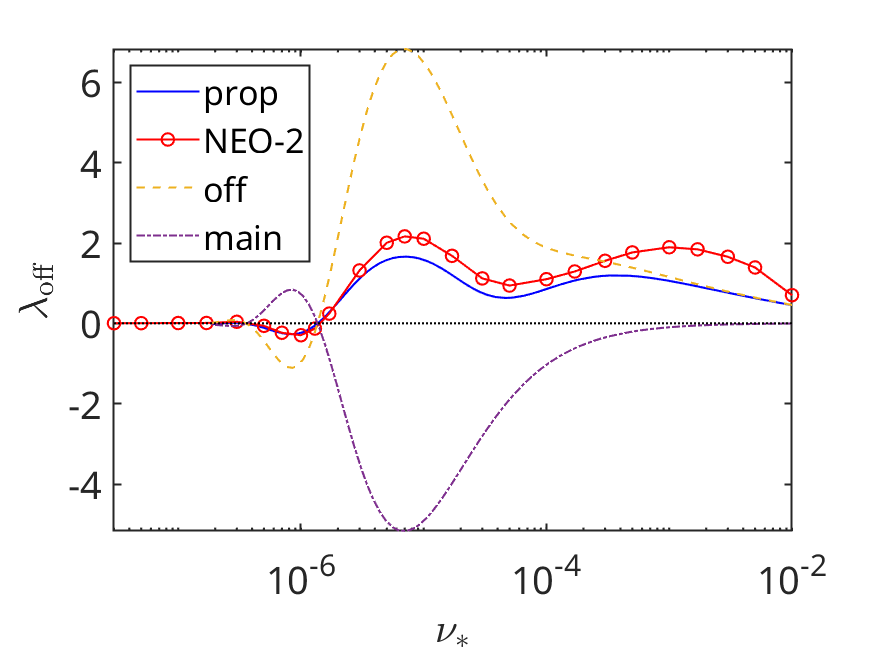}
}
\caption[]{
Off-set of the geometrical factor, $\lambda_{\text{off}}$, from the propagator method via Eq.~\eq{lambda_off} (blue)
and from NEO-2 results shown in Fig.~\ref{fig:lambda_bB_test_newtest} (red) for $\iota=1/4$ (left)
and $\iota=2/5$ (right). In case of $\iota=2/5$, also shown are summary contributions  to Eq.~\eq{lambda_off}
of the off-set domains (dashed) and of the main regions (dashed-dotted).
}
\label{fig:lambda_off}
\end{figure}
For configuration with $\iota=1/4$, off-set is fully determined by the first (and last) off-set well, $g^{(0)}_{\text{off}}$, since
$g^{(1)}_{\text{off}}=0$ and $g^{(2)}_{\text{off}}=-g^{(0)}_{\text{off}}$ due to stellarator symmetry.
For configuration with $\iota=2/5$, both, the off-set well, $g^{(0)}_{\text{off}}$, and the main region, $g^{(1)}_{\text{off}}$,
are contributing (with $g^{(3)}_{\text{off}}=-g^{(1)}_{\text{off}}$ and $g^{(4)}_{\text{off}}=-g^{(0)}_{\text{off}}$ just doubling the result
due to stellarator symmetry).
Off-set wells merge with main regions (0 with 1 and 3 with 4) at $\nuast < 10^{-4}$, therefore we used
$\eta_\text{loc}^{(j)}$ for such combined regions in individual $w^{(j)}_\text{off}$.
It can be seen that the contribution of the off-set wells is slightly larger than of the main regions,
despite much smaller toroidal extent, which is natural for
bounce-averaged drift (the longer the orbit, the smaller is this drift).
These two contributions tend to compensate each other at lower collisionalities,
which is a general trend due to the respective increase of the off-set domain length
studied in more detail in Section~\ref{ssec:asymptotic}.

As it could be expected, relatively good agreement between propagator method and NEO-2 is at very low collisionalities while at
$\nuast > 10^{-4}$ direct off-set due to boundary layer discussed by \cite{helander11-092505} becomes also
important.

\section{Bootstrap / Ware-pinch off-set at irrational surfaces}
\label{sec:irr}

As we have seen from the example of ``anti-sigma'' configuration in Section~\ref{ssec:align} Shaing-Callen limit cannot
be reached in case of perfectly aligned maxima where bootstrap / Ware pinch coefficient diverges as $1/\sqrt{\nu}$.
This, however, is the ideal case, which cannot be realized exactly. As follows from Eq.~\eq{dimoff_B}, due to the strong
($1/\nu$) scaling of maxima mis-alignment term with collisionality sooner or later small mis-alignments of local maxima
will prevail, and configuration will turn into a general one with a single global maximum. Therefore, we have to consider
this general case now.

\subsection{Long field lines}
\label{ssec:lonfieldlines}

Field line integration technique employed in analytical asymptotical models and in NEO-2 assumes that the representative field
line covers the flux surface densely enough so that the result does not depend on its length anymore.
There are at least two ways to do so, in a way the ends of the field line are identical (and the solution is periodic) at the
irrational surface. One way (standard in NEO-2) is to follow the line until a good match with the starting point, and then
slightly modify $B(\varphi)$ in the last field period for a smooth field line closure. The result can be verified then by using
another starting point at the same flux surface. Another way (employed here) is to replace $\iota$ with a rational number of high order
well approximating the original irrational $\iota$ and start from the global maximum. Thus, within stellarator symmetry,
one can always keep quasi-Liouville's theorem to be fulfilled exactly. Within this second approach, we consider now
five different field lines for the magnetic field model used in Section~\ref{ssec:fincolnum} with
$\iota = 0.44 = 11/25$, $\iota = 0.43 = 43/100$, $\iota = 0.435 = 87/200$, $\iota = 0.438 = 219/500$
and $\iota = 0.442 = 221/500$ which cover the field period with 75, 300, 600 and 1500 passes, respectively.
\begin{figure}
\centerline{
 \includegraphics[width=0.49\textwidth]{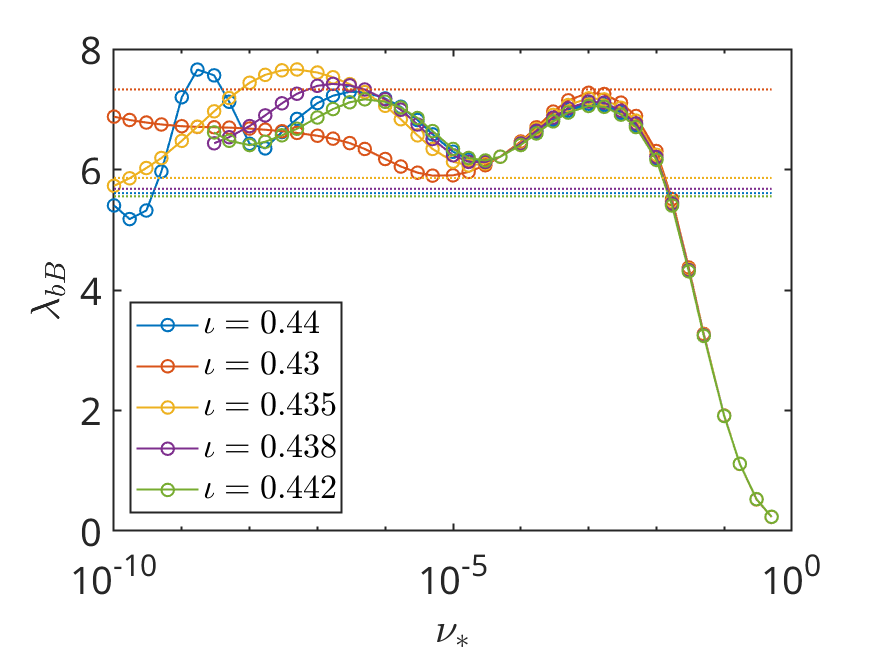}
 \includegraphics[width=0.49\textwidth]{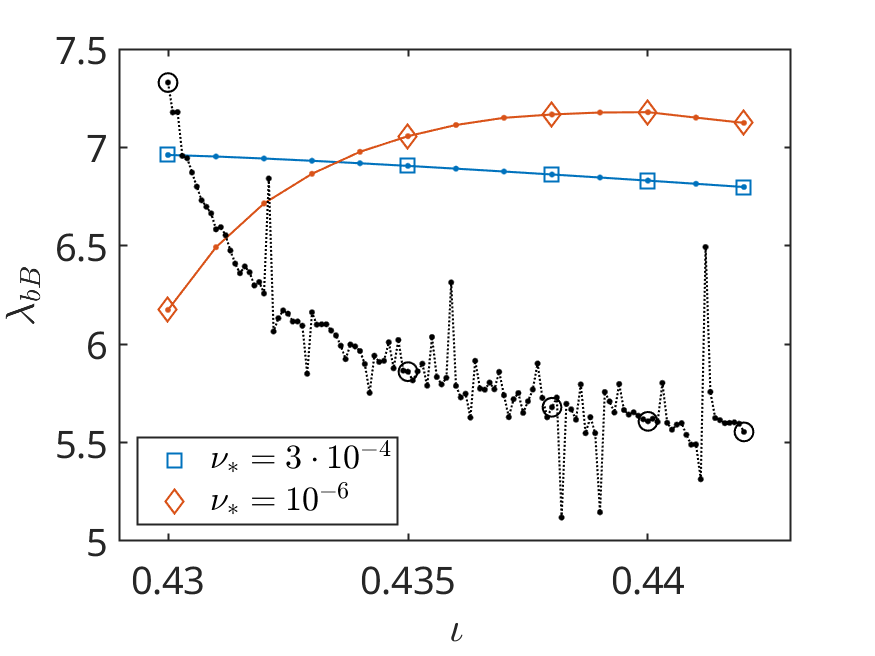}
}
\caption[]{
Normalized bootstrap coefficient~\eq{D31_lambb} for five $\iota$ values vs. normalized collisionality $\nuast$
(left) and its dependence on $\iota$ for two selected collisionalities shown in the legend (right).
Shaing-Callen limit~\eq{lambda_bB} is shown with dotted lines (data points - with dots).
Abscissa values for markers in the right plot corresponds to
the legend in the left plot.
}
\label{fig:lambb_longmfl}
\end{figure}
The resulting normalized bootstrap coefficients~\eq{D31_lambb} are shown in Fig.~\ref{fig:lambb_longmfl}.
They are weakly dependent on $\iota$ for collisionalities $\nuast > 10^{-5}$
(in the first hill), and follow this trend down to $\nuast = 10^{-6}$ in the second hill,
with the only curve, $\iota=0.43$, departing from the other four. As one can see from the $\iota$ scan in this figure,
this transition is smooth, in contrast to the Shaing-Callen limit which shows well known ``bootstrap resonances''
briefly discussed in Section~\ref{ssec:bootres}.
\begin{figure}
\centerline{
 \includegraphics[width=0.49\textwidth]{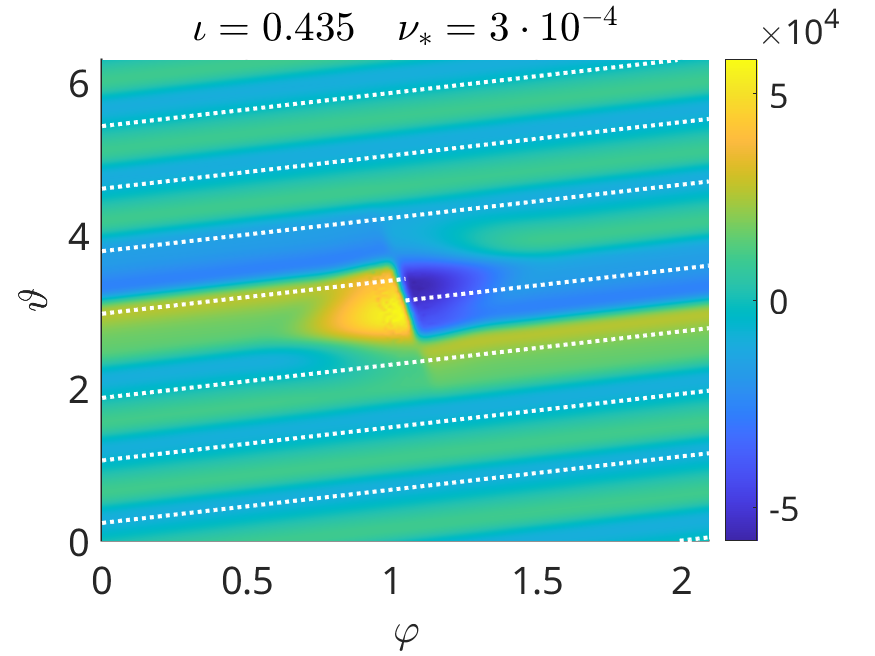}
 \includegraphics[width=0.49\textwidth]{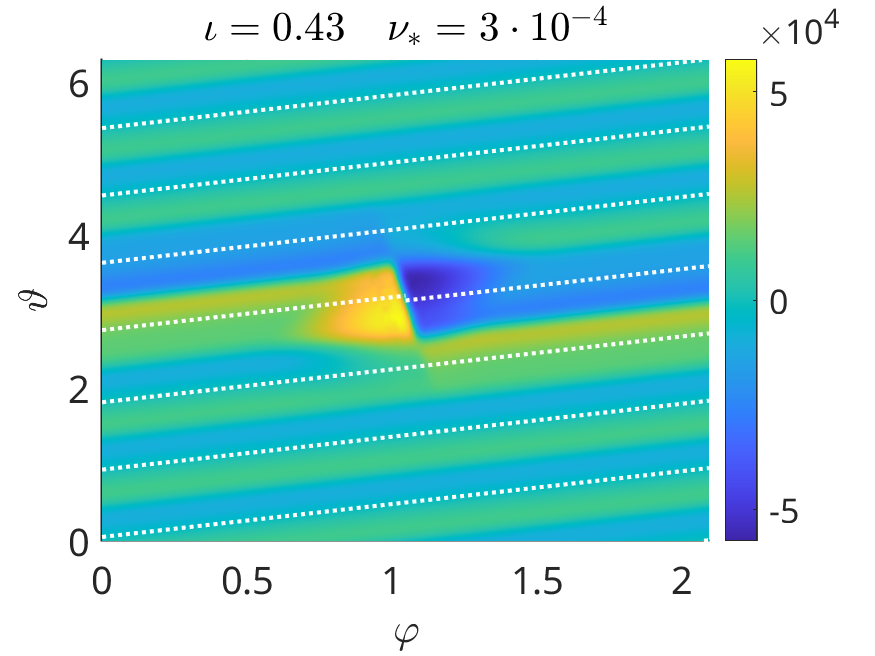}
}
\caption[]{
Generalized Spitzer function $g_{(3)}(\vartheta,\varphi,\lambda)$ for standing particles ($\lambda=0$)
at the flux surface with $\iota=0.435$ (left) and $\iota=0.43$ (right) for $\nuast = 3\cdot 10^{-4}$.
White dotted lines show the field line starting from the global maximum and making 7 toroidal turns.
}
\label{fig:g_refl_3m4}
\end{figure}
To analyze this behavior, we plot the Ware pinch distribution function $g_{(3)}(\vartheta,\varphi,\lambda)$
(mono-energetic generalized Spitzer function)
for particles with zero parallel velocity ($\lambda=0$) as function of angles for two cases of $\iota=0.435$ and $\iota=0.43$
and two different collisionalities, $\nuast = 3\cdot 10^{-4}$ (Fig.~\ref{fig:g_refl_3m4}) and $\nuast = 10^{-6}$
(Fig.~\ref{fig:g_refl_1m6}).
\begin{figure}
\centerline{
 \includegraphics[width=0.49\textwidth]{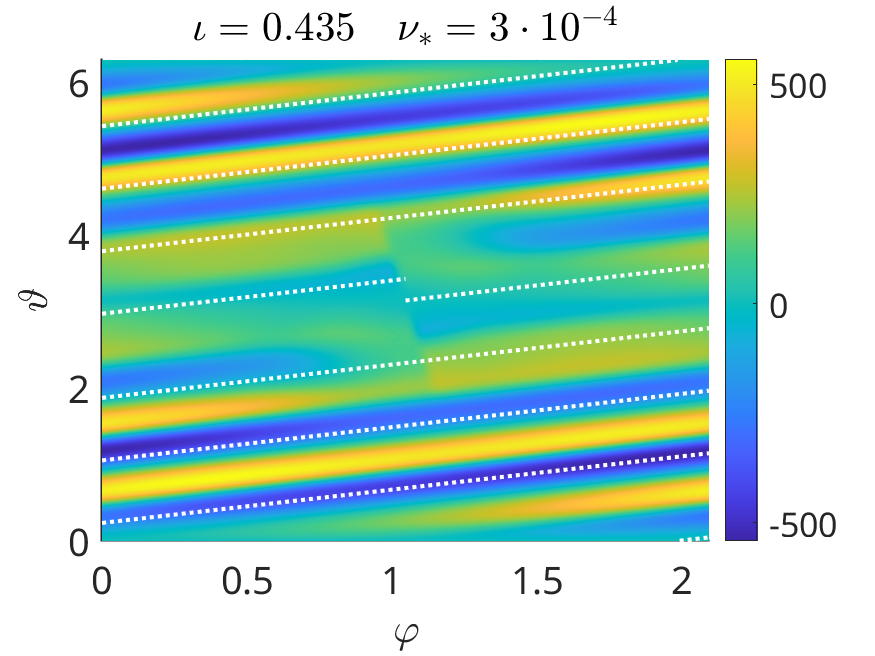}
 \includegraphics[width=0.49\textwidth]{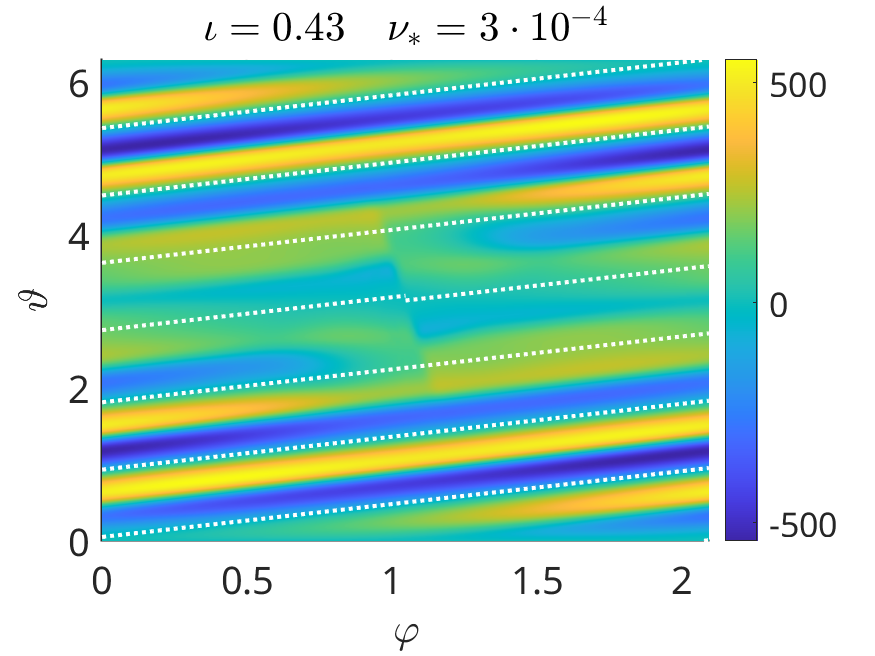}
}
\caption[]{
Normalized distribution over the angles of trapped particle radial flux $\gamma(\vartheta,\varphi)$
at the flux surface with $\iota=0.435$ (left) and $\iota=0.43$ (right) for $\nuast = 3\cdot 10^{-4}$.
White dotted lines - the same as in Fig.~\ref{fig:g_refl_3m4}.
}
\label{fig:fluxden_3m4}
\end{figure}
Since the distribution function in the trapped particle domain is almost independent of pitch parameter (equivalently of $\eta$),
the plotted quantity characterizes well the radial particle flux. Obviously, this function is stellarator anti-symmetric,
$g_{(3)}(\vartheta,\varphi,0)=-g_{(3)}(-\vartheta,-\varphi,0)$, and retains all other relevant features seen for the short field
line examples in Section~\ref{sssec:adjoint_num}.
Namely, ``hot spots'' around the global maximum ($\vartheta=\pi$, $\varphi=\pi/3$) where $g_{(3)}$ is the largest correspond
to the contact region with the boundary layer (see the left Fig.~\ref{fig:offset_neo2_pure}). These spots are responsible for the
tokamak-like off-set studied by \cite{helander11-092505}. One can clearly see in Fig.~\ref{fig:g_refl_3m4}
also the off-set domains (blue and yellow) in the ripples adjacent to the global maximum (mainly contributing
to the off-set at this collisionality) as well as widely extending ``main region'' (green). Thus, this case is rather similar
to Fig.~\ref{fig:spitf_surf} for $\nuast = 10^{-4}$.  Remarkably, results for two different $\iota$ are almost
indistinguishable, which explains close values of $\lambda_{bB}$ in Fig.~\ref{fig:lambb_longmfl} at this collisionality.
Since the radial guiding center velocity $v_g^r$ is stellarator antisymmetric as well, the resulting normalized radial trapped
particle flux distribution over the angles $\gamma$ given with good accuracy by
$\gamma(\vartheta,\varphi) = \cV (\vartheta,\varphi) g_{(3)}(\vartheta,\varphi,0)$
where $\cV$ is the sub-integrand in Eq.~\eq{w_off_booz}, is stellarator-symmetric.
It is shown for $\nuast = 3\cdot 10^{-4}$ in
Fig.~\ref{fig:fluxden_3m4} where the contribution of the off-set domain ($2 < \vartheta < 4$) is positive what agrees
with the first hill in Fig.~\ref{fig:lambb_longmfl}. Note that direct contribution of the boundary layer is already small
at this collisionality which is also seen from vanishing hot spots in Fig.~\ref{fig:fluxden_3m4}.

Off-set in adjacent toroidal ripples vanishes completely at lower collisionality case $\nuast = 10^{-6}$ shown in
Fig.~\ref{fig:g_refl_1m6}. The visible difference between $\iota=0.435$ and $\iota=0.43$ is different amplitude of straps
on $g_{(3)}$ which, in fact, is higher for $\iota=0.43$ where the off-set of $\lambda_{bB}$ is smaller
(see Fig.~\ref{fig:lambb_longmfl}). Less visible is the split of the straps into off-set domain and the main region
for $\iota=0.435$. Actually, this split is responsible for lower amplitude of the straps in this case.
\begin{figure}
\centerline{
 \includegraphics[width=0.49\textwidth]{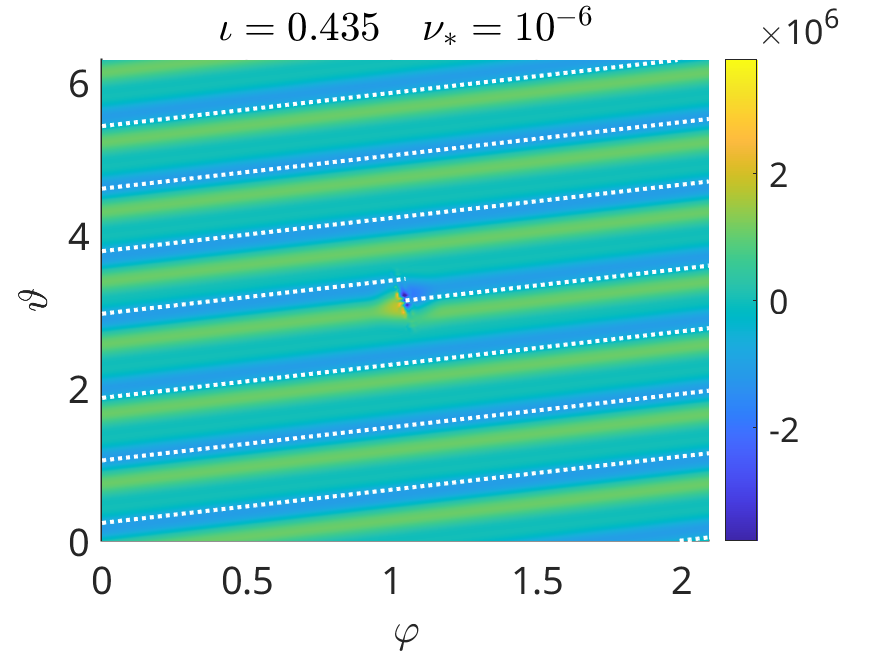}
 \includegraphics[width=0.49\textwidth]{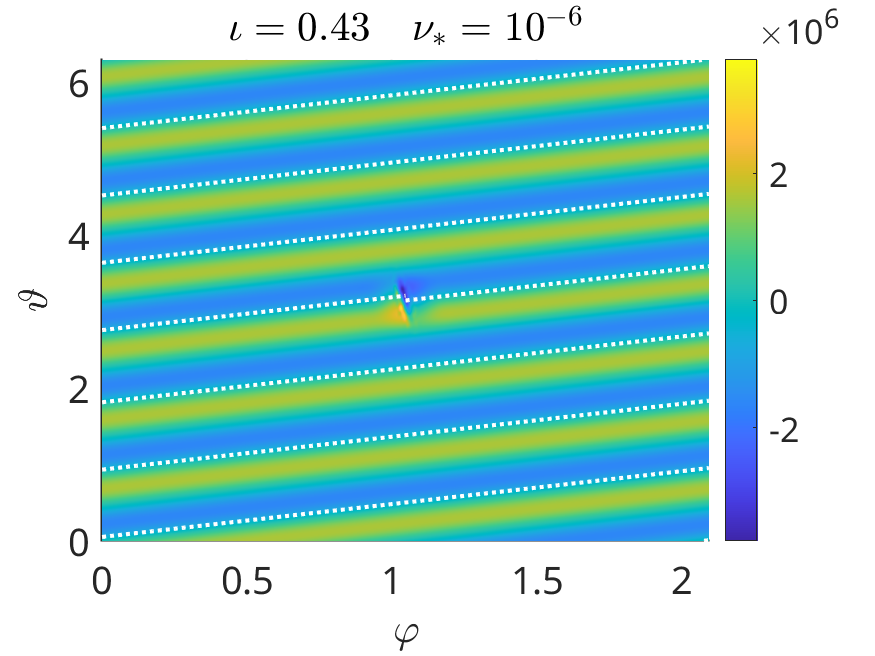}
}
\caption[]{
The same as in Fig.~\ref{fig:g_refl_3m4} for $\nuast= 10^{-6}$.
}
\label{fig:g_refl_1m6}
\end{figure}
Note that the strap structure and difference in its amplitude for two $\iota$ values transfers also to the angular
distribution of radial particle flux shown in Fig.~\ref{fig:fluxden_1m6}. Despite the factor 1.5 higher amplitude of
the flux density modulation, the value of $\lambda_{bB}$ is only 14 \% different (and is lower) for $\iota=0.43$ than
for $\iota=0.435$. The reason is in relatively low contribution of the strap structure dominating the plot
to the radial transport because this contribution is mostly averaged to zero. Namely, change of $g_{(3)}$ along the straps
(more precisely, along the field lines)
is rather slow, but not slow is the change of local guiding center velocity which, at the end, results in much smaller
bounce averaged value actually contributing to the flux.
\begin{figure}
\centerline{
 \includegraphics[width=0.49\textwidth]{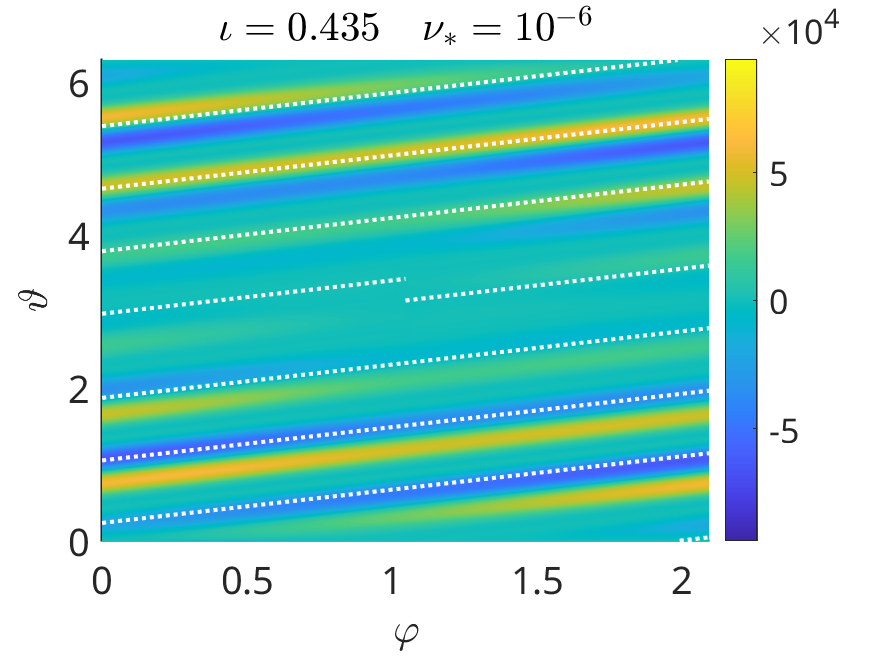}
 \includegraphics[width=0.49\textwidth]{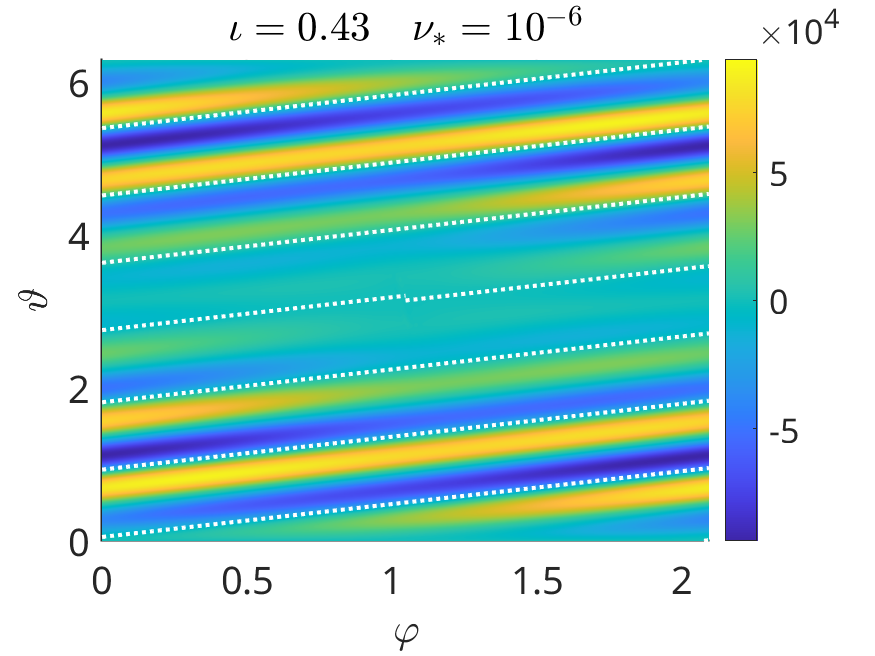}
}
\caption[]{
The same as in Fig.~\ref{fig:fluxden_3m4} for $\nuast= 10^{-6}$.
}
\label{fig:fluxden_1m6}
\end{figure}

The appearance of strap structure on generalized Spitzer function $g_{(3)}$ which only emerges in Fig.~\ref{fig:g_refl_3m4}
and becomes dominant in Fig.~\ref{fig:g_refl_1m6} is connected with a certain paradox of the asymptotic solution of \cite{helander11-092505} for trapped particles, $g_0^t$, given by Eq.~\eq{g0_adj_tr} in Section~\ref{ssec:adjprob}.
Being perfectly correct and accurately reproduced at short field lines and lowest collisionalities by NEO-2, it does not
result in smooth distributions over irrational flux surfaces, but rather in the fractal structure with an infinite amplitude.
Since two arbitrarily close points on the flux surface can be connected in the $1/\nu$ regime only along the field line,
the connecting segment of the field line is in general infinite, which results in infinite values of the integral~\eq{g0_adj_tr}.
As we saw already in Section~\ref{sssec:adjoint_num}, this paradox is resolved for finite collisionality. Namely, whenever
the field line enters the ``hot spot'' (boundary layer), it starts a new segment which should be treated independently from
the rest. Thus, the problem becomes local at finite collisionalities even at the irrational surfaces. With reducing
$\nuast$ the size of the ``hot spot'' is reduced, and, at some point there will be a transition from a 7 strap structure
to another structure with more straps. Finally, in zero collisionality limit, the fractal structure of the solution of \cite{helander11-092505} will be recovered. Of course, within this simple splitting approach combined with
asymptotic model, fractal structure will not appear anymore but the off-set effect will naturally be different from
the one in collisional model.

The actual ``splitting'' of the collisional solution for $\nuast = 3\cdot 10^{-4}$
is shown in Fig.~\ref{fig:along_mfl_3m4} where the same distribution functions
as in Fig.~\ref{fig:g_refl_3m4} are plotted along two field lines as functions of the number of toroidal turns
$(\varphi-\varphi_0)/(2\pi)$. First line is starting from
the global maximum $(\vartheta_0,\varphi_0)=(\pi,\pi/3)$, and second one starts from the close point
with $\vartheta=\vartheta_0+\delta \vartheta$.
\begin{figure}
\centerline{
 \includegraphics[width=0.49\textwidth]{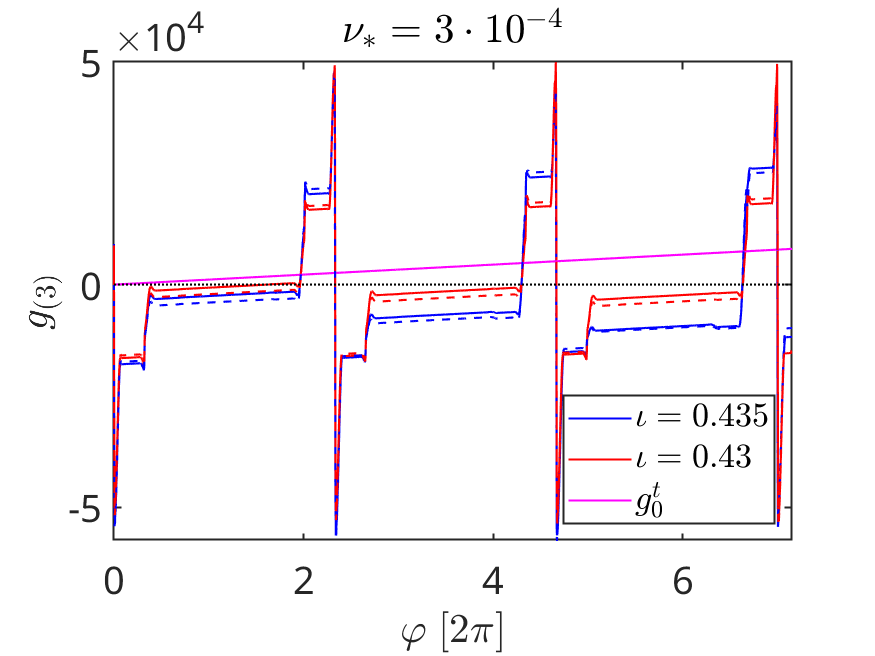}
 \includegraphics[width=0.49\textwidth]{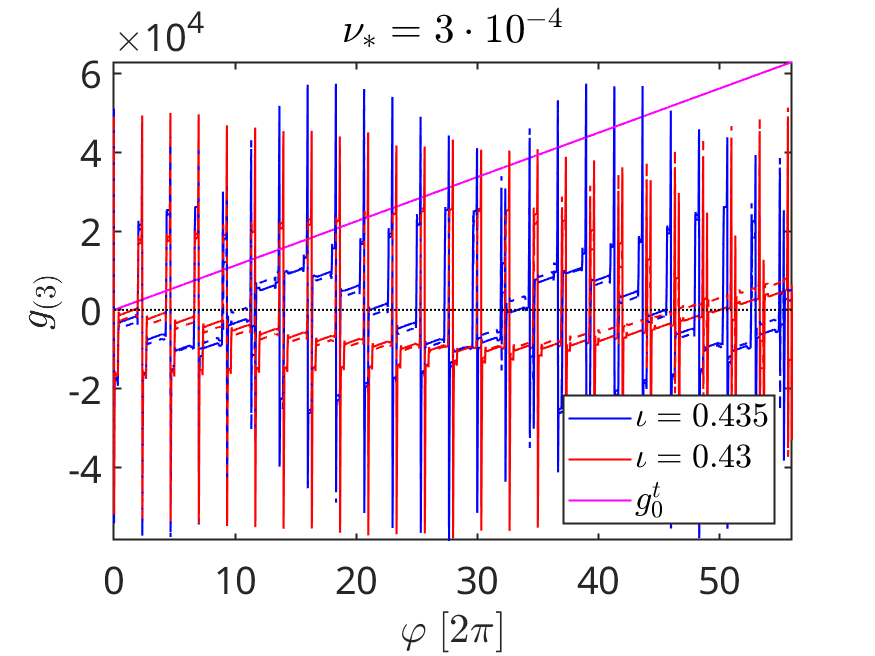}
}
\caption[]{
Generalized Spitzer function $g_{(3)}$ along the field line starting from global maximum (solid) and from the close point
displaced by $\Delta \vartheta = 0.03$ (dashed) for $\iota = 0.435$ (blue) and $\iota = 0.43$ (red)
and $\nuast = 3\cdot 10^{-4}$. The magenta line shows the
asymptotic solution $g_0^t$, Eq.~\eq{g0_adj_tr}, at the first field line.
Left and right plots show the same for 7 and 56 periods, respectively.
}
\label{fig:along_mfl_3m4}
\end{figure}
Together with solutions of NEO-2 for two $\iota$ values, also plotted is the asymptotic solution of \cite{helander11-092505}, Eq.~\eq{g0_adj_tr},
which has a form of a straight line in Boozer coordinates, $g_0^t =(\iota B_\vartheta + B_\varphi)(\varphi-\varphi_0)$.
In the left plot, one can observe the same pattern consisting of the ``main region'' and two off-set domains
(this case is similar to the one in Fig.~\ref{fig:spitf_surf}) repeated three
times, which corresponds to a ``resonance'' with $\iota = 3/7$ (note that $0.435 -3/7 \approx 0.0064$ and
$0.43 -3/7 \approx 0.0014$). One can also see that $g_0^t$ well represents the solution within each domain up to a constant
shift introduced by splitting in the boundary layers.
Spikes on $g_{(3)}$ correspond to the contact with the trapped-passing boundary layer
(``hot spots''), and one can see nearly no spikes at the transitions between the off-set domains and the main region,
which correspond to the boundary layers between trapped particle classes. One can also see slow accumulation of a different
type of off-set in the main regions, which results in a different $g_{(3)}$ patterns on the long run (see the right plot).
This off-set, however, has much smaller effect on $\lambda_{bB}$ (see Fig.~\ref{fig:lambb_longmfl}) because bounce averaged
velocity for relatively long ``main regions'' is much smaller than for the short off-set domains.

Short off-set domains fully disappear at lower collisionality $\nuast = 10^{-6}$ shown in Fig.~\ref{fig:along_mfl_1m6}.
\begin{figure}
\centerline{
 \includegraphics[width=0.49\textwidth]{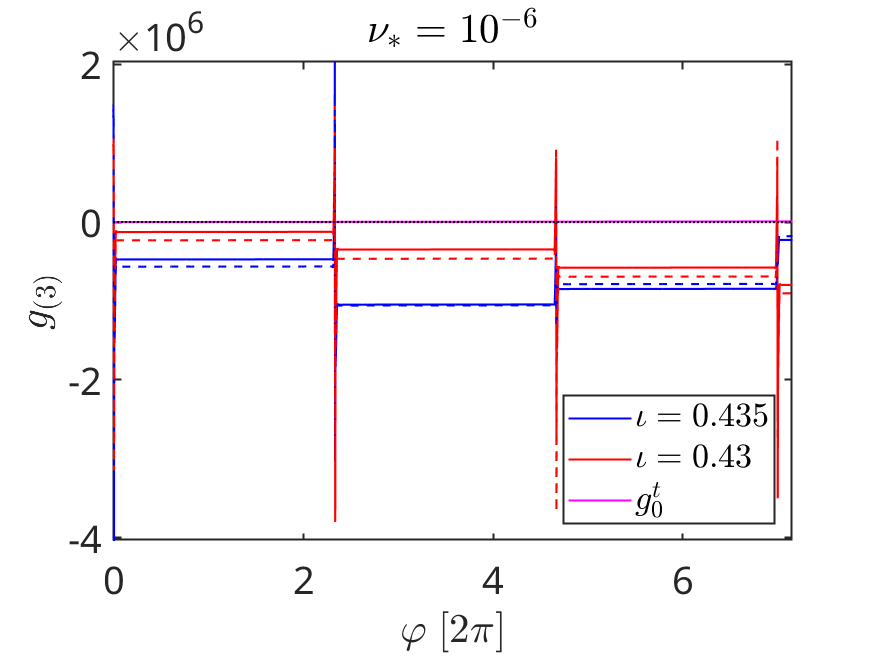}
 \includegraphics[width=0.49\textwidth]{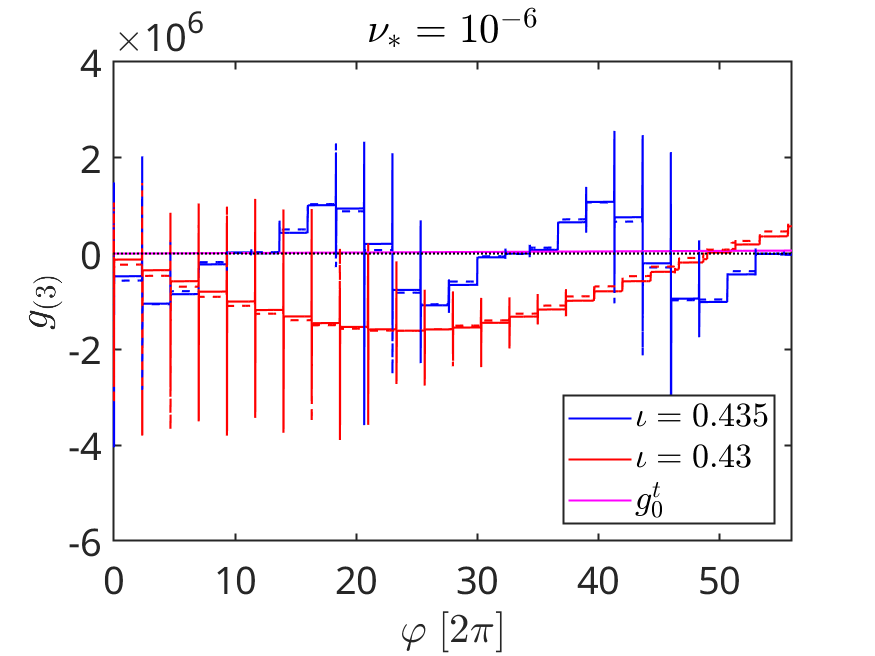}
}
\caption[]{
The same as in Fig.~\ref{fig:along_mfl_3m4} for $\nuast= 10^{-6}$ and $\Delta\vartheta = 0.01$.
}
\label{fig:along_mfl_1m6}
\end{figure}
There, the off-set pattern on the long run determines $\lambda_{bB}$, which shows then in Fig.~\ref{fig:lambb_longmfl} a
significant difference for two $\iota$ values under examination here. We see that this pattern is formed by sub-domains with
toroidal extent corresponding to a full poloidal turn of the field line.
For the case $\iota=0.435$, pattern is repeated twice in the
long run and is formed by the ``main region'' consisting of 3 sub-domains (field line misses the boundary layer during these full
poloidal turns), and there are 3 off-set domains on each side of the main region with the opposite off-set value in each triple
(field line visits the boundary layer within these sub-domains at opposite sides for each triple).
One can see also ``symmetric'' sub-domains with nearly no off-set,
which separate the patterns approximately repeated after 25 toroidal turns.
In contrast, for $\iota=0.43$, almost a harmonic structure is formed with a period determined by
$\Delta \iota = 0.44 - 0.43 = 0.01$.

Finally, the normalized trapped particle radial flux distribution over the angles $\gamma = \cV g_{(3)}$ (the same
quantity as in Fig.~\ref{fig:fluxden_1m6}) and the sub-domain average $\bar \gamma = \langle \cV \rangle_{\vartheta} g_{(3)}$
where $ \langle \cV \rangle_{\vartheta}$ are averages over the field line segments corresponding to one
poloidal turn (i.e. these are integrals in Eq.~\eq{w_off_booz} normalized by
$\varphi_{j+1}-\varphi_j$ with $[\varphi_j,\varphi_{j+1}]$ being the sub-domain)
are shown in Fig.~\ref{fig:gamma_along_mfl_1m6} for the same field lines as
in Fig.~\ref{fig:along_mfl_1m6} and $\nuast = 10^{-6}$.
\begin{figure}
\centerline{
 \includegraphics[width=0.49\textwidth]{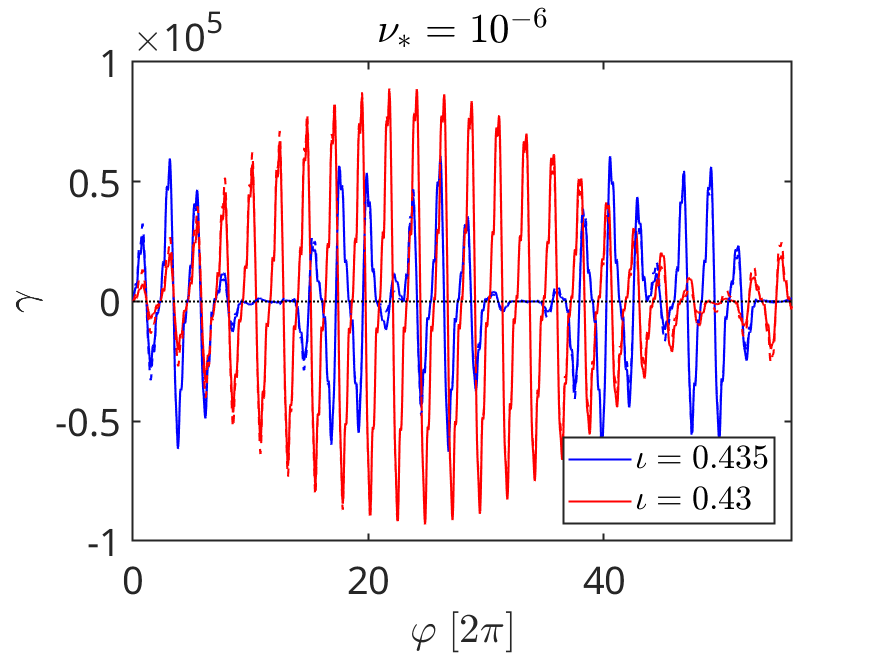}
 \includegraphics[width=0.49\textwidth]{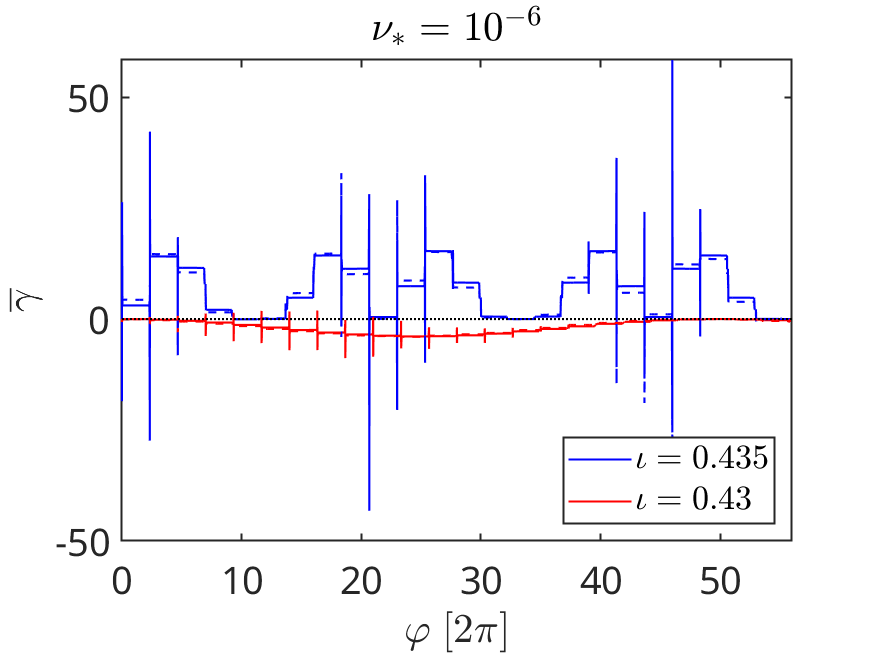}
}
\caption[]{
Normalized trapped particle radial flux angular density $\gamma$ (left) and sub-domain average $\bar\gamma$ (right)
for the distribution functions shown in Fig.~\ref{fig:along_mfl_1m6}.
}
\label{fig:gamma_along_mfl_1m6}
\end{figure}
It can be seen that large oscillations on the local flux density $\gamma$ are mostly averaged off within sub-domains,
and the resulting sub-domain average $\bar\gamma$ which is determined mainly by the bounce-averaged radial drift
shows the significant positive contribution of the off-set domains for $\iota = 0.435$ and a smaller negative
contribution for $\iota = 0.43$ (in agreement with the trend in Fig.~\ref{fig:lambb_longmfl}).

\subsection{Asymptotic behavior at irrational surfaces}
\label{ssec:asymptotic}

For estimation of $\lambda_{bB}$ trend with $\nuast \rightarrow 0$ we analyze a nearly periodic pattern seen
in right plots of Figs.~\ref{fig:along_mfl_1m6} and~\ref{fig:gamma_along_mfl_1m6} (such pattern is seen also in the left
plot of Fig.~\ref{fig:along_mfl_3m4} while the right plot shows the overlay of two patterns).
This pattern is formed if the field line starting
in close vicinity of global maximum point $(\vartheta_{\rm max},\varphi_{\rm max})$, i.e. from the point
$(\vartheta_0,\varphi_0)=(\vartheta_{\rm max}+\Delta\vartheta_0,\varphi_{\rm max})$
where $\Delta\vartheta_0 \ll \delta\vartheta$ and $\delta\vartheta$
is the poloidal boundary layer width (width of ``hot spots'' in Figs.~\ref{fig:g_refl_3m4} and~\ref{fig:g_refl_1m6}),
returns after $N$ turns back to this vicinity,
$(\vartheta_N,\varphi_N)=(\vartheta_{\rm max}+\Delta\vartheta_N,\varphi_{\rm max})$ with $\Delta\vartheta_N \ll \delta\vartheta$.
The off-set pattern is formed if within these $N$ turns at least one approximate match occurs at some
$(\vartheta,\varphi)=(\vartheta_{\rm max}+\Delta\vartheta,\varphi_{\rm max})$ with
$\Delta\vartheta_0 \ll \Delta\vartheta \le \delta\vartheta$. By the symmetry argument (one can make $N$ turns backwards),
such matches come in pairs thus forming one ``main region'' and at least two off-set domains with positive and negative
off-set respectively (see the left plot in Fig.~\ref{fig:along_mfl_3m4} and the
right plots in Figs.~\ref{fig:along_mfl_1m6} and~\ref{fig:gamma_along_mfl_1m6}
where six off-set domains are formed for $\iota=0.435$).

Expanding the magnetic field around the global maximum point,
\be{expmax}
B(\vartheta_{\rm max}+\Delta\vartheta,\varphi_{\rm max}+\Delta\varphi) \approx B_{\rm max}
+\frac{1}{2\beta}\difp{^2 B}{\vartheta^2}\left(\beta\Delta\vartheta^2+\Delta\varphi^2\right),
\qquad
\beta=\difp{^2B}{\vartheta^2}\left(\difp{^2B}{\varphi^2}\right)^{-1}.
\ee
where second derivatives correspond to the global maximum point, and
mixed derivative there is zero for our example~\eq{magfield_mod},
we can relate the distance $\Delta\eta$ between class transition boundary $\eta=\eta_c=1/B^\text{loc}_\text{max}$
and trapped-passing boundary
$\eta=\eta_b=1/B_{\rm max}$
to the poloidal distance $\Delta\vartheta$ between the respective local maximum and the global maximum as
\be{Delta_eta_Delta_theta}
\Delta\eta \approx \frac{1+\iota^2\beta}{2 B_{\rm max}^2}\left|\difp{^2B}{\vartheta^2}\right|\Delta\vartheta^2.
\ee
We have expressed here the toroidal distance between these maxima as $\Delta\varphi=-\iota\beta\Delta\vartheta$
which follows from the local maximum condition $\bh\cdot\nabla B = 0$ and Eq.~\eq{expmax}.
Assuming that rational numbers $0< m/n < 1$ with $n<N_{\max}$ are distributed uniformly in this interval (which is generally
not true but gives a correct order of magnitude estimate) we obtain for the best match at some $n=N$ that
$$
\min\limits_{m,n\le N_{\max}}|\iota n-m| \sim 1/N_{\max} \sim 1/N.
$$
Thus, a field line starting at a global maximum will be displaced from this maximum after $N$ toroidal turns by
$\Delta\vartheta_N \sim 2\pi /N$ and, according to~\eq{Delta_eta_Delta_theta}, pass through the local maximum
with respective class boundary $\eta_c$ displacement $\Delta\eta_N=\eta_c-\eta_b$ given by
\be{deta_N}
\Delta \eta_N \approx \frac{2 \pi^2 \varepsilon_t\eta_b}{N^2},
\ee
where we denoted $\varepsilon_t \equiv (1+\iota^2\beta)\eta_b\left|\partial^2 B /\partial\vartheta^2\right|$ well represented
by field modulation in axisymmetric tokamak. Trapped-passing boundary layer width corresponding to $N$ toroidal turns
is given by Eq.~\eq{delta_eta_fund} with $\varphi=\varphi_0$ and $\varphi^\prime =\varphi_0+2\pi N$ resulting in the estimate
\be{blwidth}
\delta\eta \approx 4 \pi^{-1/2}(2\varepsilon_t)^{1/4}\eta_b\left(\nuast N\right)^{1/2},
\ee
obtained here for $\varepsilon_M \ll \varepsilon_t \ll 1$ (see Eq.~\eq{magfield_mod}) and using
$\iota B_\vartheta+B_\varphi \approx R/\eta_b$.
Since the poloidal width of the boundary
layer $\delta\vartheta$ and boundary layer width $\delta\eta$ are related by~\eq{Delta_eta_Delta_theta} with replacement
$\Delta \rightarrow \delta$, best match after $N$ turns completing the off-set pattern means $\Delta\eta_N \ll \delta\eta$,
which, according to Eqs.~\eq{deta_N} and~\eq{blwidth} result in
\be{N_deta}
N \gg \frac{\pi \varepsilon_t^{3/10}}{\sqrt{2}\nuast^{1/5}},
\qquad
\delta\eta \gg 4 \eta_b \left(\varepsilon_t \nuast\right)^{2/5}.
\ee
The last expression~\eq{N_deta} agrees with the $\nuast$ scaling of the boundary layer width obtained by
a similar argument in \citep{helander17-905830206}. Since $\Delta\eta_N/\delta\eta \sim N^{-5/2}$ rapidly decays
with $N$, strong inequality~\eq{N_deta} is achieved even if one only doubles the $N$ value given by the right-hand side. Therefore,
we use Eqs.~\eq{N_deta} further, as order of magnitude estimates rather than strong conditions. Assuming for simplicity then
only one pair of off-set domains is formed with local minima
mis-match of the order of the boundary layer width in the main region,
$\Delta\eta \sim \delta\eta_{\rm ref}$ and estimating this width in the large aspect ratio limit as
$\delta\eta_{\rm ref} \sim \delta\eta \sim 4 \eta_b \left(\varepsilon_t \nuast\right)^{2/5}$, we obtain
the distribution function in the off-set domains with help of Eq.~\eq{dimoffset_def},
Eq.~\eq{last_pair_ofsreg_sols}
and estimate $\alpha^\pm_{mr} \sim 1$ due to $x_{\pm 0} \sim \Delta\eta/\delta\eta_{\rm ref} \sim 1$ as
\be{goff_est}
g_{\text{off}}^{(j)} \sim R B_0 \varepsilon_t^{-1/10} \nuast^{-3/5},
\ee
where $B_0=B_{\rm max}$, and constant~\eq{C0} has been estimated as
$C_0 \sim R B_0^2 /(\nuast \varepsilon_t^{1/2})$.
Note that up to a factor 2
scaling~\eq{goff_est} is reproduced by the relation between color ranges in Figs.~\ref{fig:g_refl_3m4} and~\ref{fig:g_refl_1m6}
and is in good agreement with collisionality dependence of the span $g_\text{off}^\text{(sp)}$ of the generalized Spitzer
function which corresponds to the maximum off-set at $\varphi=0$ line
(in particular, in Figs.~\ref{fig:g_refl_3m4} and~\ref{fig:g_refl_1m6}), see Fig.~\ref{fig:offset_trend_longmfl}.
\begin{figure}
\centerline{
\includegraphics[width=0.49\textwidth]{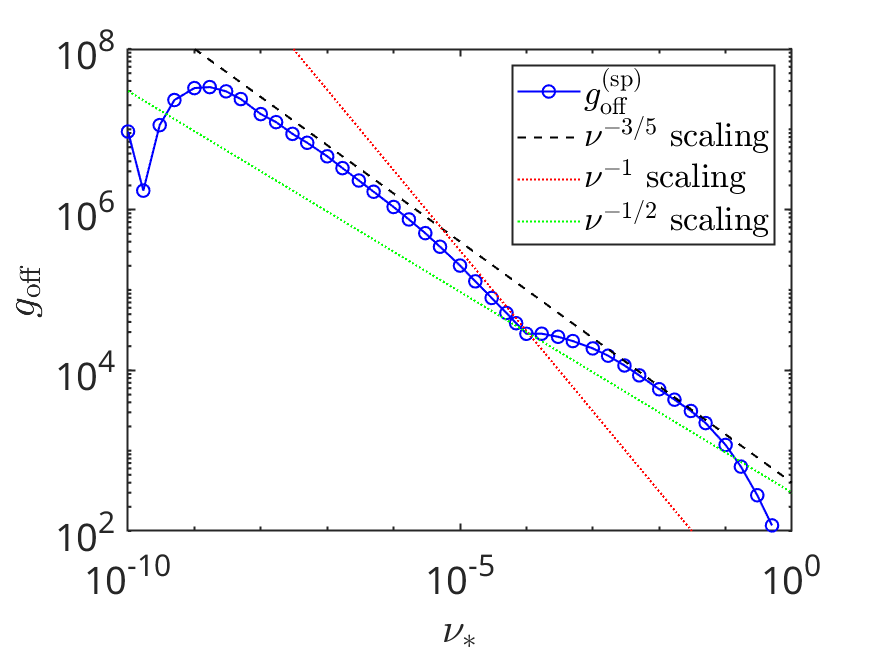}
}
\caption[]{
Span of the generalized Spitzer function
$g_\text{off}^\text{(sp)}\equiv\left(\max\left(g_\text{(3)}\right)
-\min\left(g_\text{(3)}\right)\right)/2$ for standing particles, $\lambda=0$, at $\varphi=0$ as function
of collisionality (case $\iota = 0.435$ in Fig.~\ref{fig:lambb_longmfl}).
Scalings $\nu_\ast^{-3/5}$, $1/\nu_\ast$ and $1/\sqrt{\nu_\ast}$ are shown by black dashed,
red and green dotted lines, respectively.
}
\label{fig:offset_trend_longmfl}
\end{figure}

It remains now to estimate the weighting factors $w_{\text{off}}^{(j)}$ in Eq.~\eq{lambda_off} which results for our near-periodic
pattern with two similar off-set domains in $\lambda_{\text{off}} \sim g_{\text{off}}^{(j)} w_{\text{off}}^{(j)}$ with $j=0$.
Expanding the sub-integrand $\cV$ in~\eq{w_off_booz} in Fourier series over periodic Boozer angles,
\be{fourF}
\cV(\vartheta,\varphi)=
\sum\limits_{m=1}^\infty
\sum\limits_{n=-\infty}^\infty
\cV_{mn}^s(\eta)\sin(m\vartheta+n\varphi)
\ee
with $\eta=\eta_{\text{loc}}^{(j)}$,
this series has no $m=0$ terms due to the last~\eq{cVdef} where the derivative over $\vartheta_0$ is
the same as the derivative over $\vartheta$. Here, we placed the origin of angles 
at the global $B$ maximum to
avoid cosine harmonics in stellarator symmetric fields.
Introducing the number of toroidal turns per off-set domain, $N_o \lesssim N$, we present
$\iota = \iota_o + \delta\iota$ where $\iota_o=M_o/N_o$,
$M_o$ is the number of poloidal turns and $\delta\iota \ll 1$.
Field line segment with the off-set domain starts at $(\vartheta,\varphi)=(\Delta\vartheta_0,0)$
and ends at $(\vartheta,\varphi)=(2\pi M_o+\Delta\vartheta_N,2\pi N_0)$
where $\Delta\vartheta_N=\Delta\vartheta_0 + 2\pi N_o \delta\iota$,
and the tips of the transient banana orbit are located at
\mbox{$\left(\vartheta_{(j)}^-,\varphi_{(j)}^-\right)
=\left(\Delta\vartheta_0 + \iota\delta\varphi_0,\delta\varphi_0\right)$}
and
$\left(\vartheta_{(j)}^+,\varphi_{(j)}^+\right)
=\left(2\pi M_o+\Delta\vartheta_N+\iota\delta\varphi_N,2\pi N_o+\delta\varphi_N\right)$,
where \mbox{$\delta\varphi_{0,N} \sim \max(|\Delta\vartheta_0|,|\Delta\vartheta_N||)\ll 1$}
are toroidal shifts of those banana tips from the 
global maximum points, see Fig.~\ref{fig:schematic_offset_region}.
Replacing in $I_{\cV}^{(j)}$, Eq.~\eq{w_off_booz}, the integrand $\cV$ with $\cV_b$ which is given by the same Eq.~\eq{cVdef}
with the replacement $\eta_{\text{loc}}^{(j)} \rightarrow \eta_b$,
extending there the integration interval to the whole off-set segment $0 < \varphi < 2\pi N_o$ and thus ignoring the shifts 
$\delta \varphi_{0,N}$, and
expanding $\cV_b$
around the helical line $\vartheta=\iota_o\varphi$ we get
\bea{I_cVj}
I_{\cV}^{(j)}
&=& \int\limits_{\varphi_{(j)}^-}^{\varphi_{(j)}^+}
\rd \varphi\;
\cV\left(\Delta\vartheta_0 + \iota \varphi,\varphi\right)
\\
&\approx&
\int\limits_{0}^{2\pi N_o}\rd \varphi
\left(
\cV_b(\iota_o \varphi,\varphi)
+
\cV_b^\prime(\iota_o \varphi,\varphi) (\Delta\vartheta_0+\delta\iota\, \varphi)
+
\frac{1}{2}
\cV_b^{\prime\prime}(\iota_o \varphi,\varphi) (\Delta\vartheta_0+\delta\iota\, \varphi)^2
\right),
\nonumber
\eea
where $\cV_b^\prime$ and $\cV_b^{\prime\prime}$ denote first and second derivatives of $\cV_b(\vartheta,\varphi)$
over $\vartheta$. First two approximations in Eq.~\eq{I_cVj} skip negligible small 
terms of the order of $\Delta\vartheta_{0,N}^3$.
\begin{figure}
\centerline{
\includegraphics[width=0.49\textwidth]{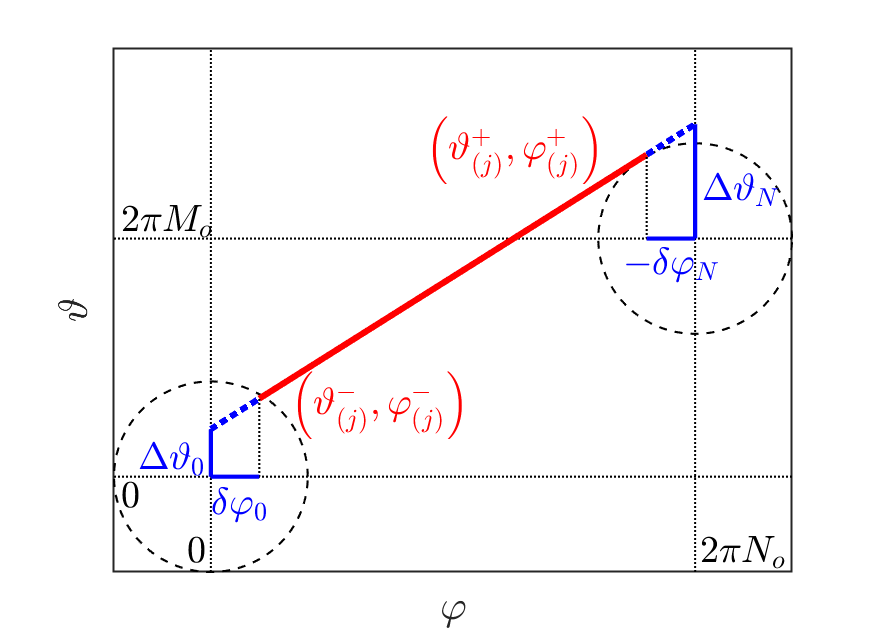}
}
\caption[]{
Field line segment containing the off-set domain (blue dotted) covered by transient banana orbit (red)
in case the lower local field maximum is on the right, $\eta_\text{loc}^{(j)}=\eta_{j+1}$. Dashed black
ellipses correspond to the contour $\eta_\text{loc}^{(j)} B(\vartheta,\varphi)=1$.
}
\label{fig:schematic_offset_region}
\end{figure}
To check that we notice that derivative $\partial B/\partial \vartheta_0$ in Eq.~\eq{cVdef} vanishes at the global maximum
and, therefore, scales in 
the additional integration intervals $0<\varphi<\varphi_{(j)}^-$ and $\varphi_{(j)}^+ < \varphi < 2\pi N_o$
as $\partial B/\partial \vartheta_0 \propto \Delta\vartheta_{0,N}$. The same scaling
is true there for the square root term $(1-\eta_b B)^{1/2} \propto \Delta\vartheta_{0,N}$.
Thus, due to the scaling $\cV \propto \Delta\vartheta_{0,N}^2$ in those intervals whose size is
$|\delta\varphi_{0,N}|$, their overall contribution has a cubic
scaling with $\Delta\vartheta_{0,N}$. 
The difference $\delta \cV=\cV-\cV_b$ can be estimated 
as $\delta\cV \propto \eta_{\text{loc}}^{(j)}-\eta_b \propto \Delta\vartheta_{0,N}^2$
in the whole integration interval including the vicinity of
its ends where $\delta\cV \sim \cV_b \propto \Delta\vartheta_{0,N}^2$. However,
in the case of stellarator symmetry, $\delta\cV$ is
essentially an odd function of $\varphi-\pi N_o$, which does not contribute to the integral,
while its even part $\delta\cV^\text{even} \propto \Delta\vartheta_{0,N}^3$ respectively provides
a negligible contribution $\propto \Delta\vartheta_{0,N}^3$.
Since in the case of stellarator symmetry
quantities $\cV_b$ and $\cV_b^{\prime\prime}$ are odd functions
of $\varphi-\pi N_o$ while $\cV_b^\prime$ is an even function, we get
\bea{I_cVj_stellsym}
I_{\cV}^{(j)}
&\approx&
\Delta\vartheta_{\text{mid}}\int\limits_{0}^{2\pi N_o}\rd \varphi
\left(
\cV_b^\prime(\iota_o \varphi,\varphi)
+
\delta\iota \cV_b^{\prime\prime}(\iota_o \varphi,\varphi) (\varphi-\pi N_o)
\right)
\nonumber \\
&=&
2\pi N_o\Delta\vartheta_{\text{mid}}\sum\limits_{m=1}^\infty
\sum\limits_{n=-\infty}^\infty
m\cV_{mn}^s(\eta_b)
\left(\delta_{\text{res}}+\frac{m N_o \delta\iota}{m M_o + n N_o}(1-\delta_{\text{res}})
\right),
\eea
where $\delta_{\text{res}}=1$ if $mM_o-nN_o=0$ and is zero otherwise,
and $\Delta\vartheta_{\text{mid}}=(\Delta\vartheta_0+\Delta\vartheta_N)/2$.
Since only the resonant harmonics with $(m,n) = \left(N_o k,-M_o k\right)$ and $k=1,2,\dots$ contribute
to the first term, one can estimate $\sum\sum m \cV_{mn}^s \delta_\text{res} \sim N_o \cV_{N_o,-M_o}^s$.
For high harmonic indices, $m,n \rightarrow \infty$, spectrum of $\cV$ is determined by the vicinity of the
global maximum,
\be{cV_spectrum}
\cV_{mn}^s(\eta_b) \approx \frac{9 \beta^{1/2} \eta_b^{7/2}}{\sqrt{8} \pi}
\left|\difp{^2 B}{\vartheta^2}\right|^{3/2} \frac{m N_\text{tor}}{(m^2 + \beta n^2)^{5/2}},
\ee
where $N_\text{tor}$ is the number of toroidal field periods and $\cV_{mn}^s=0$ for $n$ which are not
multiples of  $N_\text{tor}$,
see Appendix~\ref{sec:appendix1}.
Therefore, contribution of the first term
$N_o \cV_{N_o,-M_o}^s \sim N_o^{-3}$ is negligible small compared to that of the second term which
does not scale with $N_o$ but contains small $\delta\iota$.
Using $2\pi N_o \delta\iota = \Delta\vartheta_N-\Delta\vartheta_0$ and the definition of $\Delta\vartheta_{\text{mid}}$,
 we obtain the scaling $I_{\cV}^{(j)} \propto
(\Delta\vartheta_N^2-\Delta\vartheta_0^2) \propto N_o^{-2}  \propto N^{-2}$ and respective estimates
\be{wj_est}
w_{\text{off}}^{(j)}
\sim \frac{\varepsilon_t^{3/2}}{N^3}\frac{\rd r}{\rd \psi}
\sim \left(\varepsilon_t\nuast\right)^{3/5}\frac{\rd r}{\rd \psi},
\qquad
\lambda_{\text{off}}^\text{max} \sim \varepsilon_t^{1/2} R B_0 \frac{\rd r}{\rd \psi}
\sim \iota \lambda_{bB}^{\rm tok},
\ee
where we assumed the maximum distribution function off-set~\eq{goff_est},
used Eqs.~\eq{N_deta} and assumed $\varepsilon_M \sim \varepsilon_t$ and $\beta \sim 1$.
Thus, off-set of bootstrap coefficient $\lambda_{\text{off}}$ in stellarator-symmetric fields
does not decrease
with reducing collisionality but has an aperiodic oscillatory behavior
with oscillation amplitude $\lambda_{\text{off}}^\text{max}$ of the order
of $\lambda_{bB}$ for the equivalent tokamak~\citep{boozer90-2408,beidler2011-076001},
\be{lamequitok}
\lambda_{bB}^{\rm tok} = 1.46 \iota^{-1}\varepsilon_t^{1/2}RB_0 \frac{\rd r}{\rd \psi}
= 1.46 \iota^{-1}\varepsilon_t^{-1/2},
\ee
see Fig.~\ref{fig:lambb_longmfl} where the factor $1.46 \iota^{-1} \approx 3.4$ agrees well
with the relative off-set amplitude.

The off-set pattern corresponding to $N$ toroidal turns is replaced by the pattern with a larger number of
turns if collisionality decreases by a significant factor, and further collisionality decrease triggers another
pattern change, etc. Respectively, the off-set of the bootstrap coefficient oscillates aperiodically with changing
$\log \nu_\ast$ as seen in Fig.~\ref{fig:lambb_longmfl}. This behaviour is due to the structure of the off-set
pattern which is formed at the field line segment between two local field maxima $j=0,N$ fulfilling
strong alignment conditions $(\Delta\eta_0,\Delta\eta_N) \ll \delta\eta$ with $\delta\eta$ given by~\eq{blwidth}
and which necessarily contains one or more internal
local maxima fulfilling a weak alignment condition $\Delta\eta \lesssim \delta\eta$ (they split the
segment into off-set domains, e.g., two off-set domains and the main region in the simplest case).
With decreasing collisionality, weakly aligned internal maxima become non-aligned,
$\Delta\eta \gg \delta\eta$, what makes the variation of the distribution function off-set $g_\text{off}$
within the segment and respective contribution  of the given pattern to the off-set of $\lambda_{bB}$ exponentially small.
Consequently, the whole segment of the old pattern turns into a single off-set domain of the new, longer
pattern limited at the ends by even better aligned local maxima such that they stay well
aligned when one (or both) maxima at the ends of the former segment become only weakly aligned at some lower collisionality.
The above transition is well seen in Figs.~\ref{fig:along_mfl_3m4} and~\ref{fig:along_mfl_1m6} where
three complete off-set patterns at higher collisionality (left Fig.~\ref{fig:along_mfl_3m4}) turn into three off-set domains
of a longer pattern at lower collisionality (left Fig.~\ref{fig:along_mfl_1m6}). Note that those domains contribute
simultaneously to both off-sets, $g_\text{off}^\text{short}$ and $g_\text{off}^\text{long}$ with the first one
being dominant at higher and the second one at lower collisionality. A kink on the collisionality dependence of a summary function
$g_\text{off}^{(sp)}= \max(g_\text{off}^\text{short},g_\text{off}^\text{long})$ shown in Fig.~\ref{fig:offset_trend_longmfl}
is at the transition point $\nu_\ast = 10^{-4}$ where $g_\text{off}^\text{short}=g_\text{off}^\text{long}$. At this point,
internal maxima of the long off-set domain are still well aligned, so that $g_\text{off}^\text{long}$ scaling with
$\nu_\ast$ is similar to Eq.~\eq{dimoff_B} which includes $1/\nu$ and $1/\sqrt{\nu}$ terms.
Note that $g_\text{off}^{(sp)}$ in Fig.~\ref{fig:offset_trend_longmfl} by definition shows only
the largest off-set within the pattern
which generally contains more than two off-set domains
(see Fig.~\ref{fig:along_mfl_1m6}).
Therefore, location of $g_\text{off}^{(sp)}$ maximum at $\nu_\ast$
dependence needs not to coincide with the location of respective $\lambda_{bB}$ maximum, which is at lower $\nu_\ast$
in Fig.~\ref{fig:lambb_longmfl}. The reason is that contributions of most off-set domains within the pattern
tend to saturate and then vanish with decreasing $\nu_\ast$ earlier than $g_\text{off}^{(sp)}$ which is determined
then by the off-set domain bounded by the best aligned local maxima. One should also note that another kink
on $g_\text{off}^{(sp)}$ located in Fig.~\ref{fig:offset_trend_longmfl} at $\nu_\ast = 1.7\cdot 10^{-10}$ is
not the transition point of different off-set patterns but a point where the off-set for the same pattern changes sign
due to the mirroring within trapped-passing boundary layer discussed, in particular, in Section~\ref{ssec:efolding}.
As a result, the respective off-set of $\lambda_{bB}$ in Fig.~\ref{fig:lambb_longmfl} changes sign too
(similar to Fig.~\ref{fig:lambda_bB_test_newtest}).

According to Eq.~\eq{Delta_eta_Delta_theta},
better alignment of maxima at the off-set segment ends means smaller
poloidal displacements $\Delta \vartheta_0$ and $\Delta \vartheta_N$ of these ends, which requires
a decrease with the pattern length $N$ of the difference
$|\Delta \vartheta_N-\Delta \vartheta_0|=2\pi |\iota N_p -M|$ where
$N_p=N N_\text{tor}$ is the (integer) number of toroidal field periods within the
off-set pattern and integer $M$ is the nearest to $\iota N_p$. Thus, the increasing sequence of $N_p$
corresponds to the sequence of the best Diophantine approximations $M/N_p$ of (generally irrational) $\iota$.
These approximations are the convergents of $\iota$ representation by a simple continued fraction (see, e.g., \cite{hardy85}).
Therefore, an overall trend of the $N_p$ series $N_p^{(k)}$, $k=1,2,\dots$,
is similar to that of geometrical progression, $N_{p}^{(k+1)}=b_k N_p^{(k)}$
with $b_k > 1$ such that $b_k-1 \gtrsim 1$. Thus, due to $\nu_\ast \propto N_p^{-5}$ following from Eq.~\eq{N_deta}, a series
of collisionalities $\nu_\ast^{(k)}$ corresponding to the dominance of pattern $k$ in the off-set has similar geometrical
trend, and a significant change in the off-set of $\lambda_{bB}$ requires such a change of $\log\nu_\ast$.

Finally, in view of the off-set~\eq{wj_est} being of the same order as the collisionless asymptotic~\eq{lambda_bB}, 
let us estimate the role of the correction term driven by the source~\eq{Delta_s_bou} which has 
been ignored in the present analysis and estimated to have a vanishing effect with decreasing collisionality 
in a short field line example in Section~\ref{ssec:nonalign}. Comparison of the asymptotic and numerical
solutions in Fig.~\ref{fig:along_mfl_3m4} shows an obvious feature which cannot be described by the leading order
off-set. Namely, instead of the continuous linear growth of $g_o^t$, numerical solution tends to a nearly-periodic
behaviour with the period of the order of the off-set pattern length. In other words, finite collisionality provides
an aperiodic correction term which nearly balances the aperiodic behaviour of $g_0^t$ (while the leading order off-set
provides only a nearly periodic correction contained in the off-set domains within the pattern).
One can formally present this next order correction as a ``staircase function'', 
$\Delta g_\text{off}^\text{(tr)}=\sum_k C_k^\text{(tr)} \Theta\left(\varphi-\varphi_k^\text{(op)}\right)$,
where $\varphi_k^\text{(op)}$ are the boundaries of the off-set pattern (points of well aligned relevant maxima) and 
$C_k^\text{(tr)}=g_0^t\left(\varphi_{k-1}^\text{(op)}\right)-g_0^t\left(\varphi_{k}^\text{(op)}\right)$.
Being of the same order as $g_0^t$ which, due to its dependence on $\varphi$, makes an independent of collisionality
contribution to $\lambda_{bB}$ within each off-set pattern,
the staircase function is constant within the pattern and, according to the first estimate~\eq{wj_est}, its contribution
scales with collisionality. Namely, replacing the distribution function off-set $g_\text{off}^{(j)}$ with
$\Delta g_\text{off}^\text{(tr)} \sim N N_\text{pt}$ where $N_\text{pt}$ is the number of patterns per closed field line
(and $N$, as before, is the number of toroidal turns within the pattern),
respective contribution to the bootstrap coefficient scales as 
$\Delta \lambda_{bB} \approx w_\text{off}^{(j)} \Delta g_\text{off}^\text{(tr)} \sim N_\text{pt} N^{-2} 
\sim N_\text{pt} \nu_\ast^{2/5}$.

Another feature demonstrated for the short field line in Fig.~\ref{fig:spitf_surf_mod} is a split of the boundary layer
caused by a strong alignment of relevant maxima. This split can also be formally described by a staircase function
$\Delta g_\text{off}^\text{(bl)}
=\sum_k C_k^\text{(bl)} \Theta\left(\varphi-\varphi_k^\text{(op)}\right)\delta_\text{(b)}(\eta_b-\eta)$ where 
$\delta_\text{(b)}$ tends with vanishing collisionality to a $\delta$-function and 
is localized within the boundary layer width $\delta\eta$ where it scales as $\delta\eta^{-1}$ at finite collisionality,
whereas constants $C_k^\text{(bl)}$ are given by the integral in parentheses in Eq.~\eq{across_bou_g0_tp_int} with
the change of lower and upper limits to $\varphi_k^\text{(op)}$ and $\varphi_{k-1}^\text{(op)}$, respectively.
Since $\Delta g_\text{off}^\text{(bl)}$ is also a constant of $\varphi$
within the pattern, its contribution to $\lambda_{bB}$ can be estimated as 
$\Delta \lambda_{bB} \sim \Delta w_\text{off}^{(j)}\Delta g_\text{off}^\text{(bl)}$
where $\Delta g_\text{off}^\text{(bl)} \sim N N_\text{pt} \delta\eta^{-1}$, and the
estimate $\Delta w_\text{off}^{(j)} \sim N^{-1} \Delta I_\cV^{(j)}$ follows from~\eq{w_off_booz}
where $\Delta I_\cV^{(j)} \sim \delta\vartheta^3 \sim N^{-3}$ is the contribution of the boundary layer
to the integral $I_\cV^{(j)}$ (this estimate is similar to the estimate 
of the error due to the finite $\eta_\text{loc}^{(j)}-\eta_b$ below Eq.~\eq{I_cVj}).
Finally, we obtain $\Delta \lambda_{bB} \sim N_\text{pt}N^{-3} \delta\eta^{-1}\sim N_\text{pt}N^{-1} \sim N_\text{pt} \nu_\ast^{1/5}$
where we used~\eq{N_deta} treated again as an approximate equality instead of a strong inequality.
Thus, despite a visible effect on the generalized Spitzer function $g_{(3)}$, the ignored correction provides to
the bootstrap coefficient a contribution which is converging to zero with decreasing collisionality.

It should be noted that scaling $\Delta \lambda_{bB} \propto N_{pt}$ with the number
of off-set patterns within the closed field line is the result of the ``worst case'' estimate allowing 
for ``bootstrap resonances'' discussed below in Section~\ref{ssec:bootres}. Such a crude ``worst case'' 
estimate of the collisionless asymptotic $\lambda_{bB}^\dagger$ in the form~\eq{lambb-ware} results in a similar scaling
with the field line length which, in turn, does not contain the collisionality dependence.

\subsection{Off-set of bootstrap / Ware pinch coefficient in case of (nearly) aligned maxima}
\label{ssec:offset_imperfect}

In the advanced stellarator configurations aiming at good confinement of fusion alpha particles, global
magnetic field maxima tend to be aligned. This is the case in ideal
quasi-symmetric configurations~\citep{nuhrenberg88-113}
and also in quasi-isodynamic configurations
with poloidally closed contours $B=\rm{const}$~\citep{mikhailov02-L23,subbotin06-921}.
If these configurations are not realized exactly,
poloidal (helical in the general case of quasi-symmetry) closure of $B=\rm{const.}$ contours is destroyed, in particular,
near the field maximum where these contours split into islands centered around the global maxima
reached at one or few points instead of lines.
Local maxima $\bh\cdot\nabla B=0$, nevertheless, stay on poloidally closed contours
defined in periodic Boozer coordinates by
$\varphi=\varphi_k^\text{loc}(\vartheta)$ where $\varphi_k^\text{loc}(\vartheta)
=\varphi_0^\text{loc}(\vartheta)+2\pi k/N_\text{tor}$
with $k$ being an integer, and $\varphi_0^\text{loc}(\vartheta)$ is periodic. For simplicity, we
restrict our analysis here to the case of poloidally closed contours where also the family of anti-sigma
configurations~\eq{QI-antisigma} belongs as a sub-class with $\varphi_0^\text{loc}(\vartheta)=0$.
We assume that these local maxima are close to global such that perturbation
$\Delta B_\text{loc}(\vartheta) \equiv B(\vartheta,\varphi_0^\text{loc}(\vartheta))-\bar B_\text{max}$ of the
global maximum $\bar B_\text{max}$ achieved in the ideal (unperturbed) configuration on a line is small.
Since $(\vartheta,\varphi)=(0,0)$ is a stellarator symmetry point associated with
field maximum, function $\varphi_0^\text{loc}(\vartheta)$ is odd and $\Delta B(\vartheta)$ is even.

It is convenient now to label the field lines by the poloidal angle in the middle of the first field
period where $\varphi=\varphi_s\equiv\pi/N_\text{per}$ such that $(\vartheta,\varphi)=(0,\varphi_s)$
is a stellarator symmetry point associated with field minimum of the ideal configuration.
Namely, we label them with
$\vartheta_0 = \vartheta-\iota(\varphi-\varphi_s)=\vartheta-\iota\varphi+\iota_p$
where $\iota_p=\iota\varphi_s$.
Positions of local maxima on the so labelled field line, $\varphi_k=\varphi_k(\vartheta_0)$, being the solutions to
$\varphi_k = \varphi_k^\text{loc}\left(\vartheta_0+\iota (\varphi_k-\varphi_s)\right)$
for $-\infty < k < \infty$ can all be expressed via the left maximum of the first period $\varphi_0$
on the re-labelled field line as
$\varphi_k(\vartheta_0) = \varphi_0\left(\vartheta_0 + 2 k \iota_p\right)+2 k\varphi_s$.
Respective perturbations $\Delta B_{(k)}(\vartheta_0)$ of global maximum at these points
can be expressed via such perturbation $\Delta B_{(0)}$ of the left maximum in the first period
as $\Delta B_{(k)}(\vartheta_0) = \Delta B_{(0)}(\vartheta_0+2k\iota_p)$. Finally, integrals~\eq{bas1_defs}
where $\varphi_j^-=\varphi_j$ and $\varphi_j^+=\varphi_{j+1}$
can be mapped to the first period similarly, $I_k(\vartheta_0,\eta) = I_0(\vartheta_0+2k\iota_p,\eta)$.

Due to stellarator symmetry, function $\varphi_0(\vartheta_0)$ is odd with respect to
$\vartheta_0-\iota_p$, i.e. $\varphi_0(2\iota_p-\vartheta_0)=-\varphi_0(\vartheta_0)$,
while function $\Delta B_{(0)}$ is even. It convenient to express the latter via an even function
of the argument $\Delta B(\vartheta^\prime)$ as $\Delta B_{(0)}(\vartheta_0)=\Delta B(\vartheta_0-\iota_p)$,
such that for anti-sigma configurations  we have an identity
$\Delta B(\vartheta)=\Delta B_\text{loc}(\vartheta) = B(\vartheta,0)-\bar B_\text{max}$.
We fix $\bar B_\text{max}$ by the condition of zero average, $\overline{\Delta B}=0$ such that
an even function $\Delta\eta(\vartheta)=-\Delta B(\vartheta)/\bar B_{\rm max}^2$ has zero average too.
Using definition~\eq{xjpm} of the aspect ratio, its perturbation is given by
$\Delta A_o(\vartheta_0) = \left(I_\text{ref}/I_0(\vartheta_0,\eta_b)\right)^{1/2}-1$
being an even function with $I_\text{ref}$ determined by the condition of zero average $\overline{\Delta A_o}$.

Mapping to the first period of the distribution function off-set given by Eq.~\eq{dimoff_B}
for small $\Delta A_o \ll 1$ and $\Delta \eta \ll \delta\eta_\text{ref}$
is similar to that of the other quantities,
$g_\text{off}^{(0)}(\vartheta_0)=g_\text{off}^{(k)}(\vartheta_0-2k\iota_p)$.
Since Eq.~\eq{dimoff_B} gives the off-set in open-ended system with sources located in a single period,
solution for the closed (periodic) system with sources in each period is
an odd function given by the sum over all periods of the open ended system,
$$
g_\text{off}(\vartheta) = \sum\limits_{k=-\infty}^\infty g_\text{off}^{(k)}(\vartheta_0-2k\iota_p),
$$
which is expressed via even functions $\Delta A_o$ and $\Delta \eta$ as
\bea{dimoff_B_antisigma}
g_{\text{off}}(\vartheta_0)
&=&
\frac{l_c}{4}\left(\frac{\langle B^2\rangle}{\langle|\lambda|\rangle}\right)_{\eta=\eta_b}
\sum\limits_{k=1}^\infty
\left(
\sqrt{2} \delta\eta_{\rm ref}
\Delta^A_k \left(
\Delta A_o\left(\vartheta_0-2\iota_p k \right)
-\Delta A_o\left(\vartheta_0+2\iota_p k \right)
\right)
\right.
\nonumber \\
&+&
\left.
\Delta^B_k
\left(
\Delta\eta\left(\vartheta_0-\iota_p(2k-1) \right)
-
\Delta\eta\left(\vartheta_0+\iota_p(2k-1) \right)
\right)
\right).
\eea
Respectively, the off-set of the Ware pinch coefficient~\eq{lambda_off} is expressed in the limit of infinite
(irrational) field-line length via $g_{\text{off}}$ as follows,
\be{lambda_off_antisigma}
\lambda_{\text{off}}
\int\limits_{0}^{2\pi}\rd \vartheta_0
\int\limits_{0}^{2\pi/N_\text{tor}}\frac{\rd \varphi}{B^2}
\approx \frac{\rd r}{\rd \psi}
\int\limits_{0}^{2\pi}\rd \vartheta_0\; g_{\text{off}}(\vartheta_0) I_\cV(\vartheta_0)
=
\pi \frac{\rd r}{\rd \psi} \sum\limits_{m=1}^\infty g_m I_\cV^m,
\ee
where an odd function $I_\cV(\vartheta_0)$ is given by the second Eq.~\eq{w_off_booz} with
$\varphi_{(j)}^-=\varphi_0(\vartheta_0)$ and $\varphi_{(j)}^+=\varphi_1(\vartheta_0)$,
and we used (an odd) Fourier series expansion
\be{fourserexp_odd}
(g_{\text{off}}(\vartheta_0),I_\cV(\vartheta_0)) = \sum\limits_{m=1}^\infty (g_m, I_\cV^m)\sin(m\vartheta_0).
\ee
Coefficients $g_m$ are expressed via coefficients of (an even) Fourier series expansion
\be{fourserexp_even}
(\Delta A_o(\vartheta), \Delta B(\vartheta))
= \sum\limits_{m=1}^\infty (\Delta A_o^m,  \Delta B_m)\cos(m\vartheta)
\ee
as follows,
\be{dimoff_B_antisigma_fourier}
g_m
=
\frac{l_c}{2}\left(\frac{\langle B^2\rangle}{\langle|\lambda|\rangle}\right)_{\eta=\eta_b}
\left(
\sqrt{2} \delta\eta_{\rm ref} S_A(m \iota_p) \Delta A_o^m
-
\eta_b^2 S_B(m \iota_p) \Delta B_m
\right),
\ee
where
\be{SA_SB}
S_A(\phi) = \sum\limits_{k=1}^\infty \Delta^A_k \sin\left(2k\phi\right),
\qquad
S_B(\phi) = \sum\limits_{k=1}^\infty \Delta^B_k \sin\left((2k-1)\phi\right).
\ee
These functions have period $2\pi$ and are well approximated in the interval $-\pi<\phi<\pi$
by $S_A(\phi) \approx 0.26\, (\phi-(\pi/2)\text{sign}(\phi))$
and $S_B(\phi) \approx 1.85\,\text{sign}(\phi)$.

For the estimates, we apply general expressions~\eq{dimoff_B_antisigma}--\eq{dimoff_B_antisigma_fourier}
to the anti-sigma configuration~\eq{QI-antisigma} which corresponds to the perturbed case 3
with $\varepsilon_0=\varepsilon_M \ll 1$,
$\varepsilon_1=\varepsilon_t \ll \varepsilon_M$ and $\varepsilon_3=0$,
\be{pertcase3}
B(\vartheta,\varphi) = B_0 \left(
1+\varepsilon_M \cos(N_\text{tor}\varphi)
+\varepsilon_t \cos\vartheta \left(1- \cos(N_\text{tor}\varphi)\right)
\right)+\Delta B_1 \cos\vartheta .
\ee
In the leading order over $\varepsilon_t/\varepsilon_M$ and $\Delta B_1/B_0$,
only $A_o^1 \approx \varepsilon_t/(2 \varepsilon_M)$ harmonic
and $\Delta B_1$ harmonic contribute to $\lambda_{\text{off}}$ because of the dominance
in the $I_\cV$ spectrum of harmonic
$I_\cV^1 \approx - 8\varepsilon_t (2\varepsilon_M)^{1/2} (B_0^2 N_\text{tor})^{-1}$.
Estimating $l_c=\pi R/(2\nuast)$,
$\langle B^2 \rangle \approx B_0^2$, $\eta_b \approx 1/B_0$,
$\langle |\lambda| \rangle_{\eta=\eta_b} \approx \pi^{-1}(8 \varepsilon_M)^{1/2}$ and
$\delta\eta_{\rm ref} \sim 4 \pi^{-1/2} (2\varepsilon_M)^{1/4} (\nuast/N_\text{tor})^{1/2}B_0^{-1}$
we get
\bea{lambda_off_small}
\lambda_\text{off} &\approx&
-\varepsilon_t R B_0
\frac{\rd r}{\rd \psi}
\left(
\left(\frac{2\pi}{\nuast N_\text{tor}}\right)^{1/2}
\frac{\varepsilon_t S_A(\iota_p) }{\left(2\varepsilon_M\right)^{3/4}}
-
\frac{\pi \Delta B_1 S_B(\iota_p)}{4\nuast B_0}
\right)
\nonumber \\
&\approx& \lambda_{bB}^\text{tok}
\left(
\frac{0.42\;|\iota|\; \varepsilon_t^{3/2}}{\left(\nuast N_\text{tor}\right)^{1/2}\varepsilon_M^{3/4}}
\left(1-\frac{2|\iota|}{ N_\text{tor}}\right)
+
\frac{|\iota|\;\varepsilon_t^{1/2} \Delta B_1}{\nuast B_0}
\right).
\eea
The black dotted line showing the trend in Fig.~\ref{fig:lambb_antisigma} actually corresponds to the
result of this formula for case~3.
\begin{figure}
\centerline{
\includegraphics[width=0.49\textwidth]{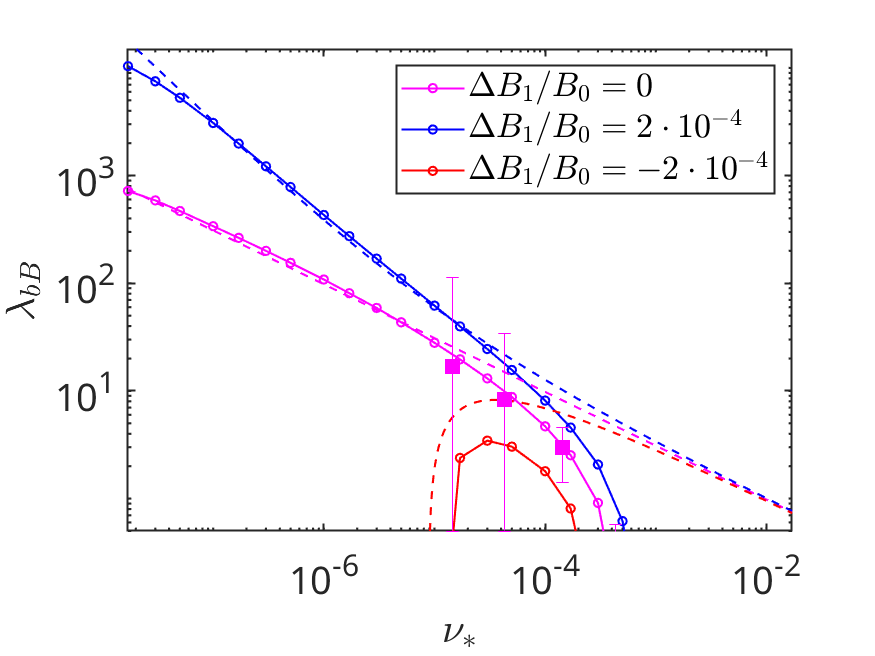}
\includegraphics[width=0.49\textwidth]{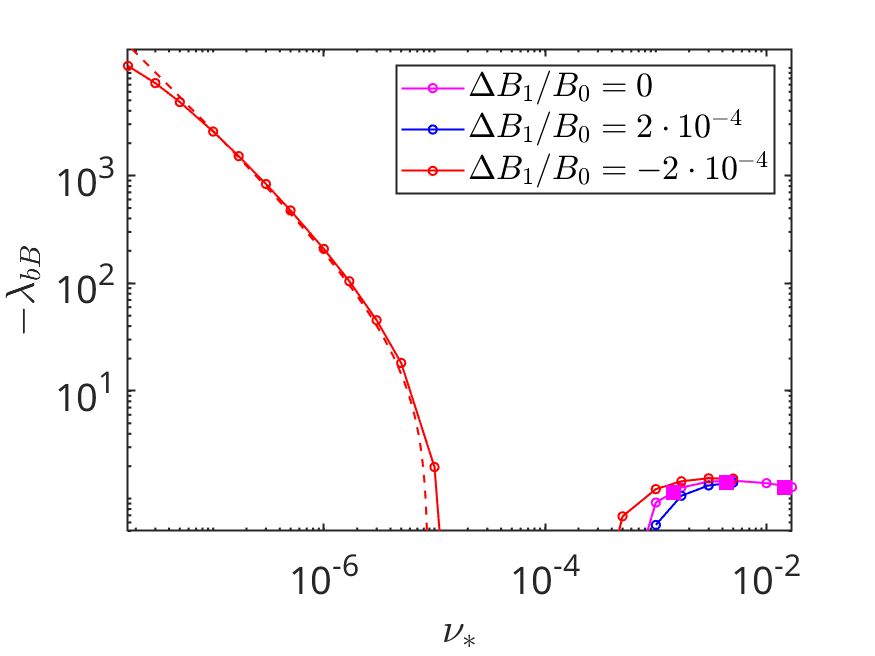}
}
\caption[]{Geometrical factor $\lambda_{bB}$ computed by NEO-2 (circles) and DKES (filled squares)
vs $\nuast$. Three different maximum-$B$ perturbations are marked with colors (see the legend).
Dashed lines correspond to the
asymptotic~\eq{lambda_off_small}. Negative $\lambda_{bB}$ are shown in the right plot.
}
\label{fig:lambda_bB_deltaB}
\end{figure}
It should be reminded here that the off-set~\eq{lambda_off_small}
due to violation of ripple equivalence
appears only for relatively strong alignment of maxima,
\mbox{$\Delta B /B_0\lesssim (\nuast / N_{tor})^{1/2} \varepsilon_M^{1/4}$},
and the validity of estimate~\eq{lambda_off_small} requires the respective strong inequality.
In the marginal case where an approximate equality is realized, the largest off-set,
$\lambda_\text{off} \sim \lambda_{bB}^\text{tok}\; \varepsilon_t^{1/2} \varepsilon_M^{1/4}
\left(\nuast N_\text{tor}\right)^{-1/2}$ occurs.
The result of a small mis-alignment of maxima due to a finite perturbation $\Delta B_1$ in Eq.~\eq{pertcase3}
is shown in Fig.~\ref{fig:lambda_bB_deltaB}.
Note that the validity condition of a linear asymptotic~\eq{lambda_off_small} becomes violated for
finite $\Delta B_1$ at
$\nuast = 1.7\cdot 10^{-8}$ where $\Delta\eta = \delta\eta_\text{ref}$, and the case of nearly
aligned maxima starts a transition to the usual case of a global maximum reached at a point.

\subsection{Bootstrap resonances}
\label{ssec:bootres}

As one can conclude from Fig.~\ref{fig:lambb_longmfl}, resonance structure of $\iota$ (equivalently, of radial)
dependence of bootstrap coefficient is absent in finite collisionality cases (see also Fig.~16 of \citep{landreman22-082501}).
This, however, does not mean that
these resonances never play a role. The reason for them is the following (see also the Appendix of \citep{boozer90-2408}). By our assumption, magnetic configuration has embedded flux surfaces everywhere
so that magnetic islands are absent at any rational magnetic surface, what is manifested by a true surface
condition~\eq{closure}. Although these configurations constitute a set of measure zero, they are physically
possible. It is straightforward to check that the first term (representing the Pfirsch-Schl\"uter current)
in Eq.~\eq{lambda_bB} is never resonant as long as conditions~\eq{closure} are fulfilled. However, this is
not the case for the second term. Up to a flux surface function, the sub-integrand of the field line integral there
is the derivative over $\eta$ of the source term $s_{(1)}$, Eq.~\eq{mono_Ak_norm}, which fulfills the quasi-Liouville's
theorem~\eq{liouv_reprp} at our representative field line, but not necessarily on the other field lines from
the same rational surface. In other words, the amplitudes of resonant harmonics fulfilling $\iota m + n = 0$
for our rational surface are generally finite, but these are harmonics of
$\sin(m\vartheta+n\varphi)=\sin(m\vartheta_0 + n\varphi_0 +(\iota m + n)(\varphi-\varphi_0))$ which are identical
zeros for the starting point $(\vartheta_0,\varphi_0)$ being a stellarator symmetry point. Would those amplitudes
be all zero, the integral over any field line would be a single-valued function fulfilling then a quasi-Liouville's
theorem~\eq{liouv_reprp} for passing particles and, as a consequence, conditions $\oint \rd l v_\parallel = \text{const}$
evaluated for passing particles along closed field lines
(see the discussion by \cite{shaing83-3315} after Eq.(44b)), which actually mean ``true surface'' conditions
for the effective magnetic field $\bB^\ast$ \citep{morozov66-201}, i.e. for drift surfaces.
The latter, however, need not have the same topology as the magnetic surfaces (see, e.g., \citep{heyn12-054010})
and, at best, can be made embedded for one particular $\eta$ value for the cost of destroyed magnetic surfaces and
drift surfaces with different $\eta$ (actually, true drift surface conditions reduce for $\eta=0$ to
true magnetic surface conditions).
Thus, drift islands are generally present at any rational drift surface (see, e.g., \citep{smirnova97-2584}),
and, due to the increased spectral width of $s_{(1)}$ near the trapped-passing boundary~\citep{boozer90-2408}
where the exponential decay of $s_{(1)}$ spectrum is replaced with a power law,
$(m B_\varphi  - n B_\vartheta)(m^2+\beta n^2)^{-3/2}$ estimated similarly to Eq.~\eq{cV_spectrum},
drift islands overlap near this boundary ($s_{(1)}=-\rd r/\rd\varphi$ along the orbits by definition~\eq{s_k}).
This
leads to barely trapped alpha particle losses observed in realistic stellarator
configurations~\citep{albert20-815860201,albert20-109065,albert23-955890301,chambliss24-arxiv}, 
which do not ideally fulfil the quasi-symmetry conditions.

Due to our choice of reference field lines, respective drift orbits correspond to invariant axes (X or O points of drift islands),
i.e. exact resonance does not destroy their closure and, therefore, does not affect collisionless $\lambda_{bB}$.
However, neighboring resonances result in near resonant contributions
to the collisionless $\lambda_{bB}\propto (\iota m - n)^{-1}$ where the denominator is rather close to zero,
$\min(\iota m - n)=m\Delta\iota \lesssim 1/n$. One should note that
exponential decay of the spectrum is restored from the power law decay
for passing particles displaced from the separatrix by $\Delta\eta=\eta-\eta_b$ and mode numbers fulfilling
$m^2+\beta n^2 > (\eta_b^2/\Delta\eta) \partial^2 B /\partial\vartheta^2$. Therefore, contribution of near-resonant modes
with high $m$ and $n$ to the integral over $\eta$ in Eq.~\eq{lambda_bB} scales with a maximum $\Delta\eta$ which still allows
the power law decay, i.e. it is
$\propto m^{-1}\Delta\iota^{-1} (m^2+\beta n^2)^{-2}(m B_\varphi  - n B_\vartheta)
\propto \Delta\iota^{-1} m^{-4}$.
Would the rational numbers be evenly distributed over the real axis as we assumed for simplicity in Section~\ref{ssec:asymptotic},
then $\Delta\iota \sim 1/m^2$, and the contribution of near-resonant modes is $\propto m^{-2}$, i.e. it is relatively small.
However, uniform distribution is not the case in reality (see, e.g., Fig.~5 in \citep{kasilov02-985}). Namely, if
$\lambda_{bB}$ asymptotic is evaluated at a high-order rational surface with $\iota m_{h}+n_h = 0$,
the distance to the low order resonance $\iota m_{l}+n_l = 0$
can be as small as $\Delta\iota \sim 1/m_h^2$ which results in contribution scaling as $m_h^2/m_l^4$, i.e.
it tends to infinity if the evaluation is at the irrational surface with $\iota$ arbitrarily close to $-n_l/m_l$.
As a result, $\lambda_{bB}$ asymptotic has a fractal dependence on $\iota$, as seen from the right plot in
Fig.~\ref{fig:lambb_longmfl}. On the contrary, if $\lambda_{bB}$ asymptotic is evaluated at low order rational field lines,
there are no other low order rationals in the neighborhood, and the resulting data points appear to belong to the smooth curve
(see the respective markers in the left plot of Fig.~\ref{fig:lambb_longmfl} which correspond to rational $\iota$ with numerators
not larger than 500 while the rest data points have numerators up to 10000).

Finite collisionality limits the effect of bootstrap resonances by modifying $\eta$ and thus moving passing particles into the
trapped particle region or deeper to the passing particle region, where the Fourier spectrum of $s_{(1)}$ decays exponentially.
Therefore, only a finite (``collisionless'') segment of its orbit corresponding to $N$ toroidal turns stays in resonance
with a given Fourier amplitude of $s_{(1)}$ limiting the resonant denominator to $\max(m\Delta\iota,(2\pi N)^{-1})$.
We can estimate $N$ equating the width of power law decay region
$\Delta \eta \sim \eta_b \varepsilon_t (1+\iota^2\beta)^{-1} m^{-2}$ (see above) to the boundary layer width
corresponding to $N$ turns, Eq.~\eq{blwidth}. Ignoring the numerical factor and $1+\iota^2\beta \sim 1$, scaling of
near-resonant contributions changes from $m^{-4}\Delta \iota^{-1}$ to
$m^{-4} (\max(\Delta\iota, m^3\nuast\varepsilon_t^{3/2}))^{-1}$. For the examples in Fig.~\ref{fig:lambb_longmfl}
where the lowest order rational number present in the $\iota$ range has $m=11$, near-resonant contributions
remain small by this estimate as long as $\nuast > 10^{-8}$, what results in a smooth $\lambda_{bB}$ dependence
on~$\iota$ in collisional cases.
Since bootstrap resonances are mostly an artifact of the collisionless asymptotic limit, they are avoided
in practical computations either by removal of respective near-resonant spectral modes,
see, e.g.~\citep{nakajima89-605} or by introducing a Gaussian
factor in the field line integration~\citep{boozer90-2408} which effectively mimics the collisional attenuation discussed above.

\section{Effect of particle precession}
\label{sec:precession}

As we saw from the previous analysis, off-set of trapped particle distribution function $g_{\text{off}}=g-g_0^t$ is determined
in the ``well trapped'' domain beyond the matching boundary $\eta_m$ introduced in Section~\ref{ssec:propmet}, $\eta > \eta_m$,
where $g_{\text{off}}$  is essentially a constant, by the collisional solution in region containing the boundary layer, $\eta < \eta_m$,
which is the consequence of the absence of a source term in the trapped particle region.
This remains the case in the presence
of particle precession caused by the magnetic drift and finite radial electric field if the typical relaxation rate
in the ``well trapped'' particle domain $\eta > \eta_m$ stays small compared to (essentially parallel) relaxation in the region
$\eta_b<\eta<\eta_m$ (case of mild rotations).
Since the latter region is rather narrow, $|\eta_b - \eta_m| \sim \delta\eta$, effect of the precession on the solution
in this trapped particle domain and in the whole passing particle region is relatively small~\citep{beidler20} as long
as Mach numbers for the $\bE\times\bB$ drift, $v_E^\ast = v_E/v = \Omega_E r / v$, are small enough,
$\Omega_E \tau_b /\delta\vartheta \sim v_E^\ast N^2 \varepsilon_t^{-3/2}
\sim v_E^\ast \varepsilon_t^{-9/10} \nuast^{-2/5} \ll 1 $,
where we estimated bounce time as the largest within the off-set pattern extending for $N$ toroidal turns~\eq{N_deta},
$\tau_b \sim N R /(v\sqrt{\varepsilon_t})$, and the width of the off-set region as $\delta\vartheta \sim 1/N$.
Since $v_E^\ast \sim \rho_L / r \ll 1$ and the respective Mach number due to magnetic drift $v_B^\ast$
is even smaller,
$v_B^\ast \sim \varepsilon_t v_E^\ast$,
this condition is fulfilled even at very low collisionalities typical for a reactor.
Therefore, we can obtain the solution in ``well trapped'' particle domain $\eta > \eta_m$ from the homogeneous problem
driven by Dirichlet boundary condition at $\eta=\eta_m$ where $g_{\text{off}}$ is determined by the solution of collisional
problem in the $1/\nu$ regime.

Since the trapped-passing boundary layer is excluded from the well trapped domain $\eta > \eta_m$, and class transition
boundary layers are rather narrow in the long mean free path regime, we can use for $g_{\text{off}}$ a homogeneous
bounce average equation. In the conservative form required for proper handling of boundary conditions at class transition
boundaries and omitting the class index $j$ for simplicity, this equation using the total energy $w=mv^2/2 + e\Phi$
and perpendicular adiabatic invariant $J_\perp = m v_\perp^2/(2\omega_c)=c m^2 v^2\eta/(2e) $ as momentum space variables
is
\be{bacons}
\difp{}{r}\left(J_b \langle v_g^r\rangle_b g_{\text{off}}\right)+
\difp{}{\vartheta_0}\left(J_b \langle v_g^{\vartheta_0}\rangle_b g_{\text{off}}\right)
=
\difp{}{J_\perp}\left(J_b \langle D^{J_\perp J_\perp}\rangle_b \difp{g_{\text{off}}}{J_\perp}\right),
\ee
where $g_{\text{off}}$ is independent of $\varphi$. The Jacobian $J_b$ and bounce-averaged pitch-angle diffusion
coefficient $\langle D^{J_\perp J_\perp}\rangle_b = \left(\partial J_\perp/\partial\eta\right)^2 \langle D^{\eta\eta}\rangle_b$ are
\be{Dee}
J_b=\frac{e}{c}\frac{\rd \psi}{\rd r}\tau_b,
\qquad
\langle D^{\eta\eta}\rangle_b=\frac{2 \eta I_j}{l_c \tau_{b}},
\ee
with $I_j$ defined in the first Eq.~\eq{bas1_defs}, and
bounce time $\tau_b$ together with bounce-averaged radial drift velocity
$\langle v_g^r\rangle_b$ are given by Eqs.~\eq{bavr} with the omission of
class index $k$.

It should be noted that due to the use of field aligned (Clebsch-like) spatial variables
both, $\langle v_g^r\rangle_b$ and $\langle v_g^{\vartheta_0}\rangle_b$ can be evaluated here along the field lines
(see \citep{shaing15-905810203,calvo17-055014,herbemont22-905880507}). On the contrary, for the non-aligned variables
(periodic flux coordinates) tangential components of the bounce-averaged drift should be evaluated over the real
bounce orbit with finite Larmor radius, which is in contrast to
$\langle v_g^r\rangle_b$ where this results only in a $\rho_L$ correction to Eq.~\eq{bavr}. Such an evaluation
is required to retain in the resulting precession frequency
the effect of magnetic shear which is of the order
one if $\Omega_E \lesssim \Omega_B$ (see \citep{shaing15-905810203,albert16-082515,martitsch16-074007}).
A way equivalent to bounce averaging velocity components
is to use Eqs.~(3.45) of \cite{morozov66-201} resulting in
$\langle v_g^r\rangle_b = J_b^{-1}\partial J_\parallel /\partial \vartheta_0$
and $\langle v_g^{\vartheta_0}\rangle_b = - J_b^{-1}\partial J_\parallel /\partial r$
where $J_\parallel = m\oint\rd l v_\parallel $ is the parallel adiabatic invariant computed along the field line,
which makes the Liouville's theorem $\partial (J_b \langle v_g^r\rangle_b)/\partial r
+ \partial (J_b \langle v_g^{\vartheta_0}\rangle_b)/\partial \vartheta_0 = 0$ obvious.

Eq.~\eq{bacons} means that off-set effect is generally non-local, i.e. radial transport due to entrapping of
passing particles at some given surface is produced in some radial range $\delta r_c$ (radial correlation length)
which in the $1/\nu$ regime is of the order $\delta r_c \sim \tau_d \langle v_g^r\rangle_b$, where
$\tau_d \sim \varepsilon_M/\nu$ is the de-trapping time, and, at lower
collisionalities, where $\langle v_g^{\vartheta_0}\rangle_b > 1/\tau_d$ and mild electric fields,
$\Omega_E \le \Omega_B$, is of the size of radial deviation of precessing trapped orbit which does not scale
with Larmor radius and is fully determined by the magnetic field geometry. Thus, local ansatz~\eq{superforces}
has a limited validity in the latter case where transport coefficients and bootstrap current need a non-local
treatment, e.g., by Monte-Carlo methods~\citep{sasinowski1995-610,satake08-S1062} (an exception is weakly perturbed
tokamak equilibria where a quasi-local ansatz~\citep{martitsch16-074007} can be used retaining the boundary layer effects).
Of course, $\delta r_c$ must
be small compared to the plasma radius for the magnetic fields intended for the reactor,
however, it is not necessarily small compared to the radial variation scale
of the off-set which is quite sensitive to small modulations of the field and $\iota$ changes. One consequence of this
non-locality is discussed in more details in Section~\ref{sec:boot_axis}.

Eq.~\eq{bacons} should be solved as is if the off-set is produced in the wide poloidal range containing the
trapped-passing boundary layer, which is the case of almost aligned maxima discussed in Sections~\ref{ssec:align},
\ref{ssec:imperfect} and~\ref{ssec:offset_imperfect}. However, in a more typical case with a distinct global maximum
(or two global maxima, which are also possible in stellarator symmetry), the width of the off-set region $\delta\vartheta$
is small (this is a typical width of a single strap in Figs.~\ref{fig:g_refl_3m4}-\ref{fig:fluxden_1m6}).
Therefore, convective term with $\langle v_g^{\vartheta_0}\rangle_b$ dominates over $\langle v_g^r\rangle_b$ which, therefore,
can be ignored in Eq.~\eq{bacons} together with the dependence on $\vartheta_0$ of
$\langle v_g^{\vartheta_0}\rangle_b \equiv \Omega^\vartheta$, $J_b$ and $D^{J_\perp J_\perp}$,
which should be taken at the global maximum point $\vartheta_0=\vartheta_{\max}$,
thus reducing Eq.~\eq{bacons} to
\be{bacons_red}
\difp{g_{\text{off}}}{\vartheta_0}
=
\frac{1}{\tau_b \Omega^\vartheta}
\difp{}{\eta}\left(\tau_b \langle D^{\eta\eta}\rangle_b \difp{g_{\text{off}}}{\eta}\right).
\ee
This can be done because $g_{\text{off}}$ is localized in $\vartheta_0$
due to essentially trivial boundary conditions over $\eta$
at $|\vartheta_0-\vartheta_{\max}| \gg \delta\vartheta$.
Therefore, solution
to Eq.~\eq{bacons_red} in the infinite domain $-\infty<\vartheta_0<\infty$ is sufficient to generalize
the results of $g_{\text{off}}$ computations in the $1/\nu$ regime for the case of finite precession frequencies (finite radial
electric fields).

For further estimates, we assume a fast precession case, $\Omega^\vartheta \tau_d \gg \delta\vartheta$
(with $\Omega^\vartheta \tau_b \ll \delta\vartheta$ staying fulfilled, what is possible at low collisionalities
where $\tau_b \ll \tau_d$), which would
be the condition of the $\sqrt{\nu}$ transport regime~\citep{galeev73}
in case $\delta\vartheta \sim 1$ but is established here for smaller
precession frequencies due to $\delta\vartheta \ll 1$. In this case, solution is localized in $\eta$ near the matching
boundary $\eta_m$, and we can ignore class transitions which occur in the well trapped particle domain and also
set $\eta=\eta_m$ in all the coefficients. Introducing the dimensionless variables as follows,
\be{rescale vars}
y=\frac{\vartheta-\vartheta_{\max}}{\delta \vartheta} {\rm sign}(\Omega^\vartheta),
\qquad
x = \left(\frac{|\Omega^\vartheta|}{\delta\vartheta\langle D^{\eta\eta}\rangle_b}\right)^{1/2}(\eta-\eta_m),
\ee
Eq.~\eq{bacons_red} is reduced to
\be{simeq}
\difp{g_{\text{off}}}{y}=\difp{^2 g_{\text{off}}}{x^2},
\ee
where $0 < x <\infty$, $-\infty < y <\infty$, and the boundary conditions are
$g_{\text{off}}(0,y)=g_{\text{off}}^{1/\nu}(\vartheta_{\max} + y\,\delta\vartheta\,{\rm sign}(\Omega^\vartheta),\eta_m)$
and $g_{\text{off}}(x,-\infty)=0$, so that solution is localized in the range $|y|\lesssim 1$.
Since the solution is also localized in the range $x \lesssim 1$, second Eq.~\eq{rescale vars} results
in the localization range $\eta-\eta_m  < \delta_E \Delta\eta_{tr}$ where $\Delta\eta_{tr} \sim \varepsilon_t^{1/2} \eta_b$
is the width of trapped particle domain and $\delta_E \ll 1$ is the attenuation factor due to the precession (we assume
here $\varepsilon_M \sim \varepsilon_t$ for simplicity).
This reduction of the $\eta$ integration range of $g_{\text{off}} s_{(1)}$ effectively contributing to the Ware pinch
coefficient~\eq{Dmono} respectively reduces its value by $\delta_E$ as compared to the $1/\nu$ regime.
Explicitly, this factor is
\be{atten_om}
\delta_E \sim \left(\frac{v\delta\vartheta}{l_c|\Omega^\vartheta|}\right)^{1/2}
\sim \left(\frac{v \nuast}{N R|\Omega^\vartheta|}\right)^{1/2},
\ee
where $N \sim 1/\delta \vartheta$ is the number of toroidal turns within a given off-set pattern.
The overall trend of the Ware pinch off-set with the reduction of $\nuast$ is obtained with the account of
switching off-set patterns and thus increasing $N$ in Eq.~\eq{atten_om} according to the first Eq.~\eq{N_deta}.
Thus, the normalized maximum off-set~\eq{wj_est} is reduced from its value in $1/\nu$ regime to
\be{desired_scaling_prec}
\frac{\lambda_{\text{off}}^\text{max}}{\lambda_{bB}^{\rm tok}}\sim \iota \delta_E
\sim \iota \varepsilon_t^{-3/20}
\nuast^{3/5}
\left(\frac{v}{R|\Omega^\vartheta|}\right)^{1/2},
\ee
i.e. $\lambda_{bB}$ converges with decreasing collisionality to the Shaing-Callen limit. It should be noted that bootstrap resonances
should still be removed from the asymptotic $\lambda_{bB}$ for the same reason as discussed in Section~\ref{ssec:bootres}.
In this case, in addition to pitch angle scattering limiting the effective orbit length contributing to the resonance
in the power low spectral decay region, also the precession de-correlates the resonant particles, effectively removing them
from that region.

In case of nearly aligned maxima, poloidal width of the off-set region in Eq.~\eq{atten_om}
is not small anymore, $\delta\vartheta \sim 1$ or, equivalently, $N\sim 1$. Thus, using the
$1/\nu$ regime estimate~\eq{lambda_off_small} we obtain the scaling for normalized off-set due to
violation of ripple equivalence as
\be{lambda_off_small_prec}
\frac{\lambda_\text{off}^\text{align}}{\lambda_{bB}^{\rm tok}}\sim
\frac{\iota\; \varepsilon_t^{3/2}}{N_\text{tor}^{1/2}\varepsilon_M^{3/4}}
\left(\frac{v}{R|\Omega^\vartheta|}\right)^{1/2},
\ee
which means that the off-set does not vanish but
saturates with reducing collisionality at a rather high level
due to the ratio of bounce frequency to the precession frequency
in the last factor.

To verify the off-set attenuation model in case of a single off-set pattern with fixed $N$ in Eq.~\eq{atten_om},
we apply this model to DKES results in Fig.~4 of \cite{helander11-092505} where the dominant pattern appears
to be unchanged in the whole range of $\nuast$.
\begin{figure}
\centerline{
\includegraphics[width=0.6\textwidth]{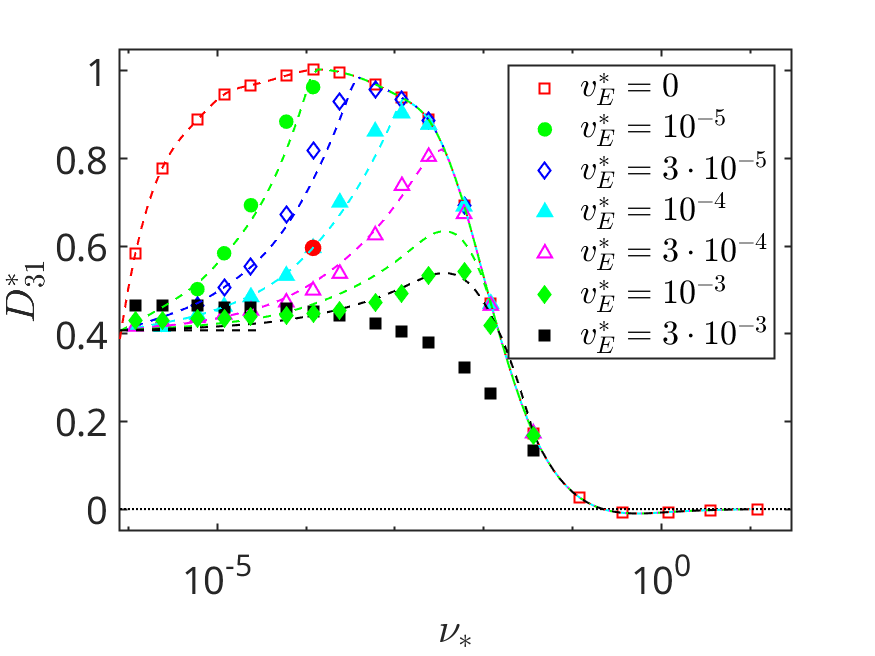}
}
\caption[]{
Attenuation test for the case in Fig.~4 of \cite{helander11-092505}.
Normalized electric field values (Mach numbers $v_E^\ast=cE_r/(vB)$) are defined in the legend.
Dashed lines show the result of scaling in Eq.~\eq{henning_plot} with
the reference point shown by a red circle.
}
\label{fig:henning}
\end{figure}
The normalized bootstrap coefficient
$D_{31}^\ast \equiv \lambda_{bB}/\lambda_{bB}^{\rm tok}$
in this figure,
where the precession is purely due to the $\bE\times\bB$ drift, $\Omega^\vartheta = \Omega_E \propto E_r$,
is represented using the scaling of attenuation factor~\eq{atten_om} with collisionality and radial electric field
$\delta_E \propto (\nuast/E_r)^{1/2}$ as follows,
\be{henning_plot}
D^\ast_{31} = D_{31}^{\rm SC}+\left(D_{31}^{1/\nu}- D_{\rm 31}^{\rm SC}\right)
\frac{D^\ast_{\text{ref}}-D_{\rm 31}^{\rm SC}}{D_{\text{ref}}^{1/\nu}-D_{\rm 31}^{\rm SC}}
\left(\frac{\nuast E_{\rm ref}}{\nu_{\rm ref} E_r}\right)^{1/2},
\ee
where $D_{31}^{\rm SC}$ is the Shaing-Callen asymptotic value of $D^\ast_{31}$,
$D_{31}^{1/\nu}$ is the actual value in the $1/\nu$ regime,
$\nu_{\rm ref}$ and $E_{\rm ref}$ are the values of $\nuast$ and $E_r$ for a single reference point
in the plot, and $D_{\text{ref}}$ and $D_{\text{ref}}^{1/\nu}$ are the values
of $D^\ast_{31}$ for $(\nuast,E_r)=(\nu_{\rm ref},E_{\rm ref})$ and $(\nuast,E_r)=(\nu_{\rm ref},0)$,
respectively.
As one can see in Fig.~\ref{fig:henning},
scaling~\eq{henning_plot} well represents the results with mild electric field
but fails for two largest values of $v_E^\ast$ where the (tokamak like) boundary layer off-set~\citep{helander11-092505}
becomes important (see also Fig.~\ref{fig:lambda_off}) and where the condition $\Omega_E \tau_b /\delta\vartheta \ll 1$
becomes violated.

\section{Off-set in the direct problem, bootstrap effect at the magnetic axis}
\label{sec:boot_axis}

We discuss now the off-set in the direct (bootstrap) problem qualitatively.
Since the off-set in the Ware pinch problem is driven by the asymmetry of the entrapping of passing particles,
off-set in the bootstrap problem is driven by the asymmetry in
the inverse process of trapped particle detrapping.
Similarly, there are two (same) reasons for this asymmetry. First is the
asymmetry of local maxima limiting ripple domains which leads to the detrapping in the direction of the lower maximum.
The second one is connected with the violation of ripple equivalence, $A_j = 1$, and, consequently, to the asymmetry
of entrapping and further detrapping in the neighboring ripples. We estimate now only the first, simpler effect.

For the equilibrium (Maxwellian) particle distribution which is constant on the flux surface,
detrapping process is balanced by entrapping of passing particles
leading to zero phase space fluxes (detailed equilibrium) and, respectively, to zero parallel current.
In the presence of radial gradient, the distribution function of trapped particles is not constant on the flux surface anymore.
It is larger in the region with positive bounce averaged radial drift,
and smaller where it is negative. In stellarator symmetry,
it is obvious that particles are detrapped in those regions in opposite directions,
thus leading to the overall shift of passing particle distribution and, respectively, to the parallel current.
Destroyed symmetry of the ripple wells is
described by the source term $s_{(1)}$, Eq.~\eq{s_k},
which generates ``anti-particles'' in the wells with $\langle v_g^r\rangle_b > 0$
and ``particles'' where $\langle v_g^r\rangle_b < 0$. In the asymptotical $1/\nu$ regime, these particles and anti-particles
produced by the source term $s_{(1)}$ within ripple wells can leave these wells only through boundary layers between classes,
where they are re-distributed and annihilated. As we saw in Sections~\ref{sssec:dirprob} and~\ref{sssec:dirprob_num},
parallel flows within these boundary layers make a contribution of the order one to the equilibrium Pfirsch-Schl\"uter current.
This changes at finite collisionality where ``relevant ripples'' appear with both limiting local maxima entering the
trapped-passing boundary layer. Re-distribution of particles and anti-particles between ``relevant'' ripples occurs then
through the trapped passing boundary layer (this layer is empty in the asymptotical $1/\nu$ regime) causing the parallel
flow there and respective asymmetry of the distribution function over the pitch parameter $\lambda$. Such an odd
distribution in the boundary layer,
$g^{\text{odd}}$,
serves as a boundary condition which shifts from zero the odd part of the
distribution function in the whole passing particle domain thus generating a bootstrap current.

We can estimate
$g^{\text{odd}}$
in the boundary layer
using the argument at the end of Section~\ref{sssec:dirprob} as
\mbox{$g_{\text{odd}} \sim H_j(\eta_{\text{loc}})/\delta\eta_{\rm ref}$} where
$H_j(\eta_{\text{loc}})$ is the collisional flux through class boundary
$\eta=\eta_{\text{loc}}\equiv\max(\eta_j,\eta_{j+1})$ of the
relevant ripple (which corresponds to the off-set domain in the adjoint problem)
with
$$
H_j(\eta)=-\int\limits_{\varphi_j}^{\varphi_{j+1}}\rd\varphi D_\eta \difp{g_{-1}}{\eta},
$$
and $g_{-1}$ being the leading order solution of direct problem~\eq{formsol_s1},
and $\delta\eta_{\rm ref}$ is the width of the trapped-passing boundary layer in the ``main region''.
Since the odd part of passing particle distribution driven by $g_{\text{odd}}$ at the trapped-passing boundary
is independent of $\eta$ and is, therefore, equal to $g_{\text{odd}}$, we can
estimate the parallel current density off-set using Eq.~\eq{pacd} as
$j_\parallel^{\text{off}} \approx C_\parallel g_{\text{odd}}$,
and the respective off-set of the normalized bootstrap coefficient in Eq.~\eq{bootaver} as
$\lambda_{\text{off}} \sim g_{\text{odd}}/\rho_L \sim H_j(\eta_{\text{loc}})/(\rho_L \delta\eta_{\rm ref})$.
Naturally, the same result is obtained for the off-set~\eq{lambda_off} in the adjoint problem
using for $w^{(j)}_{\text{off}}$ the second equality~\eq{w_off} and for $g_{\text{off}}^{(j)}$ Eq.~\eq{define_x} where $\alpha_j^\pm \sim 1$
in case of the maximum off-set, and constant~\eq{C0} is
$$
C_0 \approx
\frac{2\eta_b \langle B^2\rangle}{\delta\eta_{\rm ref}^2}\int\limits_{\varphi_0}^{\varphi_N}\frac{\rd\varphi}{B^\varphi}
\sim
\frac{B}{\delta\eta_{\rm ref}^2}\int\limits_{\varphi_0}^{\varphi_N}\frac{\rd\varphi}{B^\varphi}.
$$
Note that effect of particle precession reducing the off-set in the adjoint problem is also the same in the direct
problem where it causes re-distribution of trapped ``particles'' and ``anti-particles''
between trapping domains via the rotation within the flux surface
thus reducing the collisional fluxes into class boundaries as compared to their maximum $1/\nu$
values $H_j(\eta_{\text{loc}})$.

Recalling now definitions~\eq{bas1_defs} and~\eq{s_k} of the normalized collisional velocity space flux $H_j$
(see also Eqs.~\eq{w_off} and~\eq{w_off_viabavr}) we can approximately relate $H_j$
to the radial correlation length in the off-set well
$\delta r_{cj} \sim \tau_{dj} \langle v_g^r\rangle_{bj}$ as follows,
$H_j \sim \tau_{bj} \langle v_g^r\rangle_{bj} \Delta\eta_j \sim \delta r_{cj} \delta\eta_j^2/\Delta\eta_j$
where $\tau_{bj}$ and $\langle v_g^r\rangle_{bj}$ are bounce time and bounce averaged velocity~\eq{bavr},
$\tau_{dj} \sim \tau_{bj} (\Delta\eta_j/\delta\eta_j)^2$ is the detrapping time,
$\Delta\eta_j\sim\varepsilon_M\eta_b$
is the off-set well depth, and $\delta\eta_j$ is boundary layer width in the off-set domain
(see Eqs.~\eq{detaj_Qoff} and~\eq{delta_eta_fund}).
Thus, $g_{\text{odd}} \sim \delta r_{cj} \delta\eta_j/(A_j \Delta\eta_j)$ with the aspect ratio $A_j=\delta\eta_{\rm ref}/\delta\eta_j$,
and we can estimate current density as follows,
\be{curdensest}
j_\parallel^{\text{off}} \sim C_\parallel g_{\text{odd}}
\sim \frac{c \delta\eta_j\eta_b}{\rho_L A_j \Delta\eta_j}\left(p(r+\delta r_{cj})-p(r)\right).
\ee
This formula actually reverses local transport ansatz~\eq{superforces} which assumes infinitesimal correlation length,
and manifests the fact that variation of the distribution function on a given flux surface (along the field line) from its
equilibrium value is because its value in the ripple wells is determined by the values of equilibrium distribution
function at the neighboring flux surfaces displaced by the correlation length (strictly speaking, connection of
parallel current density to the pressure gradient is non-local, $j_\parallel^{\text{off}} = \int\rd r^\prime K(r,r^\prime) p(r^\prime)$,
with the kernel $K(r,r^\prime)$ localized within $|r^\prime-r| \lesssim \delta r_{cj}$). Thus, this variation
and resulting parallel current need not
disappear at the magnetic axis where $\rd p/\rd r = 0$. Expanding $p(r+\delta r_{cj})$ to the next order, we obtain
parallel current on axis as
\be{curdensest_axis}
j_\parallel^\text{axis}
\sim
\frac{c \eta_b \langle v_g^r\rangle_{bj}^2 \tau_{bj}^2}{\rho_L A_j}
\left(\frac{\Delta\eta_j}{\delta\eta_j}\right)^{3}\frac{\rd^2 p}{\rd r^2}
\sim
\frac{c \eta_b \rho_L \varepsilon_M^{5/4}}{A_j N_{\rm tor}^{1/2}\nuast^{3/2}}
\frac{\rd^2 p}{\rd r^2},
\ee
where we estimated $\langle v_g^r\rangle_{bj} \sim v \rho_L/R$,
$\tau_{bj} \sim R/(v N_{\rm tor} \varepsilon_M^{1/2})$, $\Delta\eta_j \sim \eta_b\varepsilon_M$
and $\delta\eta_j \sim \eta_b \left(\nuast/N_{\rm tor}\right)^{1/2}\varepsilon_M^{1/4}$,
with $N_{\rm tor}$ being the number of ripples.
Further denoting $\rd^2 p / \rd r^2 = p / L^2_{pa}$ and normalizing Eq.~\eq{curdensest_axis}
to a typical tokamak bootstrap current value
$j_b^{\rm tok} \sim c \eta_b \iota^{-1}\varepsilon_t^{-1/2}\rd p /\rd r
= c \eta_b p /(\iota\varepsilon_t^{1/2} L_p)$, we obtain
\be{curdensest_axis_norm}
\frac{j_\parallel^\text{axis}}{j_b^{\rm tok}}
\sim
\frac{\iota\varepsilon_t^{1/2} \varepsilon_M^{5/4}}{A_j N_{\rm tor}^{1/2}}
\left(\frac{L_p}{L_{pa}}\right)^2
\frac{\rho^\ast}{\nuast^{3/2}},
\ee
where $\rho^\ast=\rho_L / L_p$.
Although this expression contains a small parameter $\rho^\ast$ in the numerator, it
has another small parameter $\nuast^{3/2}$ in the denominator so that current on axis
is not vanishingly small. Estimating for AUG parameters in the presence of the helical core with
$\varepsilon_M \sim 0.01$,
$\iota \sim 0.5$, $\varepsilon_t \sim 0.25$, $A_j = 1$, $N_{\rm tor} = 2$,
and central density $n_e = 3\cdot 10^{13}$~cm$^{-3}$
and temperature $T_e = 5$ KeV resulting in $\rho^\ast=\rho_e^\ast \sim 2\cdot 10^{-4}$ and $\nuast \sim 3.5\cdot 10^{-3}$,
we get $\rho^\ast \nuast^{-3/2} \approx 1$ so that
$j_\parallel^\text{axis}/j_b^{\rm tok} \sim 5\cdot 10^{-4} (L_p/L_{pa})^2$. Thus, for peaked $T_e$ profiles
(e.g., in the presence of central ECRH) peaking factor $L_p/L_{pa} \sim 40$ is needed to make these currents comparable.
Of course, in order to achieve $j_\parallel^\text{axis}$,
special conditions on the magnetic geometry are required in order to create,
in particular, difference in correlation length $\delta r_j$ in different ripples.

Note that we have retained in the estimates~\eq{curdensest_axis} and~\eq{curdensest_axis_norm} only the off-set effect
but ignored the usual bootstrap mechanism where radial correlation length $\tau_{cj}$ corresponds to the banana width
which is much smaller for electrons than $\tau_{cj}$ resulting from the $1/\nu$ transport. However, this usual
mechanism is not small for the ions whose radial orbit extent and, respectively, radial correlation length is much
larger (with the orbits near the axis comprising a variety of classes even in an axisymmetric
tokamak~\citep{shaing15-125001,buchholz22-012012}), and, more importantly, who retain due to their high mass
nearly all the momentum within the species during Coulomb collisions.
For the axisymmetric devices, this mechanism generates the
parallel ion flow (generally different from the $E\times B$ flow) but not the current since this flow is
fully compensated by the electron flow driven by the collisional friction.
However, if the toroidal ripple is present, it brakes
the electrons leading to some steady current on axis if the ion flow is finite there.
The latter is affected by ripple too via the neoclassical toroidal viscous
torque~\citep{shaing15-905810203,albert16-082515,martitsch16-074007} which tends to bring the toroidal rotation
velocity towards the (finite) intrinsic velocity driven by the ion temperature gradient. Due to the finite radial 
orbit width, finite gradients at surrounding flux surfaces result in finite ion rotation on axis which
finally leads to a finite equilibrium current there.
Generally, once the axial symmetry of the tokamak magnetic field is violated, there appear various mechanisms resulting
in finite non-inductive toroidal current on axis which is necessary for the steady state tokamak operation, 
see, e.g., \citep{weening92-159}. At the same time, mechanisms discussed here are rather weak for that purpose which is
better achieved with help of stellarator-tokamak hybrids~\citep{ku09-082506,henneberg24-L022052,liang25-026033}.

\section{Conclusion}
\label{sec:conclusion}

To sum up, we can conclude that the bootstrap current
in arbitrary 3D toroidal fields
does not converge to the asymptotical Shaing-Callen limit in the $1/\nu$ transport regime.
The collisional current off-set being the difference between the actual bootstrap current and its
collisionless asymptotic stays of the order of the bootstrap current in the equivalent tokamak
and only oscillates aperiodically with changing $\log\nuast$.
Moreover, for a set of configurations with aligned local magnetic field maxima (such that the global maximum
is achieved at the flux surface on a line rather than at one or few points) the bootstrap current off-set
diverges as $\nuast^{-1/2}$, as long as the condition of equivalent ripples~\eq{omnigen}
is not fulfilled (this condition is naturally fulfilled in axisymmetric and exactly quasi-symmetric fields).
In turn, in the presence of orbit precession, in particular,
due to a finite radial electric field, the off-set current
converges to zero
with reducing collisionality as $\nuast^{3/5} |\Omega^\vartheta|^{-1/2}$, which is in agreement
with the results of numerical modelling by various codes~\citep{beidler2011-076001}.
Also, in the case of aligned maxima, the off-set does converge -- not to zero but to a finite value
which exceeds the equivalent tokamak current by a large factor
$\sim (v/R\Omega^\vartheta)^{1/2} \sim \left(v_E^\ast\right)^{-1/2}$
where $v_E^\ast = c E/(v B) \ll 1$ is the perpendicular Mach number.

The missing convergence of bootstrap current in the $1/\nu$ regime to the analytical collisionless
asymptotic value is the consequence of two competing limits present in the problem.
Namely, analytical derivations assume explicitly or implicitly that the field line is closed after
a finite number of toroidal turns so that the mean free path strongly exceeds the period of the
field line closure.
As a consequence, the trapped-passing boundary
layer width in velocity space $\delta\eta_{\rm ref}$ is assumed infinitesimal such that all class transition boundary layers
centered around $\eta=\eta_c=1/B_{\text{loc}}$ coming from
local field maxima $B_{\text{loc}}$ which are below the global maximum $B_{\rm max}$
are clearly separated from the trapped-passing boundary layer centered
at $\eta=\eta_b=1/B_{\rm max}$.
Obviously, this assumption cannot be fulfilled at irrational surfaces where the field line closure period is indeed infinite
but the mean free path length stays finite, and where class transition boundary layers inevitably interact with trapped-passing
boundary layer leading to the bootstrap current off-set. This contradiction between two limits is manifested in the fractal
spatial structure of the collisionless solutions for the distribution function, which is the fractal within the irrational flux surface in
the adjoint approach~\citep{helander11-092505} and has a fractal dependence on radius (leading to bootstrap
resonances) also in the direct approach~\citep{shaing83-3315,boozer90-2408}.

The off-set which appears at any collisionality at irrational surfaces
(or, equivalently, long enough rational field lines)
results from the collisional particle exchange between trapped and passing particle domains in the
velocity space. This exchange occurs through the boundary layer which, in usual configurations with
global magnetic field maximum achieved in one or few points, is localized at the flux surface
in some vicinity (contact region or ``hot spot'') around this maximum such that field line returns
to this vicinity
after $N \propto \nuast^{-1/5} \ll 1$ toroidal turns. The off-set of the distribution function
formed between two return points is the largest from all configurations,
$g_\text{off} \propto \nuast^{-3/5}$, but it does not lead to the divergence of bootstrap
current because the normalized integral of bounce averaged velocity over the long off-set domain,
$w_\text{off} \propto \nuast^{3/5} \ll 1$ is the smallest and balances the former in
resulting bootstrap coefficient $\lambda_\text{off} \propto g_\text{off} w_\text{off}$.
This compensation disappears in case of nearly aligned maxima, where boundary layer contact regions
turn into stripes with much larger area than ``hot spot'' in the usual case. These contact regions
cut the field lines into much shorter off-set domains with the length about
a single toroidal period, such that $w_\text{off}$ does not scale with collisionality anymore,
but the distribution function still does, $g_\text{off} \propto \nuast^{-1/2}$.
Although this scaling is weaker than in the usual case, missing compensation leads to $\nuast^{-1/2}$
divergence of the resulting bootstrap current. This diverging part is respectively reduced if bounce
averaged drift is minimized for trapped particles, what is naturally the case in quasi-symmetric
configurations.

For the analysis of boundary layer effects and resulting off-set of bootstrap current
we have developed a simplified propagator method, and obtained with its help, besides the condition
of equivalent ripples~\eq{omnigen} formulated earlier in~\citep{helander11-092505} as a property
of ideal quasi-isodynamic stellarators leading there to a tokamak like bootstrap effect,
also the expression~\eq{dimoff_B} for the off-set of the
distribution function in case of imperfectly aligned local maxima
and slightly violated ripple equivalence condition~\eq{dimoff_B}
together with respective estimate~\eq{lambda_off_small}
of the resulting off-set of bootstrap coefficient in case of the devices with non-optimized
bounce-averaged radial drift of trapped particles. Actually, the case of nearly aligned maxima
is the most interesting from the point of view of good fusion alpha particle confinement.
This is naturally the case in quasi-symmetric configurations and in quasi-isodynamic
configurations with poloidally closed of contours of $B$.
As it can be seen from condition~\eq{lambda_off_small}, the way to avoid the off-set of bootstrap
current in such configurations is to align the maxima accurately and to realize ripple equivalence
condition~\eq{omnigen}.
The latter condition is almost the same as the condition of being tied to the flux
surface contours of the parallel adiabatic invariant $J_\parallel$ for barely passing particles which,
once realized in case of perfectly aligned maxima, would prevent collisionless trapped-passing
particle transitions and resulting stochastic transport.
The essential difference of the respective integral $I_j$ from $J_\parallel=m\oint \rd l\,v_\parallel$
is only in an extra $1/B$ factor in the sub-integrand.
Both conditions are naturally satisfied if quasi-symmetry conditions are realized exactly, which, however,
cannot be achieved at all flux surfaces~\citep{garren91-2822}. More generally, 
these conditions are fulfilled simultaneously
for perfectly omnigeneous fields~\citep{cary97-3323,helander09-055004}
being also an idealization~\citep{cary97-3323}.

As a curious consequence of bootstrap off-set,
we have shown in Section~\ref{sec:boot_axis} a possibility of a bootstrap effect at the magnetic axis
due to rather large radial correlation length of the orbits $\delta r_c$ at low plasma collisionalities
where it scales in the $1/\nu$ regime as $\delta r_c \propto \rho_L / \nuast$ and
strongly exceeds the Larmor radius $\rho_L$.
Thus, the bootstrap current on axis being vanishingly small
in axisymmetric fields where, due to a vanishing gradient and $\delta r_c \sim \rho_L$,
it scales as $\rho_{Le}^2/L_p^2=(\rho^\ast_e)^2$,
can have much larger values in the presence of 3D perturbations (essentially, of a field ripple) on axis.

Since the bootstrap effect is often not desired in stellarators where it modifies the $\iota$ profile
and thus affects the position of an island divertor located at the plasma edge,
the minimization of the bootstrap coefficient $\lambda_{bB}$
is usually one of the goals in stellarator optimization.
The presence of a bootstrap off-set which is rather sensitive to plasma collisionality and the $\iota$ value
does not
make this task simpler, and there is little hope that the off-set effect fully vanishes at finite radial electric fields,
thus allowing to use the Shaing-Callen limit alone for the reactor optimization.
For typical reactor parameters, $B=5\cdot 10^4$~G, $n_e = 10^{14}$~cm$^{-3}$, $T_{i,e}= 10$~KeV,
$L_p=a=300$~cm and $\varepsilon_t = a/R =0.1$,
we obtain $\nuast = \pi R\nu_\perp/v$ and $\rho^\ast = \rho /L_p$ as $\nuast^e = 2\nuast^i = 0.001\div 0.1$
for the range contributing 95~\% to the convolution~\eq{enconv} of bootstrap coefficient $\bar D_{31}$,
and as $\rho^\ast_e \approx 2\cdot 10^{-5}$ and $\rho^\ast_i \approx 10^{-3}$, respectively.
Thus, using in Eq.~\eq{atten_om} $\Omega^\vartheta=\Omega_E=\rho_e^\ast v/(R\varepsilon_t)$ together with Eq.~\eq{N_deta}
we obtain the maximum value of the attenuation factor
$\delta_E \sim \left(\varepsilon_t \nuast/(N\rho^\ast)\right)^{1/2}$
as $\delta_E^{\rm max} \sim 1$ for the electrons, i.e. attenuation is practically unimportant for them and they mostly stay
in the $1/\nu$ regime, and $0.08 \le \delta_E \le 1.4$ for the ions which are, therefore, significantly affected by the
precession but still cannot be fully described by the collisionless limit.

As follows from Section~\ref{ssec:offset_imperfect}, the strongest bootstrap current
off-set due to nearly aligned maxima appears for the above parameters,
$\varepsilon_M \sim \varepsilon_t$, $\iota = 1$ and $N_\text{tor}=5$ if the mis-match
of maxima is
around
$\Delta B/B \sim 2.5 \%$ for the middle of the collisionality range
(around
$8 \%$ for the highest collisionality). According to Eq.~\eq{lambda_off_small},
bringing this off-set back to the level of bootstrap current in the equivalent tokamak would require to align
the maxima better than $\Delta B/B \sim 0.3 \%$ for the lowest collisionality and to fulfill
ripple equivalence condition to within $\Delta A_o \lesssim 50 \%$. The latter condition
does not appear to be too restrictive and can probably be achieved already when tying to the
flux surface $J_\parallel$ contours for barely passing particles.

Unfortunately, the off-set in the distribution function
cannot be further minimized by better fulfilling the above conditions
because there remains a residual off-set containing the term $\propto \log\nuast$
not accounted for in this paper.
Nevertheless, the resulting bootstrap current off-set can still be minimized
by reducing the integral bounce-averaged drift represented by factors $I_\cV$,
Eq.~\eq{w_off_booz}, which are well-defined in case of
aligned maxima by magnetic field geometry alone.
Certain minimization of these factors can be achieved
at mild collisionalities where off-set wells are relatively short
already as a by-product of minimization
of $1/\nu$ transport coefficients~\eq{dmono_epseff} via $\varepsilon_\text{eff}$.
This appears to be the case in \mbox{W-7X} ``standard'' configuration
where the off-set in $D_{31}^\ast$ shown in Fig.~3 of~\citep{kernbichler16-104001}
is of the order of Shaing-Callen value (10 \% of the equivalent
tokamak value) for collisionalities $\nu^\ast > 10^{-5}$, and where $\varepsilon_\text{eff}$ is
reduced by about an order of magnitude compared to a standard stellarator
(see Fig.~10 in~\citep{beidler2011-076001}). Strong $D_{31}^\ast$ off-set of the order one
shows up there only at very low collisionalities where it is due to the
(non-optimized) bounce-averaged drift of long bananas (off-set wells are long).

Of course, the above simple recipes assuming independent minimization of
purely geometrical quantities (asymptotic $\lambda_{bB}$ and $I_\cV$)
do not apply in the general case without strong alignment
of field maxima where multiple off-set well types contribute simultaneously
with their relative contributions depending on collisionality and precession velocity.
Minimization of bootstrap current in such cases requires direct numerical
evaluation of all transport coefficients and the resulting plasma parameter profiles
at finite plasma collisionality~\citep{geiger15-014004}.
For the evaluation of bootstrap coefficient in such cases,
computations of $\lambda_{bB}$ in the $1/\nu$ regime
would be useful if one accounts for the precession by bounce-averaged approach outlined in
Section~\ref{sec:precession}. A particular tool for this is NEO-2 which has been specially
designed for the effective evaluation of bootstrap current.
Another option would be a fast neoclassical code MONKES~\citep{escoto24-076030,escoto25-036017} which,
similarly to NEO-2, has an efficiently parallelizable algorithm.

It should be noted that the two above geometric criteria are naturally met in case of quasi-poloidal
symmetry~\citep{spong01-711}. For such an exact symmetry, $\lambda_{bB} \propto B_\vartheta = 0$ in the
absence of auxiliary current, while $I_\cV=0$ for all exact quasi-symmetries which can actually
be approached very closely~\citep{landreman22-035001}. More generally, these
criteria are met by ideal (omnigeneous) quasi-isodynamic fields~\citep{helander09-055004,helander11-092505}.
Recent realizations of a quasi-isodynamic field which are rather close to such an ideal
field~\citep{goodman23-905890504,goodman24-023010} do indeed show very low values of the bootstrap
coefficient. Note that mis-alignment of maxima in these configurations is clearly below 2.5\%.

\section*{Acknowledgements}
The authors gratefully acknowledge useful discussions with
Per Helander, Martin Heyn and Felix Parra, as well as discussions and regular help in the maintenance of \mbox{NEO-2} code from Andreas Martitsch and Rico Buchholz.
The authors are also grateful to Per Helander and co-authors for the data for Fig.~\ref{fig:henning}. We thank Andreas Hirczy for long-standing support on hardware and software for computations.

\section*{Funding}
This work has been carried out within the framework of the EUROfusion Consortium, funded by the European Union via the Euratom Research and Training Programme (Grant Agreement No 101052200 — EUROfusion). Views and opinions expressed are however those of the author(s) only and do not necessarily reflect those of the European Union or the European Commission. Neither the European Union nor the European Commission can be held responsible for them. We gratefully acknowledge support from NAWI Graz.

\section*{Declaration of interests}
The authors report no conflict of interest.

\section*{Data availability statement}
The data that support the findings of this study are openly available in Zenodo at \href{http://doi.org/10.5281/zenodo.14888543}{http://doi.org/10.5281/zenodo.14888543}, reference number 14888544.

\appendix
\section{Fourier amplitudes $\cV^s_{mn}$ of high harmonics.}
\label{sec:appendix1}
Integration by parts can be used to check that Fourier coefficients of the analytical periodic function (whose derivatives
of any order are finite) decay with harmonic index $m$ faster than any power $m^{-k}$ where $k$ is an arbitrary positive integer,
i.e. their decay is exponential. Power law decay of Fourier amplitude appears if some finite order derivative
turns into infinity at one or few points whose close vicinity mainly determines this amplitude at high $m$.
This is the case of function $\cV$ given by Eq.~\eq{cVdef} with $\eta_\text{loc}^{(j)}$ set to $\eta_b$,
so that first and higher order derivatives of $\cV$ over angles are infinite at the global maximum point. Using a simple model field
\be{simpmod}
B=B_{00}\left(1+\varepsilon_t \cos\vartheta+\varepsilon_h\cos\left(N_\text{tor}\varphi\right)\right),
\ee
which differs from~\eq{magfield_mod} by terms quadratic in $\varepsilon_t$ and $\varepsilon_M = 2 \varepsilon_h$ and shift
of the angles by $\pi$, and ignoring angular dependence of non-singular factors we present the Fourier amplitudes of $\cV$ as
\be{fouramp_cV}
\cV_{mn}^s\approx\frac{2 m}{B^2_{00}} U_{mn},
\qquad
 U_{mn} = \frac{1}{4\pi^2}\int\limits_{-\pi}^{\pi}\rd\vartheta\int\limits_{-\pi}^{\pi}\rd\varphi
\;
{\rm e}^{-im\vartheta-in\varphi}
U(\vartheta,\varphi),
\ee
where coefficients $U_{mn}$ are real due to stellarator symmetry of
\be{weprescV}
U(\vartheta,\varphi) = \left(
\varepsilon_t\left(1-\cos\vartheta\right)
+
\varepsilon_h\left(1-\cos\left(N_\text{tor}\varphi\right)\right)
\right)^{3/2}.
\ee
Integrating in Eq.~\eq{fouramp_cV} by parts over both angles and explicitly computing the resulting second
derivative $\partial^2 U /(\partial\vartheta\partial\varphi)$ we present
\be{UoverW}
U_{mn}=\frac{3\varepsilon_t\varepsilon_h N_\text{tor}}{16 mn}
\left(
W_{m+1,n+N_\text{tor}}
+
W_{m-1,n-N_\text{tor}}
-
W_{m+1,n-N_\text{tor}}
-
W_{m-1,n+N_\text{tor}}
\right),
\ee
where coefficients
\be{Wmn}
W_{m,n} =
\frac{1}{4\pi^2}\int\limits_{-\pi}^{\pi}\rd\vartheta\int\limits_{-\pi}^{\pi}\rd\varphi\;
{\rm e}^{-im\vartheta-in\varphi}
\left(
\varepsilon_t\left(1-\cos\vartheta\right)
+
\varepsilon_h\left(1-\cos\left(N_\text{tor}\varphi\right)\right)
\right)^{-1/2}
\ee
differ from zero for $n=k N_\text{tor}$ where $k$ is an integer. Reducing the integration over $\varphi$ for such $n$ to
a single field period, expanding in the sub-integrand $(1-\cos\vartheta) \approx \vartheta^2/2$ and
$1-\cos\left(N_\text{tor}\varphi\right) \approx N_\text{tor}^2\varphi^2/2$ and then extending the integration over both angles
to the infinite limits, what is justified for large $m$ and $n$, we get
\bea{Wmn_approx}
W_{m,n}
&\approx &
\frac{N_\text{tor}}{2\sqrt{2}\pi^2}
\int\limits_{-\infty}^{\infty}\rd\vartheta\int\limits_{-\infty}^{\infty}\rd\varphi\;
{\rm e}^{-im\vartheta-in\varphi}
\left(
\varepsilon_t\vartheta^2
+
\varepsilon_h N_\text{tor}^2\varphi^2
\right)^{-1/2}
\nonumber \\
&=&
\frac{N_\text{tor} I_W}{2\pi^2\sqrt{2(\varepsilon_t n^2 + \varepsilon_h N_\text{tor}^2 m^2)}},
\eea
where we changed the integration variables from $(\vartheta,\varphi)$ to polar variables $(\rho,\phi)$ as follows,
\be{polarint}
\vartheta = \left(m^2+\frac{n^2 \varepsilon_t}{\varepsilon_h N_\text{tor}^2}\right)^{-1/2}\rho\cos\phi,
\qquad
\varphi = \left(n^2+\frac{m^2 \varepsilon_h N_\text{tor}^2}{\varepsilon_t}\right)^{-1/2}\rho\sin\phi,
\ee
in order to get
\be{IW}
I_W = \int\limits_0^{\infty}\rd \rho \int\limits_{-\pi}^{\pi}\rd\phi\; {\rm e}^{-i\rho\sin(\phi+\chi)}
= 2\pi \int\limits_0^{\infty}\rd \rho J_0(\rho) =2\pi.
\ee
Here $\chi = \text{atan}\left((\varepsilon_h/\varepsilon_t)^{1/2}m N_\text{tor}/n\right)$ and $J_0(\rho)$ is Bessel function
of the first kind.
For large $m$ and $n$ finite differences in~\eq{UoverW} can be approximated by derivatives,
\be{Umnapprox}
U_{mn}
\approx
\frac{3\varepsilon_t\varepsilon_h N_\text{tor}^2}{4 mn} \difp{^2}{m\partial n}W_{m,n}
\approx
\frac{9 \varepsilon_t^2 \varepsilon_h^2 N_\text{tor}^5}
{4\pi\sqrt{2}\left(\varepsilon_t n^2 + \varepsilon_h N_\text{tor}^2 m^2\right)^{5/2}}.
\ee
Expressing here modulation amplitudes via second derivatives of the field at the global maximum,
$$
\varepsilon_t = - \frac{1}{B_{00}} \difp{^2 B}{\vartheta^2},
\qquad
\varepsilon_h N_\text{tor}^2 = -\frac{1}{B_{00}} \difp{^2 B}{\varphi^2} = - \frac{1}{\beta B_{00}} \difp{^2 B}{\vartheta^2}
$$
where $\beta$ is defined in~\eq{expmax} and approximating $\eta_b \approx 1/B_{00}$ we present~\eq{Umnapprox} as
\be{Ufin}
U_{mn}
\approx
\frac{9\beta^{1/2}\eta_b^{3/2}}{4\pi\sqrt{2}}\left|\difp{^2 B}{\vartheta^2}\right|^{3/2}
\frac{N_\text{tor}}{\left(m^2+\beta n^2\right)^{5/2}}
\ee
which, together with $\eta_b \approx 1/B_{00}$ in the first~\eq{fouramp_cV}, results in Eq.~\eq{cV_spectrum}.

\ifjpp
   \bibliographystyle{jpp}
\else
   \bibliographystyle{plainnat}
\fi

\bibliography{paper_bootstrap}

\end{document}